\documentclass[]{fairmeta}
\usepackage[dvipsnames]{xcolor}
\usepackage[most]{tcolorbox}

\usepackage{siunitx}
\usepackage{longtable}
\usepackage{array}
\usepackage{tabularx}
\usepackage{csquotes}
\usepackage{wrapfig}
\usepackage{capt-of}

\title{SWE-sweep: Can Agents Autonomously Find and Fix Bugs?}

\author[1]{Kilian Lieret}
\author[1,2]{Jeffrey Jian Ma}
\author[1]{Rahul Kindi}
\author[1]{Yuxiang Wei}
\author[1,3]{Jeremy Ma}
\author[1]{Sten Sootla}
\author[1]{Parth Thakkar}
\author[1]{Chao Beyond Zhou}
\author[4]{Pengcheng Yin}
\author[4]{Rui Hou}
\author[1]{Ofir Press}
\author[1,5]{John Yang}

\affiliation[1]{Meta FAIR}
\affiliation[2]{Harvard University}
\affiliation[3]{University of Washington}
\affiliation[4]{Meta TBD}
\affiliation[5]{Stanford University}

\abstract{
Most existing software engineering benchmarks evaluate coding agents on concrete, well-specified tasks, commonly by providing a codebase together with a user-reported issue to resolve.
However, as users delegate increasingly broad outcomes to coding agents, the natural next step is for agents to determine not only how to perform useful work, but also what useful work needs to be done.
An agent entrusted with a repository should be able to decide what is broken, which problems matter, and how to solve them before they are reported.

We introduce \bench{}, a benchmark for this open-ended setting.
Given a codebase containing many concurrent bugs and no information about their nature or location, an agent must autonomously discover and fix as many bugs as possible.
For each repository, we collect real GitHub issue--pull request pairs, then identify a single commit where the maximum number of bugs are present at the same time.
Each repair is evaluated against hidden tests from the corresponding pull request, along with the existing test suite to check for regressions.
Success requires agents to explore and understand a large codebase over long horizon work, repair bugs without introducing regressions, and manage interactions among fixes that are not independent.

\bench{} encompasses \qBenchmarkSubtaskCount{} bugs across \qBenchmarkNumLanguages{} programming languages and \qBenchmarkTaskCount{} repositories, including NumPy (Python), OpenCV (C++), and Lean 4 (Lean).
We evaluate \qEvaluatedLanguageModelCount{} LMs and find that the best agent resolves only \qBestModelMicroScore{} of all bugs, with the challenge generally lying in discovering bugs rather than repairing them.
We further study simple test-time scaling approaches and find that they are effective but yield diminishing returns.
\bench{} exposes a substantial gap between resolving reported issues and autonomously maintaining software.
}
 
\date{\today}
\correspondence{Kilian Lieret at \email{klieret@meta.com}}

\newcommand{\bench}{SWE-sweep}
\newcommand{\qEvaluatedLanguageModelCount}{7}

\definecolor{PromptBlue}{HTML}{1E5AA8}
\definecolor{PromptRed}{HTML}{C0392B}
\definecolor{PromptGreen}{HTML}{27AE60}
\definecolor{LinkBlue}{HTML}{1A6DCC}
\newcommand{\reflink}[2]{\textcolor{LinkBlue}{\href{#1}{#2}}}

\newcommand{\verbpromptboxfile}[2][PromptBlue]{\tcbinputlisting{
    colframe=#1,
    colback=#1!8,
    coltext=black,
    boxrule=0.8pt,
    arc=2pt,
    left=4pt,right=4pt,top=2pt,bottom=2pt,
    breakable,
    before skip=0.5\baselineskip,
    after skip=0.5\baselineskip,
    listing only,
    listing file={#2},
    listing options={
      basicstyle=\ttfamily\tiny\color{black},
      breaklines=true,
      breakatwhitespace=false,
      breakautoindent=false,
      breakindent=0pt,
      columns=fullflexible,
      keepspaces=true,
      literate={—}{{---}}3 {→}{{$\rightarrow$}}1,
      aboveskip=0pt,
      belowskip=0pt,
    }
  }}

\definecolor{CodeKeyword}{HTML}{0B5FA5}
\definecolor{CodeString}{HTML}{8A5A00}
\definecolor{CodeComment}{HTML}{6A737D}

\lstdefinelanguage{PythonMod}[]{Python}{morekeywords={with,as,assert,yield,None,True,False,nonlocal,async,await,
                elif,finally,raise,except,import,from,lambda,global,del},
}

\lstdefinelanguage{Go}{morekeywords={func,package,import,var,const,type,struct,interface,map,chan,
                go,defer,if,else,for,range,return,switch,case,default,break,
                continue,nil,true,false},
  sensitive=true,
  morecomment=[l]{//}, morecomment=[s]{/*}{*/},
  morestring=[b]", morestring=[b]`,
}
\lstdefinelanguage{JavaScript}{morekeywords={const,let,var,function,return,if,else,for,while,new,class,
                extends,super,this,typeof,instanceof,null,undefined,true,false,
                async,await,import,export,default,switch,case,break,continue,
                throw,try,catch,finally},
  sensitive=true,
  morecomment=[l]{//}, morecomment=[s]{/*}{*/},
  morestring=[b]', morestring=[b]",
}
\lstdefinelanguage{PlainText}{}

\lstdefinestyle{annotationdiff}{basicstyle=\ttfamily\tiny\color{black!85},
  keywordstyle=\color{CodeKeyword}\bfseries,
  stringstyle=\color{CodeString},
  commentstyle=\color{CodeComment}\itshape,
  showstringspaces=false,
moredelim=**[l][\bfseries]{+},
  morecomment=[f][\color{black!45}]{-},
  breaklines=true,
  breakatwhitespace=false,
  columns=fullflexible,
  keepspaces=true,
  literate={—}{{---}}3 {→}{{$\rightarrow$}}1,
  aboveskip=0pt,
  belowskip=0pt,
}

\newcommand{\annotationexample}[6][PromptGreen]{\tcbinputlisting{
    colframe=#1,
    colback=white,
    coltext=black,
    listing only,
    boxrule=0.8pt,
    arc=2pt,
    left=4pt,right=4pt,top=2pt,bottom=2pt,
    title={#3},
    fonttitle=\ttfamily\tiny\bfseries,
    colbacktitle=#1!14,
    coltitle=black!65,
    titlerule=0pt,
    toptitle=1.5pt,
    bottomtitle=1.5pt,
    before skip=0pt,
    after skip=3pt,
    listing file={#5},
    listing options={style=annotationdiff, language=#2},
  }{\normalfont\footnotesize\raggedright\textcolor{#1}{\textbf{\mbox{#4.}}}\enspace #6\par}}

\newcommand{\annotationrow}[2]{\par\addvspace{0.4\baselineskip}\noindent
  \begin{minipage}[t]{0.48\textwidth}\vspace{0pt}#1\end{minipage}\hfill
  \begin{minipage}[t]{0.48\textwidth}\vspace{0pt}#2\end{minipage}\par\addvspace{0.6\baselineskip}}
 \def\qBenchmarkTaskCount{100}
\def\qBenchmarkSubtaskCount{4,068}

\def\qBenchmarkMinSubtasksPerTask{4}
\def\qBenchmarkMaxSubtasksPerTask{239}

\def\qBenchmarkMinGradedSourceLocRounded{6k}
\def\qBenchmarkMaxGradedSourceLocRounded{1.9M}

\def\qBenchmarkMeanGoldPatchLines{16.3}
\def\qBenchmarkMeanTestPatchLines{37.0}
\def\qBenchmarkMeanFailToPassTests{1.6}
\def\qBenchmarkMeanPassToPassTests{50.4}

\def\qBenchmarkNumLanguages{22}

\def\qBugAnnotationPopulationCount{4,068}

\def\qBugCharacteristicsAnnotatedShare{98.9\%}

\def\qBugCharacteristicsUnannotatedShare{1.1\%}

\def\qBugTypeIncorrectOutputShare{41.3\%}

\def\qBugTypeRuntimeResourceShare{22.3\%}

\def\qBugTypeValidationDiagnosticShare{28.0\%}

\def\qBugEvidenceRepositoryGroundedCount{4,023}
\def\qBugEvidenceRepositoryGroundedShare{99.95\%}

\def\qBugEvidenceObviousOnlyCount{2}
\def\qBugEvidenceObviousOnlyShare{0.05\%}

\def\qTestExpectationSupportedShare{87.6\%}

\def\qExpectationPanelContradictionShare{12.4\%}
\def\qCompatibilityReviewedTasksCount{100}

 \def\qDiscoverabilityOutcomeModelCount{4}
\def\qDiscoverabilityOutcomeAnnotatedCount{4,121}
\def\qDiscoverabilityOutcomeAnyEditCount{1,175}

\def\qDiscoverabilityOutcomeAnyEditShare{28.5\%}
\def\qDiscoverabilityOutcomeAnyResolvedCount{252}
\def\qDiscoverabilityOutcomeAnyResolvedKnownCount{4,076}
\def\qDiscoverabilityOutcomeAnyResolvedShare{6.2\%}

\def\qBenchmarkMedianValidHunkAgeDays{481}
\def\qBenchmarkMedianValidFixLeadDays{293}

\def\qBenchmarkMedianGoldSourcePatchLines{7}

\def\qBenchmarkMedianGoldSourcePatchFiles{1}

\def\qBenchmarkSourceFileOverlapShare{56.3\%}

\def\qBenchmarkDifferentSameFileRepairContentShare{96.7\%}
\def\qBenchmarkDistantSameFileRepairShare{80.7\%}
 \def\qRepairSubtaskCount{24,268}

\def\qRepairAppliesShare{59.1\%}
\def\qRepairAcceptedShare{7.4\%}
\def\qRepairPrimaryModelShare{59.2\%}
\def\qRepairMedianTurns{15}
\def\qRepairGaveUpShare{39.3\%}

\def\qRebasedBothShare{22.7\%}
\def\qRebasedTestOnlyShare{42.7\%}

\def\qBenchmarkRebaseAnchorOnlyShare{60.4\%}

\def\qRebaseJudgeBugCount{1,806}
\def\qRebaseJudgeRejectShare{2.3\%}
\def\qRebaseResolvedShare{5.0\%}
\def\qNonRebaseResolvedShare{4.3\%}
\def\qRebaseResolvedDelta{+0.6\,pp}
\def\qRebaseResolvedCI{[-0.8, +2.2]\,pp}
 \def\qMainResultsModelCount{10}
\def\qMeasuredMainResultsVariantCount{10}
\def\qMainResultsRepositoryCount{100}
\def\qBestModelMicro{Sol 5.6 (xhigh)}
\def\qBestModelMicroScore{4.7\%}
\def\qOpusXHighMicroScore{1.3\%}
\def\qLunaDefaultMicroScore{0.5\%}
\def\qLunaHighMicroScore{1.4\%}
\def\qLunaXHighMicroScore{2.5\%}
\def\qSolXHighUnderTwentyFiveBugsMicroScore{14.9\%}
\def\qSolXHighAtLeastSeventyFiveBugsMicroScore{2.2\%}
\def\qSolXHighUnderHundredKLocMicroScore{13.4\%}
\def\qSolXHighOverFiveHundredKLocMicroScore{1.2\%}

 \def\qAblationBugCount{684}
\def\qAblationSolIssueTextsBugMicroScore{75.3\%}
\def\qAblationSolFilenamesBugMicroScore{11.0\%}
\def\qAblationLunaFilenamesBugMicroScore{4.8\%}
\def\qAblationSolLineRangesBugMicroScore{19.2\%}
\def\qAblationLunaLineRangesBugMicroScore{10.7\%}
\def\qAblationSolLineRangesSeparateBugMicroScore{39.2\%}
\def\qAblationLunaLineRangesSeparateBugMicroScore{31.3\%}

\def\qAblationLunaIssueTextsLineRangesBugMicroScore{43.8\%}
\def\qAblationSolIssueTextsLineRangesSeparateBugMicroScore{87.3\%}
\def\qAblationLunaIssueTextsLineRangesSeparateBugMicroScore{82.9\%}

\def\qTrajectoryAnyModelTargetEditShare{28.3\%}

\def\qTrajectoryOpusFilePrecision{27.9\%}

\def\qTrajectoryOpusRegionPrecision{7.1\%}

\def\qTrajectorySolFilePrecision{17.3\%}

\def\qTrajectorySolRegionPrecision{2.1\%}

 \def\qBugProgressEndpointAction{140}
\def\qBugProgressOpusEndpointFileCoverage{35.8\%}
\def\qBugProgressOpusEndpointLocCoverage{23.8\%}
\def\qBugProgressOpusLineRecovery{77.6\%}
\def\qBugProgressSolEndpointFileCoverage{39.9\%}
\def\qBugProgressSolEndpointLocCoverage{9.0\%}
\def\qBugProgressSolLineRecovery{55.6\%}

\def\qBugProgressSolEndpointFilesRead{18.5\%}
\def\qBugProgressSolEndpointFilesEdited{6.2\%}
\def\qBugProgressSolEndpointFilesFixed{1.7\%}

 \def\qAgentBehaviorTaskCount{100}
\def\qAgentBehaviorRunCount{400}
\def\qAgentBehaviorBugCount{4,165}

\def\qAgentBehaviorCheckpointCount{5,173}
\def\qAgentBehaviorUnknownFinalCount{164}

\def\qAgentBehaviorOpusReadActionShare{52.5\%}
\def\qAgentBehaviorOpusWriteActionShare{25.5\%}
\def\qAgentBehaviorOpusReadFiles{159.3}

\def\qAgentBehaviorOpusAnyEdit{6.0\%}
\def\qAgentBehaviorOpusResolved{1.3\%}

\def\qAgentBehaviorOpusRegressionRuns{6}
\def\qAgentBehaviorOpusTransientBugs{5}
\def\qAgentBehaviorOpusRunsBelowPeak{4}
\def\qAgentBehaviorOpusPostImprovement{25.9\%}

\def\qAgentBehaviorSolContactRuns{98}
\def\qAgentBehaviorSolAnyRead{38.5\%}
\def\qAgentBehaviorSolAnyEdit{21.5\%}
\def\qAgentBehaviorSolResolved{5.0\%}

\def\qAgentBehaviorSolRegressionRuns{17}
\def\qAgentBehaviorSolTransientBugs{12}
\def\qAgentBehaviorSolRunsBelowPeak{7}
\def\qAgentBehaviorSolPostImprovement{24.8\%}
\def\qAgentBehaviorTerraReadActionShare{35.9\%}
\def\qAgentBehaviorTerraWriteActionShare{37.8\%}
\def\qAgentBehaviorTerraReadFiles{48.9}
\def\qAgentBehaviorTerraContactRuns{86}
\def\qAgentBehaviorTerraAnyRead{23.4\%}

\def\qAgentBehaviorTerraRegressionRuns{5}
\def\qAgentBehaviorTerraTransientBugs{5}
\def\qAgentBehaviorTerraRunsBelowPeak{3}
\def\qAgentBehaviorTerraPostImprovement{25.9\%}

\def\qAgentBehaviorLunaRegressionRuns{14}
\def\qAgentBehaviorLunaTransientBugs{6}
\def\qAgentBehaviorLunaRunsBelowPeak{5}
\def\qAgentBehaviorLunaPostImprovement{37.6\%}
\def\qAgentBehaviorMinMedianFinalLines{14.0}

\def\qAgentBehaviorMaxMedianFinalLines{507.5}

\def\qAgentBehaviorMinMedianChurnLines{29.0}

\def\qAgentBehaviorMaxMedianChurnLines{635.0}

\def\qAgentBehaviorMinPostImprovement{24.8\%}

\def\qAgentBehaviorMaxPostImprovement{37.6\%}

\def\qAgentBehaviorMinHalfwayReadEndpoint{65.6\%}
\def\qAgentBehaviorMaxHalfwayReadEndpoint{76.6\%}
\def\qAgentBehaviorMinHalfwayEditEndpoint{12.8\%}
\def\qAgentBehaviorMaxHalfwayEditEndpoint{53.0\%}
\def\qAgentBehaviorMinHalfwayResolvedEndpoint{20.5\%}
\def\qAgentBehaviorMaxHalfwayResolvedEndpoint{52.6\%}
\def\qAgentBehaviorRegressionRuns{42}
\def\qAgentBehaviorRegressionEvents{93}
\def\qAgentBehaviorTransientBugs{28}
\def\qAgentBehaviorRunsBelowPeak{19}
\def\qAgentBehaviorNoModelCount{3,849}
\def\qAgentBehaviorNoModelShare{93.4\%}
\def\qAgentBehaviorExactlyOneModelCount{159}

\def\qAgentBehaviorAllFourModelsCount{8}

 \def\qSequentialTaskCount{5}
\def\qSequentialAgentCount{10}
\def\qSequentialSolScore{2.1\%}
\def\qSequentialLunaXHighOne{1.6\%}
\def\qSequentialLunaXHighFinal{3.5\%}
\def\qIndependentLunaXHighFinal{4.1\%}
\def\qSequentialLunaDefaultOne{0.1\%}
\def\qSequentialLunaDefaultFinal{0.9\%}
\def\qIndependentLunaDefaultFinal{0.6\%}

\def\qStoppingStepLimit{1000}
\def\qStoppingWallHours{12}
\def\qStoppingMaxActions{648}
\def\qStoppingMaxHours{5.5}
\def\qStoppingMaxWallShare{45.6\%}
\def\qStoppingSubmitted{400}
\def\qStoppingGitCheckRuns{385}
\def\qStoppingOpusMedActions{324}  \def\qStoppingOpusMaxActions{502}
\def\qStoppingOpusMedHours{1.3}    \def\qStoppingOpusMaxHours{3.9}
\def\qStoppingSolMedActions{322}   \def\qStoppingSolMaxActions{648}
\def\qStoppingSolMedHours{1.6}     \def\qStoppingSolMaxHours{5.5}
\def\qStoppingTerraMedActions{110} \def\qStoppingTerraMaxActions{226}
\def\qStoppingTerraMedHours{0.3}   \def\qStoppingTerraMaxHours{1.9}
\def\qStoppingLunaMedActions{214}  \def\qStoppingLunaMaxActions{334}
\def\qStoppingLunaMedHours{1.5}    \def\qStoppingLunaMaxHours{3.1}

\def\qStoppingBudgetCiting{0}

\def\qCoverageOpusRuns{98}       \def\qCoverageOpusMedian{3.8\%}   \def\qCoverageOpusMax{81.8\%}
\def\qCoverageSolRuns{100}       \def\qCoverageSolMedian{6.9\%}    \def\qCoverageSolMax{77.8\%}
\def\qCoverageTerraRuns{100}     \def\qCoverageTerraMedian{2.6\%}  \def\qCoverageTerraMax{63.6\%}
\def\qCoverageLunaRuns{100}      \def\qCoverageLunaMedian{4.4\%}   \def\qCoverageLunaMax{62.2\%}

\def\qSearchImportSubcohort{29}
\def\qSearchImportRateLow{79\%}
\def\qSearchImportRateHigh{86\%}
\def\qSearchTestSuiteShareLow{5\%}
\def\qSearchTestSuiteShareHigh{28\%}

 \def\qHumanReviewBugCount{50}

\def\qHumanReviewVerdictCount{200}

\def\qHumanReviewChangedToAcceptCount{12}

\def\qHumanReviewInitialMajorityAcceptCount{47}
\def\qHumanReviewInitialMajorityAcceptShare{94\%}

\def\qHumanReviewInitialRejectVerdictCount{11}

\def\qHumanReviewInitialUnsureVerdictCount{11}

\def\qHumanReviewMajorityAcceptCount{49}
\def\qHumanReviewMajorityAcceptShare{98\%}

\def\qHumanReviewRejectVerdictCount{5}

\def\qHumanReviewUnsureVerdictCount{5}

\def\qHumanReviewInitialDiscoverableUnanimousAcceptCount{42}
\def\qHumanReviewInitialDiscoverableMajorityAcceptCount{49}

\def\qHumanReviewDiscoverableMajorityAcceptCount{50}

\def\qHumanReviewDiscoverableRejectVerdictCount{3}
\def\qHumanReviewDiscoverableUnsureVerdictCount{1}

\def\qHumanReviewSingleBugUnanimousAcceptCount{49}
\def\qHumanReviewSingleBugMajorityAcceptCount{50}

\def\qHumanReviewAppropriateTestsUnanimousAcceptCount{47}
\def\qHumanReviewAppropriateTestsMajorityAcceptCount{50}

\def\qHumanReviewAppropriateTestsRejectVerdictCount{3}

\def\qHumanReviewPreservesExpectationsUnanimousAcceptCount{50}

\def\qHumanReviewMinTotalHours{0.9}
\def\qHumanReviewMaxTotalHours{2.5}
 
\begin{document}

\maketitle

\section{Introduction}
Language Models (LMs) have become remarkably capable at software engineering. Given a concrete task, today's agents can navigate large codebases and produce substantial functional changes. 
Typical tasks include fixing bugs, developing features, and completing tasks in a terminal environment~\citep{jimenez2024swebenchlanguagemodelsresolve,merrill2026terminalbenchbenchmarkingagentshard}.
As LMs' coding abilities have improved, they have been applied to increasingly demanding tasks, including implementing significant features~\citep{research2026composer2technicalreport,huang2026deepswe}, performing project-scale work~\citep{desai2026swemarathonagentsautonomouslycomplete}, and writing entire programs from scratch~\citep{ding2026nl2repobenchlonghorizonrepositorygeneration,yang2026programbenchlanguagemodelsrebuild}.

But increasing the amount and complexity of defined work that agents are asked to carry out is only one dimension of making them more useful as tools for developers.
A complementary direction is to ask agents to operate with increasingly little specification and identify useful work themselves.
For example, rather than waiting for a developer to notice, diagnose, and turn a software failure into an open bug report, such an agent would proactively inspect the codebase, find issues, and repair them.

Unlike issue-conditioned repair, much of the difficulty of this setting arises before implementation begins. 
Agents must systematically explore a large codebase and distinguish genuine defects from unusual but intended behavior, all before the more familiar task of repairing them can begin.
The capabilities exercised in this setting, including reasoning about user intent, identifying subtasks, and working on interrelated problems throughout a long trajectory, also transfer to any open-ended software engineering tasks where the agent pursues broad objectives.

\begin{figure}[tb!]
\centering
\includegraphics[width=\textwidth]{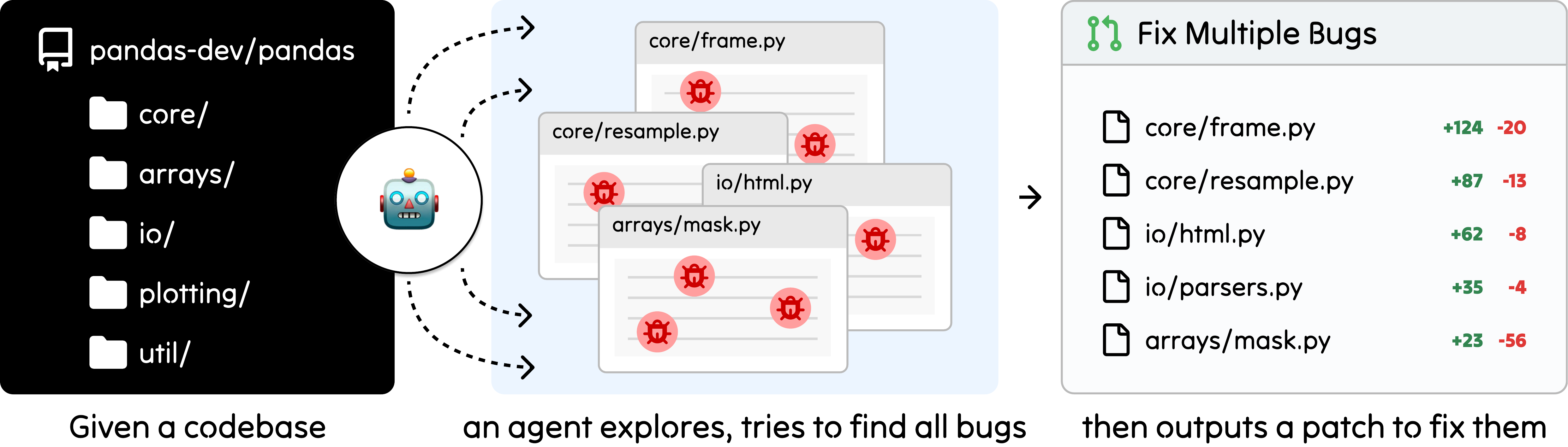}
\captionof{figure}{
\bench{} evaluates coding agents on their ability to identify and fix \emph{unspecified} bugs within a software repository.
An agent is asked to scour a codebase for issues and resolve them without \textit{any} information about specific faults. The agent is then graded using a suite of hidden tests.
}
\label{fig:preview}
\end{figure}

We introduce \bench{} to study this setting.
In \bench{}, an agent is given an entire codebase and asked to identify and resolve as many bugs as possible with no hints about the nature or location of any bugs. The agent must do this without introducing regressions.
We score the submitted patch against a reference set of previously identified bugs, counting how many the agent successfully repairs. Unlike our zero information setting, prior work~\citep{li2026active,liu2026testexplora,lee2026secbenchprolanguagemodels} typically provides agents with substantial information about the bugs it needs to fix, such as failure types, failure signals, or where the bugs occur.

\bench{} tasks are constructed by crawling and filtering repositories from GitHub.
Per repository, we collect all pull requests (PRs) with a linked issue, then retain those that ship tests and fix a bug.
We then identify a commit of the repository where the most bugs are concurrently present, and use it as the starting state of a task instance.
Each bug's PR tests are used to verify whether it has been fixed. 
Besides many quality filters shared with other benchmarks, we apply extensive filtering to retain only bugs that can be discovered from reading the repository alone.
While any history-derived benchmark necessarily under-counts the bugs present in a repository, restricting \bench{} to defects confirmed by an upstream fix provides strong evidence that the observed behavior was considered erroneous by the repository maintainers.

\bench{} is made up of \qBenchmarkTaskCount{} repositories that contain a total of \qBenchmarkSubtaskCount{} bugs.
The final test set covers 22 unique programming languages and includes notable libraries such as pandas, Lean 4, and scikit-learn.
We evaluate \qEvaluatedLanguageModelCount{} LMs, including GPT-5.6 Sol and Claude Opus 5.
Even the strongest model, GPT-5.6 Sol (xhigh), resolves only \qBestModelMicroScore{} of all bugs, leaving substantial room for improvement in autonomous bug discovery and repair.
Across several analyses, agents are considerably more successful at repairing discovered bugs than at finding them.

We open source \bench{} to enable the community to reproduce and build on our investigations.
 \section{\bench}
\label{sec:benchmark}

We first formalize the \bench{} task in \S\ref{sec:benchmark:formulation}.
We then discuss the construction process in \S\ref{sec:benchmark:construction}, and finally review benchmark statistics in \S\ref{sec:benchmark:statistics} and features in \S\ref{sec:benchmark:features}.

\subsection{Task Formulation}
\label{sec:benchmark:formulation}
Given a code repository, a coding agent is asked simply to discover and fix all bugs in the repository.
The agent is not given \textit{any} additional information about any specific bug's location, description, or tests.
The agent is expected to modify the codebase such that all bugs are resolved.

\textbf{Evaluation.} To evaluate correctness, we run the agent's submitted codebase against two sets of tests.
First, we restore the repository's original test suite and run it to verify no existing behavior was broken. 
Second, for each bug, we run a set of hidden tests; at least one of these tests fails on the unmodified codebase and passes once the bug is fixed (fail-to-pass).
A bug is considered resolved if all its hidden tests pass and the original suite still passes.

\begin{figure}[t]
    \centering
    \begin{minipage}[t]{0.66\linewidth}
        \vspace{0pt}
        \centering
        \includegraphics[width=\linewidth]{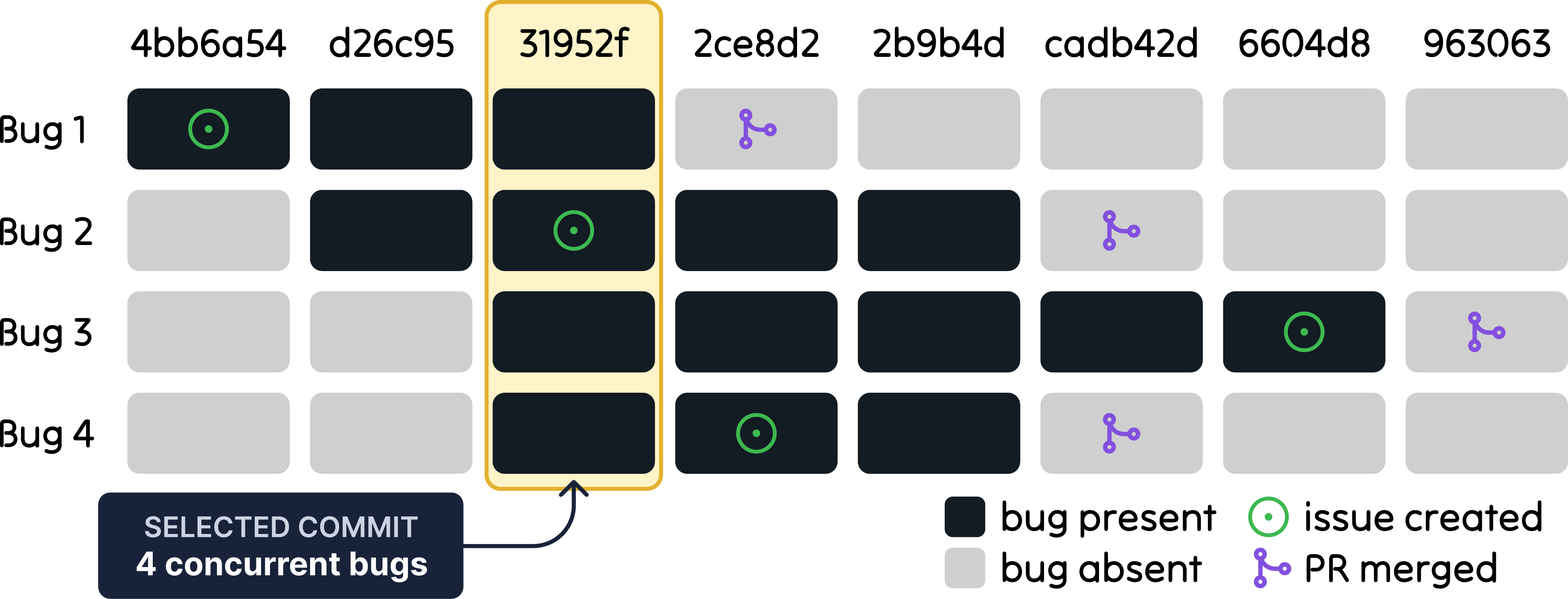}
        \captionof{figure}{ 
        \bench{} tasks are constructed from a repository's \emph{buggiest} commit.
        Rows represent bugs, columns are commits; dark cells indicate a bug is present at a commit.
        The \texttt{31952f} codebase becomes the task start state (\emph{base commit}).
        }
        \label{fig:construction-matrix}
    \end{minipage}
    \hfill
\begin{minipage}[t]{0.31\linewidth}
        \vspace{0pt}
        \centering
        \footnotesize
        \setlength{\tabcolsep}{3pt}
        \captionof{table}{
\bench{} construction pipeline.
        Bug candidate counts correspond to the total number of bugs retained per stage across all 100 repositories represented in the final evaluation dataset.
        }
        \label{fig:construction-funnel}
        \begin{center}
            \begin{tabular}{@{}l
    S[table-format=6.0,group-separator={,},group-minimum-digits=4]
    @{\,}r@{}}
\toprule
\textbf{Stage} & \multicolumn{2}{r@{}}{\textbf{Bug Candidates}} \\
\midrule
Candidates & 704386 & (100\%) \\
Pre-filter & 146520 & (20.8\%) \\
Patches apply & 19470 & (2.8\%) \\
Reproduces & 9057 & (1.3\%) \\
Quality control & 4068 & (0.6\%) \\
\bottomrule
\end{tabular}
 \end{center}
    \end{minipage}
\end{figure}

\subsection{Construction}
\label{sec:benchmark:construction}
For each repository, construction proceeds in three stages: First, we collect bugs from a real-world software project. Second, we select a suitable base commit that maximizes the number of concurrent bugs. Third, we apply additional repair and filter operations.
Full details appear in \S\ref{appx:benchmark:collection}.

\textbf{Identify usable repositories.}
For conversion to be feasible, a GitHub repository must have a test suite along with an ample ledger of issues and pull requests.
We select 100 codebases that satisfy these requirements and collectively represent a diverse mix of programming languages.

\textbf{Source bugs from real-world issue--PR pairs.}
Per repository, we crawl all closed issues and merged pull requests (PRs).
Echoing SWE-bench's collection strategy~\citep{jimenez2024swebenchlanguagemodelsresolve}, we identify \emph{bug candidates} as PRs with an associated issue, then check each PR for a meaningful fix (\emph{gold fix}) and accompanying tests (\emph{hidden tests}).
Using an LM, we filter the collected PRs to retain those explicitly described as bug fixes in the associated descriptions or issue texts.
Table~\ref{fig:construction-funnel} summarizes candidate attrition across the construction stages.

\textbf{Estimate bug lifetime.}
To find commits in which many bugs coexist, we estimate each bug's lifetime using the SZZ algorithm~\citep{sliwerski2005}.
A bug's lifetime ends at the merge commit of the corresponding PR that resolves it.
To estimate where it begins, we \texttt{git blame} the lines modified by the PR to identify the commit that last touched them.
Aggregating these estimated lifetimes produces a coverage curve over repository history,
indicating how many candidate bugs may be present at each commit.
Figures~\ref{fig:repository-szz-01-50} and~\ref{fig:repository-szz-51-100}
show these curves for all benchmark repositories.

\textbf{Select a base commit with the maximum number of concurrent bugs.}
Given each bug's lifetime, we can create a timeline of how many bugs exist simultaneously at any commit, as visualized in Figure~\ref{fig:construction-matrix}.
From this timeline, we identify several base commits with many bugs.

Per base commit, we check how many of the bugs are reproducible: the gold fix must apply, fail-to-pass hidden tests fail when the gold fix is not applied, and all hidden tests must pass when the gold fix is applied.
Frequently, the patches for bugs do not apply cleanly to a base commit; we use an LM to rebase such patches onto the base commit without changing their behavior (details in \S\ref{appx:benchmark:patch-rebasing}).
Per repository, we then select one base commit with the largest number of reproducible bugs.

\textbf{Additional quality control.} 
We then run additional quality control checks on the bug instance candidates to ensure that each corresponds to a single unit of work, that its bug is discoverable from reading the repository alone, and that its tests are not overly specific.
We furthermore ensure that solving the bug instances does not contradict other bug instances or any of the base commit's tests.
This quality control is performed with a mix of deterministic checks and several stages of LM annotations. For all annotations, we employ Gemini 3.1 Pro, Opus 5 (high), and Sol 5.6 (high) and require unanimous agreement between the three models.
Appendices~\ref{appx:benchmark:subtask-criteria} and~\ref{appx:benchmark:quality-control} give the full criteria and quality-control procedure; Appendix~\ref{appx:benchmark:validation} reports supporting validation studies.
We acknowledge it is certainly possible these filters end up excluding meaningful bugs; our checks do not assess the importance or difficulty of a bug.
That said, we intentionally trade away some coverage to ensure identified bugs can be graded fairly.

\textbf{What makes a bug \textit{discoverable}?}
We define a bug as behavior that diverges from the software's expected behavior~\citep{ieee1990glossary,iso2017vocabulary,avizienis2004basic}, and call it \emph{discoverable} when this expectation can be inferred from the repository alone, such as from contracts defined in the documentation, types, tests, referenced standards, and well-defined mathematical or scientific semantics~\citep{herzig2013its,antoniol2008bug}.

To ensure that every bug in \bench{} is discoverable, we operationalize these definitions via a two-stage annotation effort.
First, we query three LMs (Gemini 3.1 Pro, Opus 5, Sol 5.6) three times each to judge if a bug is discoverable, only keeping those that are unanimously accepted.
This filter rejects 40\% of bug candidates; examples of accepted and rejected bugs are in Appendix~\ref{appx:benchmark:discoverability}.
Second, four human reviewers independently audit a random sample of 50 bugs from the final benchmark.
For 98\% of sampled bugs, 3 of 4 reviewers agreed with the LM's determination that the bug is discoverable.
Details about the annotation effort parameters and outcomes can be found in Appendix~\ref{appx:benchmark:validation}.

\subsection{Statistics}
\label{sec:benchmark:statistics}
\begin{figure}[t]
    \centering
    \begin{minipage}[t]{0.48\linewidth}
        \vspace{0pt}
        \centering
        \small
        \includegraphics[width=\linewidth]{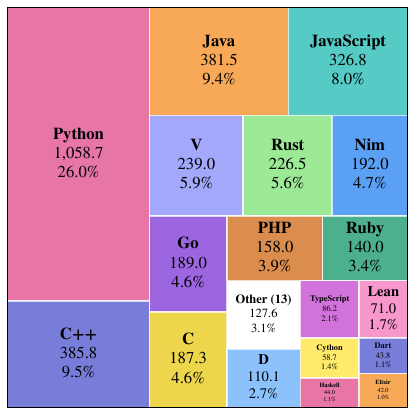}
        \captionof{figure}{
Bug counts and ratios by language, measured from the files modified by gold fixes.
        Details in Table~\ref{tab:subtask-languages}.
        }
        \label{fig:subtask-language-treemap}
    \end{minipage}
    \hfill
    \begin{minipage}[t]{0.48\linewidth}
        \vspace{0pt}
        \centering
        \small
        \captionof{table}{Repository statistics. The lower part of the table is measured at the base commit.}
        \label{tab:benchmark-task-stats}
        \begin{center}
            \begin{tabular}{lrrr}
\toprule
 & \textbf{Min} & \textbf{Mean} & \textbf{Max} \\
\midrule
GitHub stars & 819 & 25.4k & 137k \\
Contributors & 10 & 690.8 & 5,605 \\
\midrule
\# bugs & 4 & 40.7 & 239 \\
LoC & 6k & 292k & 1.9M \\
Files & 10 & 1,026 & 6,861 \\
\# bugs / 100k LoC & 0.8 & 46.3 & 254.5 \\
\bottomrule
\end{tabular}
 \end{center}

        \vspace{2ex}

        \captionof{table}{Bug statistics. See Section~\ref{sec:benchmark:formulation} for definitions and Figure~\ref{fig:benchmark-bug-cumulative} for distributions.
        }
        \label{tab:benchmark-subtask-stats}
        \begin{center}
            \begin{tabular}{lrrr}
\toprule
 & \textbf{Min} & \textbf{Mean} & \textbf{Max} \\
\midrule
\# F2P tests & 1 & 1.6 & 268 \\
\# P2P tests & 0 & 50.4 & 19,230 \\
Gold fix lines & 1 & 16.3 & 3,905 \\
Test patch lines & 1 & 37 & 2,533 \\
\bottomrule
\end{tabular}
 \end{center}
    \end{minipage}
\end{figure}

\bench{}'s final test set consists of \qBenchmarkTaskCount{} tasks created from open-source GitHub repositories spanning \qBenchmarkNumLanguages{} languages (Figure~\ref{fig:subtask-language-treemap}).
The repositories covered represent a range of utilities, from small single-purpose libraries (e.g., date/time handling in \texttt{dayjs}, \texttt{pendulum}) to web and application frameworks (\texttt{rails}, \texttt{quarkus}), scientific computing (\texttt{numpy}, \texttt{astropy}, \texttt{qutip}), database engines (\texttt{duckdb}, \texttt{trino}), developer tooling (\texttt{babel}, \texttt{mypy}, \texttt{ruff}), and programming language toolchains (\texttt{go}, \texttt{roslyn}, \texttt{lean4}, \texttt{z3}).
Repositories vary significantly across a number of dimensions, such as in size, from \qBenchmarkMinGradedSourceLocRounded{} to \qBenchmarkMaxGradedSourceLocRounded{} non-test source code lines
(Table~\ref{tab:benchmark-task-stats}).
The test set represents a total of \qBenchmarkSubtaskCount{} bugs, where each repository contains between \qBenchmarkMinSubtasksPerTask{} and \qBenchmarkMaxSubtasksPerTask{} bugs (Table~\ref{tab:benchmark-task-stats}).
The gold fixes are similarly assorted in scope: a median fix edits \qBenchmarkMedianGoldSourcePatchLines{} lines in  \qBenchmarkMedianGoldSourcePatchFiles{} file, but the mean of \qBenchmarkMeanGoldPatchLines{} lines indicates a long tail of far larger repairs.
Appendix~\ref{appx:task_index} gives the full distributions and lists all repositories.

\subsection{Features}
\label{sec:benchmark:features}

\textbf{Bugs lie dormant for a long time, well before their upstream fixes.} 
At the repository state given to agents, the lines eventually repaired had already been
unchanged for a median of \qBenchmarkMedianValidHunkAgeDays{} days, and the upstream fix
did not arrive for another \qBenchmarkMedianValidFixLeadDays{} days.

\textbf{Different bugs lead agents to overlapping files but widely separated repair
locations.}
More than half of bugs have a gold fix that modifies at least one file also modified by the gold fix for another benchmark bug. Yet same-file repairs almost always change distinct content and are usually more than 100 lines apart (Appendix Table~\ref{tab:repair-file-overlap}). This structure reflects \bench{}'s multi-bug design: agents can reuse
repository context, but cannot solve the benchmark by repeating one type of repair.

\textbf{Bugs are diverse in defect type and are discoverable from a variety of sources.}
We use an LM agent to review each bug in the context of its repository and annotate both its observable failure mode and the evidence showing that the behavior is incorrect.
To avoid biases, the agent is only handed the hidden-test patch introduced in the PR corresponding to the bug, but no other information such as the issue description.
The most frequent bug type causes the affected repository to directly produce incorrect outputs (\qBugTypeIncorrectOutputShare{}), followed by issues related to validating various types of input (\qBugTypeValidationDiagnosticShare{}), and resource or runtime failures (\qBugTypeRuntimeResourceShare{}).
For \qBugEvidenceRepositoryGroundedShare{} of labeled bugs, the agent finds evidence in the repository that shows that the observed behavior is incorrect.
These repository contracts come from sources such as documentation, types, existing tests, and consistency within the repository.
The remaining two bugs correspond to failures that are undesirable independent of the context (a race condition and an out-of-bounds read).
Appendix~\ref{appx:evidence_expected} reports the full breakdown of defect types.
 \section{Experiments}
\label{sec:experiments}
\begin{figure*}[t]
    \centering
    \begin{minipage}[t]{0.53\textwidth}
        \vspace{0pt}
        \centering
        \small
        \captionof{table}{Main results, ordered by the percentage of bugs resolved. Cost, agent turns, and output tokens (\emph{out tok}) are arithmetic means across repositories.
}
        \label{tab:main-results}
        \begin{center}
            \begin{tabular}{l>{\bfseries}r>{\scriptsize}r@{\hspace{3pt}}>{\scriptsize}r@{\hspace{3pt}}>{\scriptsize}r}
\toprule
\textbf{Model} & \textbf{\% bugs} & \multicolumn{3}{c}{\scriptsize\hspace{3pt}\textcolor{gray}{\textbf{Usage per repo}}\hspace{-3pt}} \\
 & \textbf{resolved} & \textcolor{gray}{\textbf{\$\phantom{00}}} & \textcolor{gray}{\textbf{turns}} & \textcolor{gray}{\textbf{out tok}} \\
\midrule
Sol 5.6 (xhigh) & 4.7\% & \textcolor{gray}{72.30} & \textcolor{gray}{233} & \textcolor{gray}{105k} \\
Luna 5.6 (xhigh) & 2.5\% & \textcolor{gray}{2.24} & \textcolor{gray}{204} & \textcolor{gray}{61k} \\
Terra 5.6 (xhigh) & 1.5\% & \textcolor{gray}{3.57} & \textcolor{gray}{76} & \textcolor{gray}{47k} \\
Luna 5.6 (high) & 1.4\% & \textcolor{gray}{0.28} & \textcolor{gray}{75} & \textcolor{gray}{20k} \\
Opus 5 (xhigh) & 1.3\% & \textcolor{gray}{53.63} & \textcolor{gray}{323} & \textcolor{gray}{193k} \\
Kimi K3 (max) & 0.6\% & \textcolor{gray}{24.51} & \textcolor{gray}{337} & \textcolor{gray}{153k} \\
Luna 5.6 & 0.5\% & \textcolor{gray}{0.04} & \textcolor{gray}{22} & \textcolor{gray}{4k} \\
GPT-5.4 Mini (high) & 0.5\% & \textcolor{gray}{1.22} & \textcolor{gray}{75} & \textcolor{gray}{41k} \\
GPT-5.4 Mini & 0.2\% & \textcolor{gray}{0.05} & \textcolor{gray}{13} & \textcolor{gray}{2k} \\
Gemini 3.5 Flash Lite & 0.1\% & \textcolor{gray}{0.06} & \textcolor{gray}{34} & \textcolor{gray}{5k} \\
\bottomrule
\end{tabular}
 \end{center}
    \end{minipage}
    \hfill
    \begin{minipage}[t]{0.43\textwidth}
        \vspace{0pt}
        \centering
        \includegraphics[width=\linewidth]{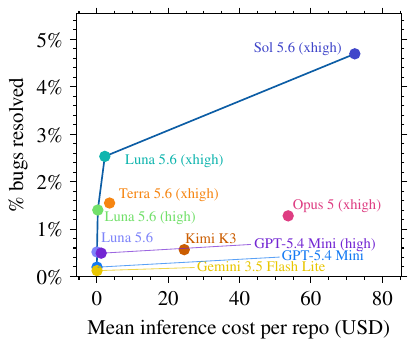}
        \captionof{figure}{Bugs resolved versus inference cost; the line marks
        the Pareto frontier.}
        \label{fig:score-cost-pareto}
    \end{minipage}
\end{figure*}
\textbf{Models.}
We evaluate \qEvaluatedLanguageModelCount{} language models in \qMainResultsModelCount{} configurations that have shown strong performance on current leading benchmarks including DeepSWE, Terminal-bench 2.1, and ProgramBench.
Our list features models from Anthropic (Opus 5), OpenAI (GPT 5.6 Sol/Terra/Luna, GPT 5.4 mini), Google (Gemini 3.5 Flash Lite), and Moonshot AI (Kimi K3).
Unless explicitly specified, we use the default reasoning strength.

\textbf{Agent system.}
We use \texttt{mini-SWE-agent}, a simple coding agent scaffold that enables models to interact with a terminal environment via bash commands~\citep{yang2024sweagentagentcomputerinterfacesenable}.
To focus on measuring \textit{model} performance and reduce confounds between model capability and harness design, prior works including DeepSWE, Terminal-bench, ProgramBench, and CodeClash use \texttt{mini-SWE-agent} as the primary harness~\citep{huang2026deepswe,merrill2026terminalbenchbenchmarkingagentshard,yang2026programbenchlanguagemodelsrebuild,yang2026codeclashbenchmarkinggoalorientedsoftware}.
A model attempts to repair each repository within a Docker container provisioned with 16 vCPUs and 64~GiB of memory.
An inference run is constrained to 1,000 steps and 12 hours of wall-clock time.
Appendix~\ref{appx:exp_setup} details the inference setting, and Appendix~\ref{appx:prompts:inference} reproduces the exact prompt.

\textbf{Metrics.}
The primary metric is \textbf{\% bugs resolved}, the percentage of bugs from all repositories that are resolved (i.e., bugs for which all bug-specific hidden tests and all base version tests pass).
 \section{Results}

\label{sec:results}
Table~\ref{tab:main-results} reports the main results; across all evaluated models, \bench{} remains extremely challenging.
The best model, \qBestModelMicro{}, resolves \qBestModelMicroScore{} of bugs in the dataset for a total of \$7k.
The second leaderboard position is held by Luna 5.6 (xhigh), which resolved about half as many bugs for considerably less spending (only \$224).
We also observe a clear progression in scores across the three Luna runs: default \qLunaDefaultMicroScore{}, high \qLunaHighMicroScore{}, and xhigh \qLunaXHighMicroScore{}.
Across the three effort levels, we also observe a scaling of turns and output tokens.
However, compared with other models, high output tokens or step counts do not directly translate to higher benchmark scores.
For example, Opus 5 (xhigh) has the highest output token count (and second highest step count), but is only ranked in the middle of the leaderboard (\qOpusXHighMicroScore{}).
The Pareto frontier with respect to cost is shown in Figure~\ref{fig:score-cost-pareto}.
A similar figure in terms of output tokens is shown in Appendix Figure~\ref{fig:score-tokens-pareto}.

Figure~\ref{fig:task-score-survival} shows repository-level scores. Sol 5.6 (xhigh) is consistently strongest across the full distribution, even though its median score is only 3.0\% and it scores less than 30\% on all but 13 repositories. Its highest score on any repository is 55\%.
Table~\ref{tab:score-by-bug-count} shows that the percentage of bugs resolved is lower among repositories with more bugs.
Among repositories with fewer than 25 bugs, Sol 5.6 (xhigh) resolves \qSolXHighUnderTwentyFiveBugsMicroScore{}, while its resolution rate falls to \qSolXHighAtLeastSeventyFiveBugsMicroScore{} for repositories with at least 75 bugs.
Similarly, Table~\ref{tab:score-by-source-loc} shows that the percentage of bugs resolved is lower among repositories with more lines of code, with the best rates ranging from \qSolXHighUnderHundredKLocMicroScore{} for repositories with fewer than 100k lines of code to only \qSolXHighOverFiveHundredKLocMicroScore{} for repositories with more than 500k lines of code.

Appendix~\ref{appx:results} provides alternative aggregations and detailed per-language and per-repository results.
\begin{figure*}[t]
    \centering
    \begin{minipage}[t]{0.48\textwidth}
        \vspace{0pt}
        \centering
        \includegraphics[width=\linewidth]{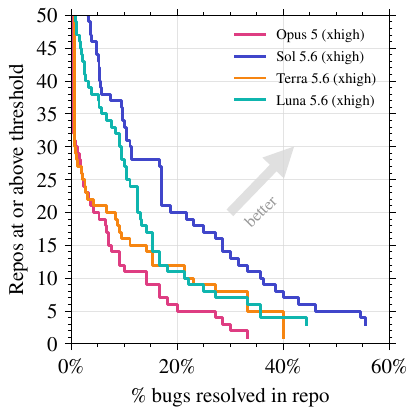}
        \captionof{figure}{
            Distribution of repository-level bug-resolution scores.
            The endpoint of each line indicates the best repository-level score.
        }
        \label{fig:task-score-survival}
    \end{minipage}
    \hfill
    \begin{minipage}[t]{0.48\textwidth}
        \vspace{0pt}
        \centering
        \small
        \setlength{\tabcolsep}{2.5pt}
        \captionof{table}{Percentage of bugs resolved within subsets of repositories, grouped by bug count per repository.}
        \label{tab:score-by-bug-count}
        \begin{center}
            \begingroup
\setlength{\tabcolsep}{1.8pt}
\begin{tabular}{@{}lrrrr@{}}
\toprule
\textbf{\# bugs/repo} & \textbf{\# repos} & \shortstack{\textbf{Opus 5}\\\textbf{(xhigh)}} & \shortstack{\textbf{Sol 5.6}\\\textbf{(xhigh)}} & \shortstack{\textbf{Luna 5.6}\\\textbf{(xhigh)}} \\
\midrule
$<25$ bugs & 56 & 4.2\% & 14.9\% & 9.3\% \\
25--74 bugs & 27 & 0.8\% & 3.2\% & 1.0\% \\
$\geq 75$ bugs & 17 & 0.6\% & 2.2\% & 1.2\% \\
\bottomrule
\end{tabular}
\endgroup
 \end{center}
\captionof{table}{Percentage of bugs resolved within subsets of repositories, grouped by the number of lines of code in the repository (excluding test files).}
        \label{tab:score-by-source-loc}
        \begin{center}
            \begingroup
\setlength{\tabcolsep}{1.8pt}
\begin{tabular}{@{}lrrrr@{}}
\toprule
\textbf{Repo size} & \textbf{\# repos} & \shortstack{\textbf{Opus 5}\\\textbf{(xhigh)}} & \shortstack{\textbf{Sol 5.6}\\\textbf{(xhigh)}} & \shortstack{\textbf{Luna 5.6}\\\textbf{(xhigh)}} \\
\midrule
$<100$k LoC & 44 & 3.2\% & 13.4\% & 8.2\% \\
100k--500k LoC & 37 & 0.9\% & 3.3\% & 1.5\% \\
$>500$k LoC & 19 & 0.6\% & 1.2\% & 0.5\% \\
\bottomrule
\end{tabular}
\endgroup
 \end{center}
    \end{minipage}
\end{figure*}

\subsection{Ablations}
\label{sec:results:ablations}
To identify the main sources of difficulty in \bench{}, we ablate the information provided about each bug and evaluate alternative agent scaffolds.

Table~\ref{tab:ablations} shows how the percentage of bugs resolved changes when we give the agent more information about the bugs or allow for separate work per issue.
We use the first 20 repositories of Table~\ref{tab:task-results} (a total of \qAblationBugCount{} bugs).
Appendix~\ref{appx:prompts:inference} shows the exact prompts.

\textbf{Identifying the bug is the challenging part.}
When changing the prompt to include information from the original GitHub issue reports, similar to the setting used by SWE-bench (except that many bugs need to be fixed in a single run), the percentage of bugs resolved increases dramatically, with Sol 5.6 (xhigh) jumping to a \qAblationSolIssueTextsBugMicroScore{} solve rate.
This strongly suggests that identifying and localizing the bug is the most challenging part of the benchmark, while the actual repair is likely comparable in difficulty to SWE-bench-style benchmarks, which are nearing saturation \citep{anthropic2026mythos}.

\textbf{Giving more information about the bug location makes the tasks easier.}
Since bug identification is the challenging part, we also test whether providing more information about the bug location improves performance.
We test two different types of information: In the \emph{filenames} ablation, we provide all filenames changed by the bugs' gold fixes. This doubles performance for both tested models (to \qAblationSolFilenamesBugMicroScore{} for Sol and \qAblationLunaFilenamesBugMicroScore{} for Luna). In the \emph{line ranges} ablation, we also provide the corresponding edited line ranges, leading again to a doubling of performance (\qAblationSolLineRangesBugMicroScore{} for Sol and \qAblationLunaLineRangesBugMicroScore{} for Luna).

\textbf{Handling multiple bugs in one trajectory adds to the challenge of the benchmark, particularly for smaller models.}
In the ablation where line ranges are provided for every bug, we can also isolate the additional difficulty of addressing many bugs within a single trajectory by allowing the model to solve each bug in a separate run.
This doubles the performance of Sol (\qAblationSolLineRangesBugMicroScore{} $\rightarrow$ \qAblationSolLineRangesSeparateBugMicroScore{}) and nearly triples the performance of Luna (\qAblationLunaLineRangesBugMicroScore{} $\rightarrow$ \qAblationLunaLineRangesSeparateBugMicroScore{}).
In the setting where both issue statements and line ranges are provided, the performance of Sol increases slightly, while Luna still almost doubles (\qAblationLunaIssueTextsLineRangesBugMicroScore{} $\rightarrow$ \qAblationLunaIssueTextsLineRangesSeparateBugMicroScore{}).
While the default setting of the benchmark does not allow a similar experiment to be conducted, we present test-time scaling results in the next section.

\textbf{Additional agent scaffolds.} We further evaluated additional harnesses on the set of 20 tasks of Table~\ref{tab:task-results} and the 10 repositories with the most bugs.
Sol 5.6 (xhigh) scores 1.6\% with the native Codex harness, significantly lower than with \texttt{mini-SWE-agent} (2.9\%).
Opus 5 (xhigh) scores 1.1\% with the native Claude Code harness, only slightly higher than with \texttt{mini-SWE-agent} (1.0\%).
We therefore conclude that our choice of harness is reasonable for the baseline results of this benchmark.

 \section{Analysis}
\label{sec:analysis}
\begin{figure}[tb]
    \centering
    \includegraphics[width=\textwidth]{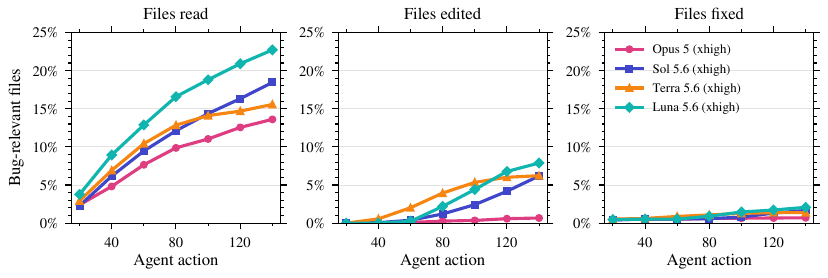}
    \captionof{figure}{Agents reach far more bug-relevant files than they fully fix.
    We show every file changed by a gold fix and report whether
    each model has read or edited it and whether every bug whose gold fix changes that file is resolved. Read means an agent explicitly read the file; edited means an agent edited a file involved in a gold fix;
    a fixed file requires all accepted bugs related to the file to pass together at a checkpoint with no
    correctness test regression.}
    \label{fig:bug-progress-by-action}
\end{figure}

We study \bench{} trajectories to investigate how different models approach \bench{} tasks and how performance scales with more test-time compute.
Appendix~\ref{appx:agent-behavior} provides additional trajectory analysis results.

\textbf{Agents spend much of their trajectories searching, although models differ in
exploration breadth.}
Reading is the largest observable action category for three of four models
(Appendix Figure~\ref{fig:action-frequency}). Opus explicitly reads an average of
\qAgentBehaviorOpusReadFiles{} files per repository, more than three times Terra's
\qAgentBehaviorTerraReadFiles{}, whereas Terra is the only model that writes more often than it
reads. Thus, models differ not only in outcomes but also in how broadly they explore before editing.

\textbf{Agents often read or edit files changed by gold fixes without resolving the
corresponding bugs.}
As shown in Figure~\ref{fig:bug-progress-by-action}, by action \qBugProgressEndpointAction{}, Sol has read \qBugProgressSolEndpointFilesRead{} of these
files and edited \qBugProgressSolEndpointFilesEdited{}, but only
\qBugProgressSolEndpointFilesFixed{} have every corresponding bug resolved. In our evaluations,
one or more models edit a file changed by the gold fix for \qTrajectoryAnyModelTargetEditShare{} of bugs. For bugs with completed annotations, Appendix Table~\ref{tab:discoverability-outcomes} breaks
these results down by source of evidence. Reading or editing a file changed by a gold fix does not
by itself mean that the corresponding bug is resolved.

\textbf{Most agent-edited files are not bug-relevant.}
Across models, only \qTrajectorySolFilePrecision{}--\qTrajectoryOpusFilePrecision{} of files
changed in submitted patches are bug-relevant. This measures overlap with known repairs rather
than correctness, since a valid repair may change other code (Appendix~\ref{appx:agent-behavior}).

\textbf{Agents keep revising their patches, but later work does not consistently improve bug
recovery.}
Across runs from four analyzed models, \qAgentBehaviorRegressionRuns{} trajectories lose at least one
previously resolved bug after a patch change. In total, \qAgentBehaviorTransientBugs{} bug repairs
pass at an intermediate checkpoint but not at the end, and \qAgentBehaviorRunsBelowPeak{} runs
finish with fewer bugs resolved than at their best intermediate checkpoint
(Appendix Table~\ref{tab:agent-behavior-dynamics-main}). Among runs that resolve at least one bug,
models spend a median \qAgentBehaviorMinPostImprovement{}--\qAgentBehaviorMaxPostImprovement{} of
their actions after the final increase in bugs resolved. Later edits can therefore undo successful
repairs.

\begin{figure*}[t]
    \centering
    \begin{minipage}[t]{0.53\textwidth}
        \vspace{0pt}
        \centering
        \includegraphics[width=0.95\linewidth]{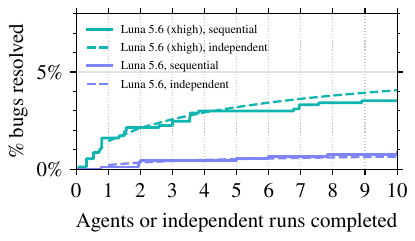}
        \captionof{figure}{
        Repeated attempts fix additional bugs, but gains diminish.
        Results on the \qSequentialTaskCount{} repositories with the most bugs.
        Solid curves show sequential agent runs, each starting from the preceding agent's patch while dashed curves show the union of bugs resolved by independent attempts.
        }
        \label{fig:sequential-score-by-step}
    \end{minipage}
    \hfill
    \begin{minipage}[t]{0.43\textwidth}
        \vspace{0pt}
        \centering
        \small
        \setlength{\tabcolsep}{2pt}
        \captionof{table}{Ablation results for the first 20 repositories (in the order of Table~\ref{tab:task-results}). Conditions provide fault-localization hints (gold fix filenames or line ranges), issue texts, or both.
        The settings marked \emph{per bug} run a separate agent for each bug.
        }

        \label{tab:ablations}
        \begin{center}
            \begin{tabular}{@{}lrr@{}}
\toprule
\textbf{Ablation} & \textbf{\shortstack{Sol 5.6\\(xhigh)}} & \textbf{\shortstack{Luna 5.6\\(xhigh)}} \\
\midrule
No information (default) & 4.8\% & 2.0\% \\
Filenames & 11.0\% & 4.8\% \\
Line ranges (LR) & 19.2\% & 10.7\% \\
LR (per bug) & 39.2\% & 31.3\% \\
Issue texts & 75.3\% & 36.8\% \\
Issue texts + LR & 78.1\% & 43.8\% \\
Issue texts + LR (per bug) & 87.3\% & 82.9\% \\
\bottomrule
\end{tabular}
 \end{center}
    \end{minipage}
\end{figure*}

\textbf{Inference-time scaling through repeated attempts improves bug recovery, but gains
diminish.}
We evaluate two simple inference-time scaling strategies on the \qSequentialTaskCount{} repositories with the most bugs: sequential runs, where we run repeated agents, each starting from the preceding agent's patch, and multiple independent agent runs, each starting from the initial task state.
In the latter case, we report the union of bugs resolved by all independent attempts.
The results are shown in Figure~\ref{fig:sequential-score-by-step}.
The gains for Luna (default reasoning) are marginal, but for Luna (xhigh), sequential attempts raise bug recovery from \qSequentialLunaXHighOne{} (first agent) to \qSequentialLunaXHighFinal{} (after 10 agents), while the union of \qSequentialAgentCount{} independent attempts
reaches \qIndependentLunaXHighFinal{}.

 \section{Related Work}
\textbf{SWE-bench-style tasks.}
The SWE-bench benchmark piloted the use of GitHub issue and pull request pairs to construct software-engineering-style tasks in which models are asked to modify the corresponding codebase such that the original issue is resolved~\citep{jimenez2024swebenchlanguagemodelsresolve}.
Subsequent work extends the original task format along many dimensions, including more programming languages~\citep{yang2024swebenchmultimodalaisystems,yang2025swesmithscalingdatasoftware,rashid2025swe,zan2025multiswebenchmultilingualbenchmarkissue}, increased difficulty~\citep{deng2025swebenchproaiagents,desai2026swemarathonagentsautonomouslycomplete,huang2026deepswe,ma2026swefficiencylanguagemodelsoptimize}, better alignment between task specification and test coverage~\citep{chowdhury2024introducing}, and solving multiple associated task instances in succession~\citep{deng2026swemilestoneevaluatingaiagents,lam2026swe}.
Across the many variations to the original task format, the task design choice to provide models with a clear specification has remained constant.
A handful of prior works, notably Ambig-SWE and SWE-together, explored feeding models vague problem statements to study how well models can solicit feedback from a simulated user~\citep{vijayvargiya2026ambig,wu2026swe}.
To our knowledge, \bench{} is the first to investigate whether models can surface issues in large repositories on their own, with no specification or hints given.

\textbf{Fault Localization.}
\bench's key research questions most closely follow the line of research around fault localization: identifying the code responsible for erroneous behavior~\citep{wong2016survey}.
Classical techniques typically ingest execution evidence, such as failing tests, test-coverage reports, and mutated programs, then rank how suspicious different programmatic entities are~\citep{wong2016survey}.
With LMs, more recent approaches reason directly over the source code~\citep{yang2023largelanguagemodelstestfree,Kang_2024,qin2025agentflscalingllmbasedfault}.
However, prior work generally still assumes a known bug and provides evidence to guide the system towards it~\citep{liu2019fixfindinvestigationfault}.
Our definition and operationalization of bug discoverability discussed in Section~\ref{sec:benchmark:construction} and Appendix~\ref{appx:benchmark:discoverability} obviate the need for explicit guidance that is typically provided in conventional fault localization evaluations.

\textbf{Bug Detection.}
LMs have also been used, without a report, to find bugs or vulnerabilities in fuzzing~\citep{deng2023largelanguagemodelszeroshot,Xia_2024} and cybersecurity benchmarks~\citep{yang2023language,zhang2025cybenchframeworkevaluatingcybersecurity,wang2026cybergymevaluatingaiagents,zhang2025bountybenchdollarimpactai,lee2026secbenchprolanguagemodels}.
However, these settings target failures that are wrong regardless of context, such as crashes or exploits, whereas most \bench{} bugs can only be recognized by inferring the intended behavior from the repository.
Concurrent work has begun to place more emphasis on proactive bug discovery but remains narrower in scope than \bench.
In TestExplora, the agent is asked to generate a test that exposes a bug, not a fix~\citep{liu2026testexplora}.
Active-SWE and TIDE directly fold bug discovery and repair into a single task as well, but are smaller in scale and inform the model which files or functions are defective~\citep{jeong2026tideproactivemultiproblemdiscovery,li2026active}.
In contrast, \bench{} evaluates open-world, unguided debugging over a whole GitHub repository.

\textbf{Long-Horizon Code Benchmarks.}
Recent code evaluations have primarily increased difficulty by challenging models to solve long-horizon tasks that take many turns and multiple hours to solve, such as writing a program from scratch~\citep{yang2026programbenchlanguagemodelsrebuild}, out-coding opponents in competitive programmatic environments~\citep{yang2026codeclashbenchmarkinggoalorientedsoftware}, and manually curated collections of arduous tasks~\citep{merrill2026terminalbenchbenchmarkingagentshard,zhao2024commit0librarygenerationscratch,zhu2026edgebench}.
\bench{} demonstrates the effectiveness of bug discovery and repair as a long-horizon task.
 \section{Discussion}
\label{sec:discussion}

\textbf{Limitations and Future Work.} As discussed in \S\ref{sec:benchmark:construction}, we recognize that \bench{} is constructed purely from the bugs that were explicitly identified with an issue and fixed by a corresponding pull request.
Consequently, we omit bugs that were reported but were not fixed, or were never reported to begin with.
\bench{} therefore necessarily under-counts the total bugs present in a codebase.
That said, we intentionally keep our focus on explicitly addressed bugs because the upstream fix signals the maintainer's confirmation that a bug is legitimate and provides the tests for evaluating the repair objectively.
Extending \bench{} to cover unreported bugs would be a natural next step.
To tackle this successfully, it would be necessary to validate a proposed defect without a ground-truth fix to check against.

\textbf{Conclusion.} We introduce \bench{} to measure models' ability to proactively identify and repair bugs, without explicit specification of any bug's location or errant behavior.
Our construction pipeline allows us to automatically synthesize GitHub repositories into tasks with a wide range of bugs.
The experiment results suggest that models still struggle to exhaustively identify defects.
We observe empirically that bug discovery is more difficult than repair.
That said, the gradual increase in bugs resolved across more turns demonstrates promise that more capable models will be more adept at surfacing bugs.
We release the \bench{} dataset and evaluation harness to advance coding agents towards autonomous repository maintainers.
\newpage
 \subsection*{AI Use Statement}
In this work, we used generative AI tools to clean and reformat the dataset, to implement methods, and to support qualitative and thematic data analysis through experimental annotation, while the authors interpreted the results and drew the conclusions. Specifically, the benchmark is built from public GitHub data, and AI tools were used to reformat, adapt, and filter this data; AI tools were also used to write code implementing our methods. We have not used generative AI tools to generate synthetic datasets, to develop theoretical models or conceptual frameworks, to propose or refine hypotheses, to design or provide feedback on research methodology or experiments, or to interpret results. Formulating mathematical claims, providing ingredients for proofs, assisting in the writing of proofs, and translation are not applicable to this work. Additionally, we used generative AI tools to create and edit software code. The authors have reviewed all AI-assisted work and all AI-generated code was reviewed and tested for correctness by the authors. We take responsibility for the final content of this work, including text, claims, or artifacts produced with the aid of generative AI.

\subsection*{Ethics Statement}
To the best of our efforts, we have worked to ensure that \bench{} complies with the principles outlined in the ICLR Code of Ethics.
No private data is included in the benchmark; \bench{} is constructed using bugs whose solution was previously published publicly, within publicly-facing repositories. All of our experiments run within Docker containers that do not have internet access. 
We believe that our work will lead to more resilient and higher-quality software libraries, driven by the autonomous bug finding and fixing systems that \bench{} could be used to build.
\subsection*{Reproducibility Statement}
We open-source the \bench{} dataset and evaluation harness\footnote{\url{https://anonymous.4open.science/r/swe-sweep-iclr-2027-C504/README.md}}, along with rollouts for all models reported in the main results (Table~\ref{tab:main-results}) to ensure that all metrics and analyses discussed in our work can be recovered from those artifacts.
The additional assets we use for carrying out evaluation, particularly \texttt{mini-SWE-agent}, are themselves open source.
The codebase includes detailed instructions on how to run evaluation.
To support understanding of the LM-based annotation effort that was involved in the benchmark, in Appendix~\ref{appx:benchmark:collection} provides examples, as well as human validation studies.

\clearpage
\beginappendix
\section{The benchmark}
\label{appx:benchmark}
This appendix supplements the main task description in \S\ref{sec:benchmark} with inference settings (\S\ref{appx:exp_setup}), a formal task definition (\S\ref{appx:benchmark:subtask-criteria}), and feasibility studies (\S\ref{appx:benchmark:feasibility}).
Appendix~\ref{appx:benchmark:collection} describes the task collection process.
\subsection{Inference settings}
\label{appx:exp_setup}
\begin{wrapfigure}{r}{0.33\textwidth}
    \vspace{-0.8\baselineskip}
    \textbf{Task scope}
    \verbpromptboxfile{assets/inference_prompt_task_scope_excerpt.txt}

    \textbf{Regression constraint}
    \verbpromptboxfile{assets/inference_prompt_regression_excerpt.txt}
    \caption{Excerpts from the inference prompt.}
    \label{fig:inference-prompt-safeguards}
\end{wrapfigure}

This section details the exact setup used for the experiments described in
section~\ref{sec:experiments}. To ensure a fair evaluation, agents are given
enough information to understand what is expected of them and how their
solutions are scored, while preventing access to information that would
circumvent the intended discovery process.

In addition to the task (\enquote{fix all bugs}), the agent receives
\begin{enumerate}
    \item The codebase at the base commit (which naturally includes the base commit tests).
    \item Information about the scoring of its work, in particular the requirement not to break (or change) any base commit tests.
    \item The selection criteria defining discoverable bugs, which were used during the construction of the benchmark (see \S\ref{appx:benchmark:discoverability}). Providing these criteria reduces effort spent investigating behaviors that would not qualify as benchmark bugs.
\end{enumerate}

Figure~\ref{fig:inference-prompt-safeguards} shows excerpts of the prompt to the agent.

Appendix~\ref{appx:prompts:inference} includes the complete inference prompt
and all ablation-specific changes.

\textbf{Precautions against reward-hacking behavior.}
Agents do not have access to the internet during evaluation. Web access
could reveal the original issue reports, pull requests, or later commits
containing the reference fixes.
Furthermore, the environment does not provide the \texttt{git} history of the repository
(in particular, this prevents the agent from discovering commits after the base commit, which would reveal all fixes).
\subsection{Feasibility}
\label{appx:benchmark:feasibility}

In our experiments, the best agent (\qBestModelMicro{}) resolves only \qBestModelMicroScore{} of bugs.
This raises the question of whether a perfect agent could resolve all bugs, or whether some bugs cannot be resolved at all.
To resolve all bugs, an agent must find each bug, write a fix that passes its hidden tests, keep all fixes compatible, and finish within its budget.
Below, we address the possible concern for each of these steps.

\textbf{Possible concern: The agent cannot distinguish between expected behavior and a bug (i.e., the bug is not discoverable).}
Agents do not see issue texts, so the expected behavior must be apparent from the repository.
We detail our more precise definition of discoverable bugs that is built around \emph{repository contracts} (i.e., expected behavior inferable from content of the repository) in \S\ref{appx:benchmark:discoverability}.
All bug candidates are checked for these discoverability criteria by 3 independent LMs during construction.
We only keep bugs where all three LMs agree that the bug is discoverable.

We recognize that this is the concern that is most specific to the \bench{} setting.
We therefore validate this part of the filtering process with two studies:

First, in a human annotation study (\S\ref{appx:benchmark:human-annotation}), we ask four human annotators to validate the discoverability of the bugs that pass all LM-based filters.
Before seeing the LM judges' reasoning, the reviewers accept that \qHumanReviewInitialDiscoverableMajorityAcceptCount{} of \qHumanReviewBugCount{} bugs are discoverable by at least a 3/4 majority, including \qHumanReviewInitialDiscoverableUnanimousAcceptCount{} unanimously. After seeing the reasoning, several reviewers changed their verdicts and also the last bug was considered to be discoverable.

Second, we run a validation study where the test patches of all bugs that pass the construction filters are given to an agent. The agent is tasked to find evidence within the repository that supports the expected behaviors that the tests assert.
The agent does not see issue texts, gold fixes, or similar information.
In this study, the agent finds a repository-defined contract for \qBugEvidenceRepositoryGroundedShare{} of bugs (\S\ref{appx:evidence_expected}).
The most common evidence is a related, pre-existing test (\num{88.1}\%), followed by internal invariants (\num{82.4}\%) and docstrings (\num{57.9}\%).
The 2 bug instances where the agent does not find a direct repository-defined contract are explicitly shown in \S\ref{appx:benchmark:discoverability}.
Both bugs are still deemed discoverable because they concern race conditions and memory-safety issues, which are both considered to be inherently undesirable behavior.

Finally, the results show large headroom.
Scores differ by more than $10\times$ between the evaluated models (Table~\ref{tab:main-results}).
In the ablations, localization hints alone raise performance by about $8\times$ for Sol 5.6 (xhigh) and $16\times$ for Luna 5.6 (xhigh) over the default setting, when the agent gets the line ranges of each bug and one run per bug (Table~\ref{tab:ablations}).
These hints do not reveal the expected behavior (or any new information that is not apparent from the repository); they only influence where agents spend their time.
Likewise, separate runs per bug add no information and only spare the agent from tracking many bugs in one long context.
Better discovery alone could therefore raise scores several-fold, which leaves ample room for agents to improve on \bench{}.

\textbf{Possible concern: Some bugs may not exist at the base commit, or may have no working fix.}
Every bug passes executable checks at the base commit.
At least one hidden test fails without the gold fix, and all hidden tests pass with it.
We run the hidden tests 10 times with the gold fix applied and keep only bugs whose tests pass every time.
To make sure that the regression tests are not flaky, we similarly rerun all base commit tests 10 times and keep only tests that pass every time.
So every bug has a reliable test and a known fix, taken from the upstream pull request.

\textbf{Possible concern: A correct fix may still fail the hidden tests.}
This concern is shared with most other coding benchmarks that use unit tests for validation, including SWE-bench, Terminal-Bench, and ProgramBench~\citep{jimenez2024swebenchlanguagemodelsresolve,merrill2026terminalbenchbenchmarkingagentshard,yang2026programbenchlanguagemodelsrebuild}.
Hidden tests can be too strict, for example if they check the exact wording of an error message that the gold fix introduced.
We reject such bugs with the LM annotations detailed in \S\ref{appx:benchmark:appropriate-tests}.
Three LMs have to agree that the tests are appropriate for a bug to be part of the benchmark.
We further validate the appropriate-tests judgments in the human study described in \S\ref{appx:benchmark:human-annotation}.
Before seeing the LM judges' reasoning, a 3-of-4 majority of reviewers accepts all sampled bugs under this criterion.

Furthermore, when Sol 5.6 (xhigh) receives the issue texts for all bugs in each repository, it resolves \qAblationSolIssueTextsBugMicroScore{} of bugs across the runs in Table~\ref{tab:ablations}, increasing to \qAblationSolIssueTextsLineRangesSeparateBugMicroScore{} if it also receives the line ranges of each bug and runs separately for each bug.
This puts a tight bound on the number of tests that might suffer from the above concern (note that this ablation merely frames the bug in a SWE-bench-style setting that removes the difficulty of discovering the bug; solving the bug might still be challenging for an LM, which is why a performance of 100\% is not necessarily expected).

We also conduct a stricter source-blind validation in which an LM judge sees only the base repository and test patch and checks whether the repository supports every material test expectation. The judge finds direct support for every expectation in \qTestExpectationSupportedShare{} of cases (\S\ref{appx:benchmark:grounding-all-test-expectations}), which also puts a bound on the number of tests that might suffer from this concern.

\textbf{Possible concern: Fixing one bug may break another bug or a base commit test.}
If two bugs required opposite behavior, no agent could resolve both.
We rule out conflicts with the base commit tests by execution: we apply each gold fix alone, run all base commit tests, and remove every bug whose gold fix breaks one.
We also remove base commit tests that do not pass in all of 10 runs, so a flaky test cannot count as a regression.
For conflicts between bugs, an LM judge reviews all test patches of a repository together.
It reports no conflicting pair in any of the \qCompatibilityReviewedTasksCount{} repositories.
The issue-text ablation also sets a clear upper bound on any such problem: across the tested repositories, Sol 5.6 (xhigh) fixes \qAblationSolIssueTextsBugMicroScore{} of bugs at the same time.

\textbf{Possible concern: The budget is too small to fix every bug.}
Agents stop long before they reach their limits (Table~\ref{tab:agent-behavior-stopping}).
The longest run uses \qStoppingMaxActions{} of \qStoppingStepLimit{} steps and \qStoppingMaxHours{} of \qStoppingWallHours{} hours.
We do not limit cost or tokens per run and do not observe any failures due to models exceeding their context window.
Every run ends because the agent submits (i.e., declares the task to be done), and no run states that it ran out of budget.
More steps are also not always helpful, since agents spend a median of \qAgentBehaviorMinPostImprovement{}--\qAgentBehaviorMaxPostImprovement{} of their actions after their last gain in bugs resolved (\S\ref{sec:analysis}).
Appendix~\ref{appx:agent-behavior} discusses stopping behavior and the full trajectory analysis in detail.

\subsection{Other concerns}
\label{appx:benchmark:other-concerns}

\textbf{Possible concern: The selection of benchmark bugs might systematically disadvantage certain models.}
Two models may fix a similar number of genuine bugs overall, while differing in the kinds of bugs they tend to identify. If one model's bug distribution happens to more closely match the bugs represented in \bench{}, it could receive a substantially higher benchmark score despite comparable overall performance.
However, agents are given the same bug-discoverability criteria that we use when selecting bugs for \bench{}, reducing potential mismatch between the benchmark definition and the task presented to the agent.
In addition, we evaluate this concern directly using a substantially looser definition of a bug in \S\ref{appx:benchmark:potentially-fixed}. For each repository, an LM judge counts the total number of bugs that each model attempts to fix, including bugs outside the \bench{} reference set. The attempted-fix count gives the same model ordering as the benchmark score, which is consistent with benchmark performance reflecting models' broader tendency to identify and fix bugs.

\textbf{Possible concern: The benchmark might favor models that simply attempt more changes, rather than identifying the most relevant bugs.}
A model that modifies more files or attempts more fixes may have more opportunities to match one of the benchmark bugs, even if much of its additional work is unnecessary.
However, our strict bug-selection process and requirement for maintainer-confirmed ground truth already limit this effect: models receive credit only for fixing known bugs that were validated by the PR merge by the repository maintainers.
In addition, models are limited by the edits they can make by the base commit tests that check for any regression.
Repositories in \bench{} generally have high test coverage, which should severely limit the impact of any possible introduced issues.

We nevertheless observe (\S\ref{appx:benchmark:potentially-fixed}) that the model with more attempted fixes also has the higher benchmark score. Penalizing models for additional work, however, would introduce a more fundamental problem. We cannot reliably determine whether work outside the reference set is unnecessary: it may address code quality improvements, or address genuine bugs that are absent from our ground truth, including previously unknown defects. We therefore avoid penalizing additional fixes and restrict the benchmark score to the set of bugs described. We also believe that models which surface more bug candidates can still be useful systems for human users: even when some candidates are incorrect or lower priority, they can be ranked or filtered downstream, whereas models that fail to identify potential bugs in the first place provide no such opportunity.
\subsection{Validation studies}
\label{appx:benchmark:validation}
The construction filters (\S\ref{appx:benchmark:quality-control}) rely on LM judges that see the
issue text and the gold fix. This leaves two questions. First, do the LM judges apply our criteria
as a human would? Second, is the expected behavior still clear with only the information an agent
has? We address the first question with a human annotation study on 50 randomly sampled
bugs (\S\ref{appx:benchmark:human-annotation}). We address the second with two blind LM studies in
which an agent sees only the base repository and the test patch. The first finds the repository
evidence for each bug's main failure (\S\ref{appx:evidence_expected}). The second checks that the
repository supports every material test expectation
(\S\ref{appx:benchmark:grounding-all-test-expectations}). 
Finally, we present a side study that compares benchmark scores with the total number of potential bugs that LMs attempt to fix (\S\ref{appx:benchmark:potentially-fixed}).
\subsubsection{Human annotation validation of rubrics used in construction}
\label{appx:benchmark:human-annotation}
We use a human study to check the four LM-based quality rubrics.

\textbf{Four human reviewers judge 50 of the included bug instances.}
 We draw a seeded uniform random sample of 50 bugs from the \qBenchmarkSubtaskCount{}  bugs.
Every reviewer judges the same fixed sample.
Four authors review each sampled bug separately. 

Reviewers are presented with an annotation interface that contains the raw GitHub issue, GitHub pull request, gold fix, and test patch. The interface first shows an LM-written summary of these artifacts to aid understanding. This summary was produced by GPT-5.6 Terra (high), and was neither instructed about the rubrics nor saw the LM judges' verdicts.

Reviewers can mark a bug as accept, reject, or unsure based on the following criteria:

\begin{enumerate}
    \item Is this a genuine bug whose expected behavior can be found from the
    repository without the issue?
    \item Do the tests represent one defect, or several symptoms of one shared
    root cause?
    \item Would any reasonable fix for the bug pass the added tests?
    \item Does the test patch preserve all expectations already recorded in the
    test suite?
\end{enumerate}

These criteria correspond to the LM annotations that are used in benchmark construction (\S\ref{appx:benchmark:subtask-criteria}).

To reject a bug, a reviewer must name at least one failed criterion and explain the rejection in writing. A reviewer who cannot decide may instead mark the bug unsure, optionally naming the criteria in doubt. Reviewers cannot skip a bug.
If the reviewer accepts the bug, the verdict is saved and the interface moves to the next bug. 

If the reviewer rejects the bug or selects unsure, the first verdict is locked. 
The interface then shows a summary of the LM judges' decision for each criterion, followed by the full verdicts and reasoning from the individual models. 
These reasoning traces often contained mentions of specific domain knowledge and considerations which were not previously directly available to the reviewers as part of the issue or PR texts or summaries thereof. 
Reviewers were then asked to cast a second vote (\emph{final vote}), either confirming or changing their judgment.
A written explanation was required for a change of judgment.

\begin{figure}[p]
    \centering
    \includegraphics[width=0.95\linewidth]{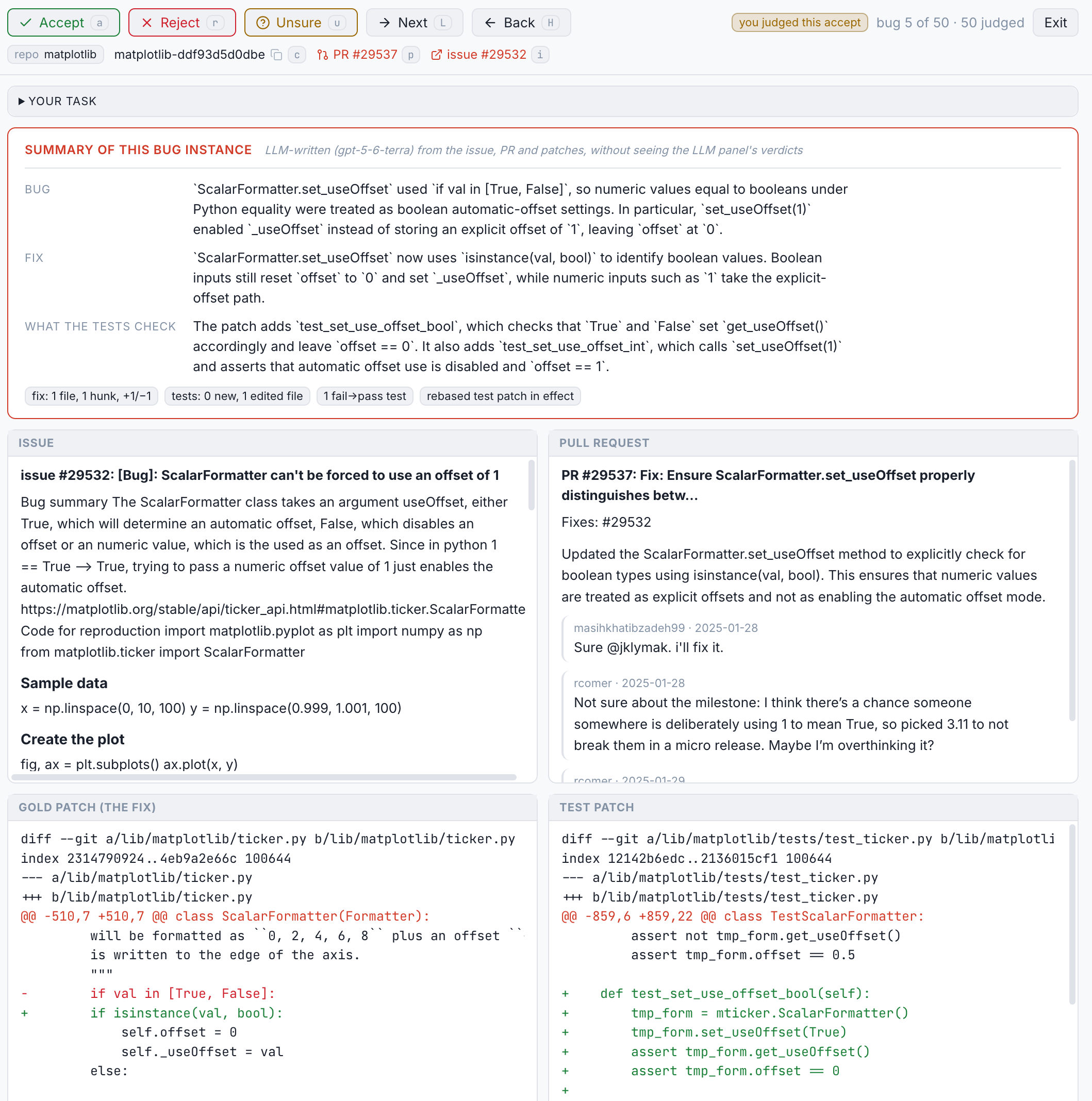}
    \captionof{figure}{Human-annotation interface. In this example of a bug, the \texttt{ScalarFormatter} in \texttt{matplotlib} cannot be set to have an offset of 1. This is due to a bug in handling boolean values (in Python \texttt{1==True}, so \texttt{1 in [True, False]} evaluates as true, causing the integer 1 to be handled as the boolean \texttt{True}).}
    \label{fig:human-annotation}
    \vspace{1.5em}
    \includegraphics[width=0.95\linewidth,trim={5px 0 0 0},clip]{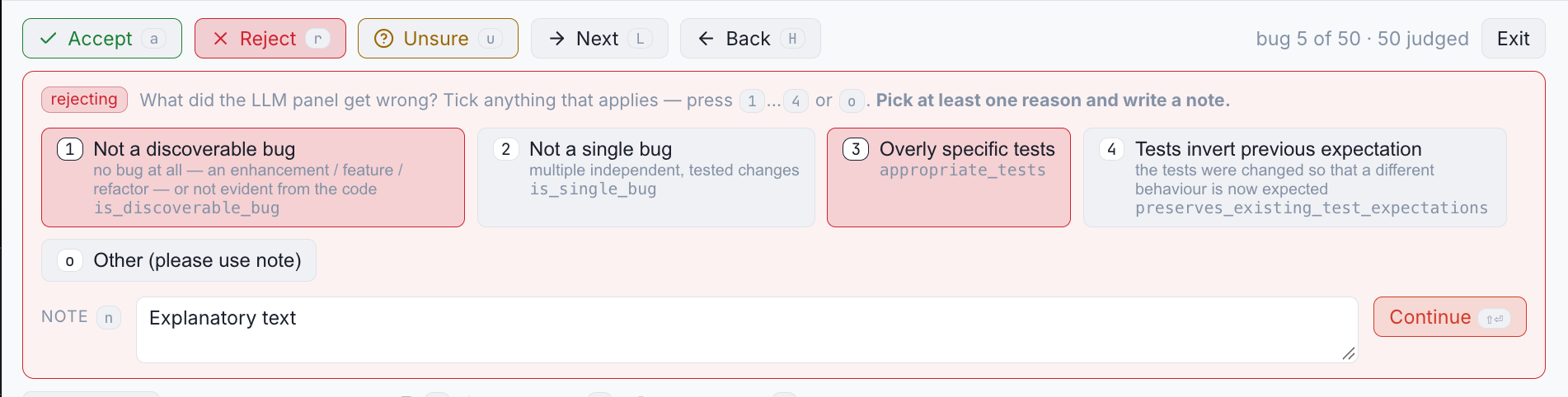}
    \captionof{figure}{Rejection step in the rubric-study interface. After selecting reject,
    the reviewer names the failed criteria and writes an explanation.}
    \label{fig:human-annotation-reason-prompt}
\end{figure}

The interface is shown in Figures~\ref{fig:human-annotation} and \ref{fig:human-annotation-reason-prompt}.

\textbf{Human reviewers accept \qHumanReviewInitialMajorityAcceptShare{} of bugs by at least a 3/4 majority before seeing the LM judges' reasoning.}
Reviewers spent between \qHumanReviewMinTotalHours{} and \qHumanReviewMaxTotalHours{} hours on the review task.
Table~\ref{tab:human-review-summary} summarizes the verdicts. 
In their first votes (before seeing the reasoning of the LM judges), \qHumanReviewInitialMajorityAcceptCount{} of the \qHumanReviewBugCount{} bugs were accepted by a 3/4 majority (counting any unsure votes as rejects).
Reviewers therefore strongly agreed with the LM judges that \qHumanReviewInitialMajorityAcceptShare{} of the reviewed bugs belong in the benchmark.
Of the \qHumanReviewVerdictCount{} votes that were cast, 178 were accept verdicts, \qHumanReviewInitialRejectVerdictCount{} were reject verdicts, and \qHumanReviewInitialUnsureVerdictCount{} were unsure verdicts.

\textbf{After seeing the LM judges' reasoning, reviewers accept 49 of 50 bugs by a 3/4 majority; no consensus is reached on the final bug.} For the reject and unsure votes, the reviewers were then shown the LM judges' reasoning and given the option to change their opinions.
\qHumanReviewChangedToAcceptCount{} changed to accept, while \qHumanReviewRejectVerdictCount{} rejects and \qHumanReviewUnsureVerdictCount{} unsure verdicts were kept.
With the final verdicts, \qHumanReviewMajorityAcceptCount{} of the \qHumanReviewBugCount{} bugs (\qHumanReviewMajorityAcceptShare{}) were accepted by at least a 3/4 majority.
The remaining bug received two accept votes, one unsure vote, and one reject. The latter votes were cast by the two most critical judges, who did not agree on why the bug might be considered problematic (discoverability vs overly specific tests).

\textbf{Split by individual rubrics, agreement is even higher.} We further break down the final verdicts by rubric in Table~\ref{tab:human-review-summary}.
The discoverability rubric is the only rubric with a single bug that was not initially accepted by at least a 3/4 majority; however, it was accepted unanimously after seeing the LM judges' reasoning.
The \emph{human validation} paragraphs in Appendix~\ref{appx:benchmark:collection} discuss agreement on each rubric with examples.

\begin{table}[tb]
    \centering
    \small
    \caption{\textbf{Human verdicts on the \qHumanReviewBugCount{} sampled bugs with non-accept votes split by rubric.}
    Reviewers cast a single accept/reject/unsure vote on each bug and name any rubric(s) they reject or are unsure about. For this table, unsure votes count as rejects for each rubric they name. \emph{First} is the verdict before seeing the LM judges' reasoning; \emph{Final} is the verdict after, which only reject and unsure verdicts reach.}
    \label{tab:human-review-summary}
    \begin{center}
        \begin{tabular}{@{}lrrrrrr@{}}
\toprule
 & \multicolumn{2}{c}{Majority accept} & \multicolumn{2}{c}{Majority reject} & \multicolumn{2}{c}{Split judges} \\
\cmidrule(lr){2-3}\cmidrule(lr){4-5}\cmidrule(lr){6-7}
 & First & Final & First & Final & First & Final \\
\midrule
Discoverable bug & 98\% & 100\% & 0\% & 0\% & 2\% & 0\% \\
Single bug & 100\% & 100\% & 0\% & 0\% & 0\% & 0\% \\
Appropriate tests & 100\% & 100\% & 0\% & 0\% & 0\% & 0\% \\
Preserves expectations & 100\% & 100\% & 0\% & 0\% & 0\% & 0\% \\
\bottomrule
\end{tabular}
     \end{center}
\end{table}

\subsubsection{Using LM agents to collect evidence for expected behavior}
\label{appx:evidence_expected}

During construction, a panel of 3 LM judges checks a bug is discoverable by reading the issue, test patch and gold fix (\S\ref{appx:benchmark:discoverability}).
Each judge is a single LM call, and all 3 judges must unanimously agree that a bug is discoverable using the given info.

In this validation study, we collect stronger evidence for this criterion. We give each bug to an LM agent instead of a single LM call, so it can search and read the repository over many steps. 
Rather than just asking for a boolean result, the agent must find and cite the repository contract (e.g., a piece of documentation) that shows the expected behavior that the actual behavior contradicts.

The agent sees only the base repository and the hidden tests. It does not see the issue, the gold fix, later versions of the code, or the construction annotations. In particular, it is not told which source lines the fix
changes. These lines are often close to the repository contract, so the agent must find the
contract on its own.

We run mini-SWE-agent with GPT 5.6 Sol at high reasoning effort in a read-only,
network-disabled copy of the repository at the base commit without git history. 
We run this study on all \qBugAnnotationPopulationCount{} bugs.

The annotation returned a valid label for \qBugCharacteristicsAnnotatedShare{} of selected bugs.
The remaining \qBugCharacteristicsUnannotatedShare{} of reviews exceeded run limits or produced invalid submissions and are excluded from these percentages.
We group each completed annotation into one of two broad
categories. Of these annotations, \qBugEvidenceRepositoryGroundedCount{}, or
\qBugEvidenceRepositoryGroundedShare, cite a repository-defined contract.
Only \qBugEvidenceObviousOnlyCount{} bugs, or \qBugEvidenceObviousOnlyShare, relied solely on
a repository-independent failure such as a crash, hang, or unintended data loss.
These cases are discussed in detail in \S\ref{appx:benchmark:discoverability} and are found to be reasonable inclusions, as they involve a race condition and an out-of-bounds read.

Table~\ref{tab:quality-characteristics} further separates the repository-defined contracts into
detailed sources, which may overlap for the same bug.
Appendix~\ref{appx:prompts:bug-characteristics} gives the annotation
prompt.

\par\smallskip
\noindent\begin{minipage}{\textwidth}
    \centering
    \small
    \setlength{\tabcolsep}{5pt}
    \captionof{table}{\textbf{Defect types and evidence for expected behavior.}
    Defect and headline evidence shares use completed post-filtering annotations. Detailed source
    shares use bugs with repository-defined support. Each bug has one primary observable
    failure and one broad evidence category. Detailed sources may overlap.}
    \label{tab:quality-characteristics}
    \begin{center}
        \begin{minipage}[t]{0.41\linewidth}
\vspace{0pt}
\begin{tabularx}{\linewidth}{@{}>{\raggedright\arraybackslash}Xrr@{}}
\toprule
\multicolumn{3}{@{}l}{\textbf{Observable failure}} \\
\midrule
\textbf{Category} & \textbf{Bugs} & \textbf{Share} \\
\midrule
Incorrect output & 1,662 & 41.3\% \\
Runtime or resource failure & 899 & 22.3\% \\
Validation or diagnostic failure & 1,129 & 28.0\% \\
Incorrect state or side effect & 209 & 5.2\% \\
Missing or spurious behavior & 126 & 3.1\% \\
Other & 0 & 0.0\% \\
\bottomrule
\end{tabularx}
\end{minipage}\hfill
\begin{minipage}[t]{0.56\linewidth}
\vspace{0pt}
\begin{tabularx}{\linewidth}{@{}>{\raggedright\arraybackslash}Xrr@{}}
\toprule
\multicolumn{3}{@{}l}{\textbf{Evidence basis}} \\
\midrule
\textbf{Category} & \textbf{Bugs} & \textbf{Share} \\
\midrule
Repository-defined contract & 4,023 & 99.95\% \\
Repository-independent failure only & 2 & 0.05\% \\
\addlinespace
\midrule
\multicolumn{3}{@{}l}{\textbf{Detailed repository evidence source (may overlap)}} \\
\midrule
Documentation & 1,918 & 47.7\% \\
Docstring & 2,329 & 57.9\% \\
Type annotation & 309 & 7.7\% \\
Pre-existing test & 3,544 & 88.1\% \\
Caller or example & 334 & 8.3\% \\
Identifier or assertion & 1,144 & 28.4\% \\
Internal invariant & 3,315 & 82.4\% \\
Sibling implementation & 1,581 & 39.3\% \\
Referenced standard & 447 & 11.1\% \\
Mathematical or scientific definition & 383 & 9.5\% \\
\bottomrule
\end{tabularx}
\end{minipage}
 \end{center}
\end{minipage}
\par\smallskip
\subsubsection{Using LM agents to ground all test expectations}
\label{appx:benchmark:grounding-all-test-expectations}
This section conducts a validation study similar to that in the previous section (\S\ref{appx:evidence_expected}), but goes even further.
Rather than only checking for evidence of the expected behavior that conflicts with the bug, the agent is tasked to check for evidence of every single assertion that is made by the tests of a bug instance.
This addresses concerns not only about discoverability, but also about whether the hidden tests could be overly specific.

The construction-stage annotation (\S\ref{appx:benchmark:appropriate-tests}) sees the issue, test patch, and
gold fix and asks whether any reasonable fix for the bug would pass the tests.
Seeing the gold fix and original issue description could have introduced a bias.
We therefore apply a stricter post-filtering validation study using the same agent setup from \S\ref{appx:evidence_expected}:
The judging agent sees only the base repository and test patch (as opposed to seeing details about the issue or gold fix, as in the construction-stage annotation).
The agent must decide whether every material test expectation is supported by the repository or by a repository-independent failure such as a crash or data loss (see our definitions of discoverability in \S\ref{appx:benchmark:discoverability}).

Among bugs with unanimous
acceptance from the original discoverability annotation, the stricter annotation flagged
only about \qExpectationPanelContradictionShare{} as potentially requiring more evidence.
However, this validation task is itself challenging for LM agents, and some flagged cases may therefore reflect limitations of the judge rather than genuine problems with the tests. We consequently interpret these results only as an upper bound on the share of potentially problematic test cases.
The exact prompts used are shown in Appendix~\ref{appx:prompts:test-expectation-support}.

\subsubsection{Benchmark scores vs. number of bugs attempted}
\label{appx:benchmark:potentially-fixed}

\begin{figure}[tb!]
    \centering
    \begin{minipage}[t]{0.48\linewidth}
        \vspace{0pt}
        \centering
        \includegraphics[width=\linewidth]{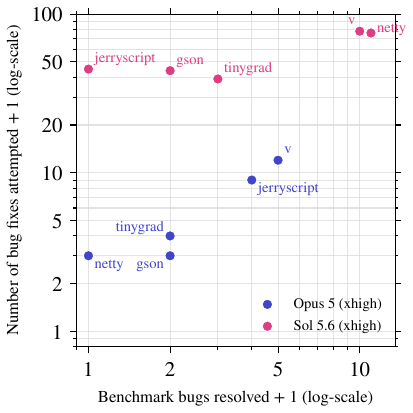}
        \captionof{figure}{Benchmark bugs resolved and bug fixes attempted for several repositories.}
        \label{fig:offref-verified}
    \end{minipage}
    \hfill
    \begin{minipage}[t]{0.48\linewidth}
        \vspace{0pt}
        \centering
        \small
        \setlength{\tabcolsep}{2.5pt}
        \captionof{table}{\textbf{\bench{} score vs bugs attempted, summed over all repositories.}
        \emph{Bugs attempted} is the number of distinct bugs that an LM judge finds in the final patch, including bugs that are not part of \bench{}. Each of two judges (Opus 5 and Sol 5.6 (high)) is run three times; we report the range [min, max] over the six runs. }
        \label{tab:potentially-fixed}
        \begin{center}
            \begin{tabular}{lrr}
\toprule
Model & Score & Bugs attempted \\
\midrule
Sol 5.6 (xhigh) & 4.7\% & [3843, 4427] \\
Opus 5 (xhigh) & 1.3\% & [387, 393] \\
\bottomrule
\end{tabular}
         \end{center}
    \end{minipage}
\end{figure}

In this section, we provide data that helps address possible concerns mentioned in~\S\ref{appx:benchmark:other-concerns}.
Both concerns are related to whether the benchmark score reflects how many bugs a model attempts to fix in general, or only how well its fixes match the distribution of bugs in \bench{}.
To answer this, we count all bugs that a model attempts to fix (based on rather loose filtering with LM judges), including bugs that are not part of \bench{}, and compare this count to the benchmark score.

\textbf{Setup.}
We take the final patches of Opus 5 (xhigh) and Sol 5.6 (xhigh) on all \qMainResultsRepositoryCount{} repositories.
For each patch, an LM judge lists every distinct bug that the patch attempts to fix.
This is an inherently noisy decision, as many edits can be related.
We adopt a similar prompt to that given for the construction-stage annotation on \emph{single unit of work} (\S\ref{appx:benchmark:other-quality}; full prompt in \S\ref{appx:prompts:is-single-bug}), which focuses on the number of distinct root causes.
The judge only sees the changes that the agent made to source files and the lines around them.
We further instruct the LM to filter based on the definition of \emph{bugs} from the inference prompt (\S\ref{appx:prompts:inference}).
In particular, changes that do not fix a bug are not counted, for example refactors, new features, formatting, or checks for conditions that cannot occur.
However, the judge does not decide whether a fix is correct or complete.
In contrast to the bugs in \bench{}, a counted bug does not need an upstream fix or a hidden test, and it is not checked by execution.
We use two LMs as judges, Opus 5 (high) and Sol 5.6 (high), each run three times, and report the results of both.

\textbf{The benchmark score and attempted-fix count give the same model ordering.}
Table~\ref{tab:potentially-fixed} shows each model's \bench{} score next to the number of bugs that it attempts to fix, summed over all repositories. For bugs attempted, it reports the range over six runs of the two judges.
The \bench{} score and count of attempted fixes give the same order of models for both judges.
Sol 5.6 (xhigh) attempts around 4k bug fixes according to the LM judges, a similar order of magnitude to the total number of bugs tracked in \bench{}, while Opus~5 attempts significantly fewer fixes in the first place.
Figure~\ref{fig:offref-verified} further visualizes this relationship per repository five repositories.

 \clearpage
\section{Task collection}
\label{appx:benchmark:collection}

In this section, we describe benchmark construction in detail, expanding on
\S\ref{sec:benchmark:construction}.

\subsection{How are tasks defined?}
\label{appx:benchmark:subtask-criteria}
Before describing the construction pipeline, we define more formally what tasks contain and the requirements that make them valid.

The central unit of \bench{} is a \emph{task}. A task represents a repository at a fixed point in its history, called the \emph{base commit}.
A \bench{} task consists of a repository at a particular commit, which we call the \emph{base commit}, and a collection of \emph{bug instances}.

\label{appx:benchmark:formulation}
\textbf{What conditions must a bug instance fulfill?}
Every bug instance is formalized as a test patch and a \emph{gold fix}.
The tests introduced or changed by the test patch are referred to as the \emph{hidden tests} of the bug instance.
To be a valid bug instance for the purposes of \bench{}, it must fulfill the following criteria:
\begin{itemize}
    \item[\hypertarget{appx:benchmark:criterion-r1}{R1}] The test patch and gold fix independently apply to the base commit.
    \item[\hypertarget{appx:benchmark:criterion-r2}{R2}] All hidden tests pass after the gold fix has been applied to the base commit, but at least one hidden test fails on the base commit (\emph{F2P tests}). There might be additional tests that pass before and after the gold fix has been applied (\emph{P2P tests}).
    \item[\hypertarget{appx:benchmark:criterion-r3}{R3}] The F2P tests reveal a single \emph{discoverable bug} in the base commit (see \S\ref{appx:benchmark:discoverability} for more information on how we define discoverable bugs).
    \item[\hypertarget{appx:benchmark:criterion-r4}{R4}] The target tests are not \emph{overly specific}; any reasonable fix to the discoverable bug will pass the target tests.
    \item[\hypertarget{appx:benchmark:criterion-r5}{R5}] The target tests do not contradict the base commit tests.
    \item[\hypertarget{appx:benchmark:criterion-r6}{R6}] Target tests of different bug instances do not contradict each other.
\end{itemize}
\subsection{Repository selection}

We build \bench{} tasks from permissively licensed open-source GitHub repositories.
We filter for repositories that link each bug fix to a bug report, e.g., projects that merge pull requests closing tracker issues, or that name tickets in their commit messages.
Repositories must have sufficiently good build systems that we can build and test them at a revision years in the past.
Building and executing tests must work on a 64-bit Linux system and complete in reasonable time (preferably less than 1 hour).
The number of GitHub stars together with the number of contributors and the age of the repository is a good heuristic to select projects that are likely to have a good test suite and strong standards for issues and pull requests.
A strong heuristic for bug instance yield is the number of issues in the project's tracker. 

\subsection{Constructing bug instance candidates}

This stage turns a repository's history into a population of bug instance
candidates.

We start with a large pool of merged pull requests that have an associated issue.
Occasionally, to handle repositories that do not use pull requests as a main editing mechanism or are public mirrors of a private repository, we mine candidates from commit messages paired with reports from their custom issue trackers instead.
Each bug instance begins with a commit or pull request linked to a bug report.
We divide the associated changes into two artifacts: a gold fix containing changes outside the test suite, and a test patch containing changes to the repository's test suite.

In order to do so, we build two sets of regular expressions for path names:
one for the graded source code and one for the test suite and the in-tree test infrastructure the tests import.
Anything matching neither of these (e.g., documentation or CI configuration) is dropped. The glob sets must also cover build-system files for new tests; otherwise, a test patch may add a test that is never executed.

A bug instance candidate therefore is specified by the following three items:
\begin{itemize}
    \item A \emph{test patch}, which changes the repository's test suite. All tests that are changed are referred to as the \emph{hidden tests} of the bug instance candidate.
    \item A \emph{gold fix}, which changes code outside the test suite.
    \item Additional metadata: Description of the bug that the bug instance candidate fixes. This can be taken
    from the PR description or comments, 
    from the issue or comments of any linked issue, or
    information directly from commit messages. 
    We also collect additional metadata such as GitHub tags or labels. 
    This additional information is not used for the evaluation or the main inference setting. 
    However, it is used for some of the LM annotations, as well as some of the ablations.
\end{itemize}
We then apply a series of pre-filters to narrow down the number of bug instance candidates.
\begin{enumerate}
    \item \textbf{Patches do not exceed size limits.} The gold fix and test patch together must touch fewer than
    \num{100} files and change fewer than \num{10{,}000} lines. 
    \item \textbf{Patches are not duplicates.} Fixes might be cherry-picked onto several branches. To exclude duplicates, we therefore remove candidates with an identical gold fix or an identical test patch.
    \item \textbf{Adds tests.} The test patch must be non-empty. 
    \item \textbf{Has a source fix.} The gold fix must be non-empty.
    \item \textbf{Is a bug fix.} We only consider changes that can reasonably be considered a bug fix.
    For this, we use Gemini 3.5 Flash Lite as a judge, chosen for throughput and price point due to the large number of candidates that need to be processed.
    The LM sees the PR/issue titles, labels, descriptions, and comments, and decides whether the change is a bug fix at all.
    This pre-filter maximizes recall measured against the stricter downstream filters while significantly reducing the number of candidates that need to pass through the more expensive later stages.
    In a limited study of repositories with well-maintained labeling of PR types, the results of this stage are largely identical to labeling based on said labels.
\end{enumerate}
\begin{figure}
  \centering
  \begin{minipage}[t]{0.48\linewidth}
      \centering
      \includegraphics[width=\linewidth]{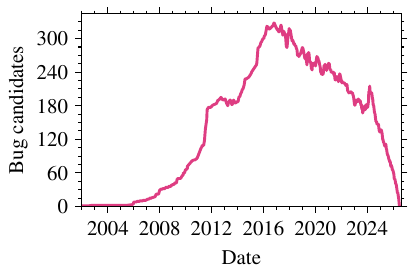}
      \captionof{figure}{Number of concurrent bug instance candidates across the history of NumPy.}
      \label{fig:numpy-szz}
  \end{minipage}\hfill
  \begin{minipage}[t]{0.48\linewidth}
      \centering
      \includegraphics[width=\linewidth]{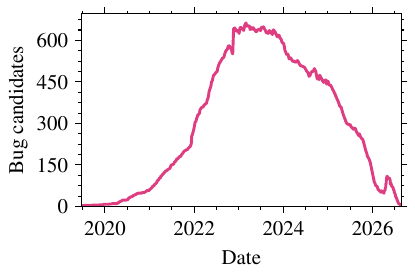}
      \captionof{figure}{Number of concurrent bug instance candidates across the history of V.}
      \label{fig:v-szz}
  \end{minipage}
\end{figure}
\subsection{Determining the base commit for the repository}

The goal of this stage is to find a repository commit containing many bug instance
candidates at the same time. For each bug instance candidate, we estimate the period during
which the bug was present as
$[\text{introduction time},\ \text{fix time})$.
We then count, for every commit, how many such intervals overlap it. 

The fix time
is given by the fix's mainline commit. When a fix appears only as a backport, we
map it to the corresponding mainline change; if no mainline placement can be
established, we discard the candidate.

To estimate when a bug was introduced, we apply the SZZ algorithm~\citep{sliwerski2005}. We
blame the source lines that the fix removes or modifies after excluding tests,
documentation, benchmarks, and examples, and select the latest commit identified
by blame. When a fix consists only of added lines, which leaves no removed lines
to blame, we apply blame to the surrounding context. We locate interval boundaries
using committer dates rather than positions along the first-parent history, so
that introducing commits made on pull-request branches can still be represented.

The resulting curves (e.g., Figures~\ref{fig:numpy-szz} and~\ref{fig:v-szz}) are only estimates.
Around the peak of this curve, we select up to 10 base-commit candidates to test, focusing on release tags as they are typically more stable to build and spaced around larger changes.

For each shortlisted base commit candidate, we then build a Docker image and ensure that tests can be executed inside the image.
This requires writing test runner/adapter code for each test framework.
Base commit candidates that cannot be built or have tests that cannot be executed are excluded.

We then count the number of bug instances whose patches apply
(\hyperlink{appx:benchmark:criterion-r1}{R1}) and whose bugs can be reproduced
(\hyperlink{appx:benchmark:criterion-r2}{R2}).
This leads to a more precise estimate of the number of concurrent bug instance candidates than the initial estimate based on the SZZ algorithm.

Based on these counts, we then select the final base commit.
\subsection{Making more patches apply}
\label{appx:benchmark:patch-rebasing}

Patch application is the single largest source of attrition in construction: at a
base commit years older than most fixes, the great majority of candidate patches
no longer apply. We therefore task \texttt{mini-SWE-agent}~\citep{yang2024sweagentagentcomputerinterfacesenable}
with rebasing them. The agent starts in the base-commit image with the repository
checked out at the base commit, and receives the two original patches and the issue
description. 
It writes a rebased gold fix and a rebased test patch.
Repaired candidates are checked again for applying patches
(\hyperlink{appx:benchmark:criterion-r1}{R1}) and reproduction
(\hyperlink{appx:benchmark:criterion-r2}{R2}).

Two examples of rebased patches are shown below:

\annotationrow
  {\annotationexample[PromptBlue]{JavaScript}{Gold fix (original)}{Original: \reflink{https://github.com/babel/babel/pull/11417}{babel~PR:\#11417}}
    {assets/rebase_example_babel_original.txt}
    {The upstream fix replaces a condition that also tests for \texttt{let}.}}
  {\annotationexample[PromptGreen]{JavaScript}{Gold fix (rebased)}{Rebased: \reflink{https://github.com/babel/babel/pull/11417}{babel~PR:\#11417}}
    {assets/rebase_example_babel_rebased.txt}
    {The \texttt{let} check does not exist at the base commit, so only the replaced line differs. The inserted code is identical.}}

\annotationrow
  {\annotationexample[PromptBlue]{PlainText}{Test patch (original)}{Original: \reflink{https://github.com/idris-lang/Idris2/pull/2010}{Idris2~PR:\#2010}}
    {assets/rebase_example_idris_original.txt}
    {The upstream patch registers a new test at the end of the list.}}
  {\annotationexample[PromptGreen]{PlainText}{Test patch (rebased)}{Rebased: \reflink{https://github.com/idris-lang/Idris2/pull/2010}{Idris2~PR:\#2010}}
    {assets/rebase_example_idris_rebased.txt}
    {The same test is registered. The base commit's list is shorter because \texttt{interactive038} does not exist yet.}}

For most rebasing attempts (\qRepairPrimaryModelShare{}), Claude Opus 4.8 was used, while Gemini 3.5 Flash Lite and GPT-5.6 Luna were substituted for later runs after they showed similar performance at lower cost.
The median agent run takes \qRepairMedianTurns{} steps.
The agent self-reports failure on
\qRepairGaveUpShare{} of attempts (indicating that the codebase has shifted in such a way that the patches cannot easily be adapted while being functionally equivalent).

Of the \qRepairSubtaskCount{} bug instance candidates we attempt to repair,
\qRepairAppliesShare{} yield a patch that applies, and \qRepairAcceptedShare{} pass the
remaining criteria and enter the benchmark.
Out of these, \qRebasedBothShare{} have both patches repaired, while
\qRebasedTestOnlyShare{} have only the test patch repaired.

\textbf{Most rebased patches only differ in hunk position.}
We check whether the rebased patches are semantically identical to the originals in two
ways. First, in \qBenchmarkRebaseAnchorOnlyShare{} of rebased patches the added and
removed lines are identical to the original and only the hunk position moved.
Second, across the \qRebaseJudgeBugCount{} bug instances in the benchmark that depend on
a rebased patch, we ask three LMs to compare each rebased patch against its original and
judge whether the same defect is repaired and still exercised by the test, using the
panel and the unanimous-acceptance rule of the annotation stages described below. 
Only \qRebaseJudgeRejectShare{} of patches do not pass this test.
The corresponding bug instances were excluded from the benchmark.

\textbf{Human validation.}  The authors manually reviewed 50 of the accepted rebased patches.
This resulted in no further rejections of any bug candidates.
Corroborating the above finding, most reviewed patches only differed in the hunk positions of the changes, while most of the remaining patches were adapted because of surrounding cosmetic changes or changed function signatures.
Only four out of the 50 patches were materially changed. Three of these simply removed unrelated changes that were present in the original source patch but were irrelevant to the bug at hand.
The one remaining materially changed patch adapted to a refactor of surrounding logic but applied the identical fix.

\textbf{Resolve rates are not significantly different between rebased and non-rebased patches.}
Repair could also leave the bug intact but make it easier or harder to find and fix. To
test this, we compare agent success on rebased and non-rebased bug instances (Table~\ref{tab:rebase-split}): \qRebaseResolvedShare{}
against \qNonRebaseResolvedShare{}, a difference of \qRebaseResolvedDelta{}.
This difference is not statistically significant, as the 95\% confidence interval
overlaps with zero (\qRebaseResolvedCI{}).
The confidence interval was computed using cluster bootstrapping over repositories.
We therefore find no evidence that rebasing changes how hard a bug instance is.

\begin{table}[tb]
    \centering
    \small
    \caption{\textbf{Rebasing a patch does not change how hard its bug instance
    is.} Share of bug instances that GPT-5.6 Sol resolved,
    split by whether or not the test patch or gold fix was rebased to make it apply to the base commit.
    }
    \label{tab:rebase-split}
    \begin{center}
        \begin{tabular}{@{}lrrr@{}}
\toprule
& \textbf{Gold fix rebased} & \textbf{Not rebased} & \textbf{Any gold fix} \\
\midrule
\textbf{Test patch rebased} & 5.0\% & 5.8\% & 5.5\% \\
\textbf{Test patch not rebased} & 4.0\% & 4.3\% & 4.3\% \\
\textbf{Any test patch} & 4.4\% & 4.7\% & 4.6\% \\
\midrule
\textbf{Any patch rebased} & & & 5.0\% \\
\textbf{Neither patch rebased} & & & 4.3\% \\
\bottomrule
\end{tabular}
 \end{center}
\end{table}

\subsection{Quality control}
\label{appx:benchmark:quality-control}

At this stage of the construction pipeline, we now have a collection of bug instance candidates that apply to the base commit (\hyperlink{appx:benchmark:criterion-r1}{R1}) and can be reproduced (\hyperlink{appx:benchmark:criterion-r2}{R2}).

To obtain a reliable final set of bug instances, we next filter out flaky tests and candidates whose fixes break behavior covered by the base test suite.

After the base commit has been chosen, we first run all base commit tests 10 times.
We exclude all base commit tests that do not pass every single time (for simplicity, we use \emph{base commit tests} to refer only to these passing tests below).
We then ensure that \hyperlink{appx:benchmark:criterion-r1}{R1} and
\hyperlink{appx:benchmark:criterion-r2}{R2} are \emph{reliably} satisfied for all bug instance
candidates by executing the hidden tests.
For every bug instance candidate, we execute all hidden tests 10 times and exclude all bug instance candidates that have hidden tests that do not pass every single time with the gold fix applied.
We further ensure that \hyperlink{appx:benchmark:criterion-r5}{R5} is satisfied as
follows: For every bug instance candidate, we apply its gold fix,
then run all base commit tests. Any bug instance candidate that breaks a base
commit test is excluded.

The remaining criteria (\hyperlink{appx:benchmark:criterion-r3}{R3}, \hyperlink{appx:benchmark:criterion-r4}{R4}, and \hyperlink{appx:benchmark:criterion-r6}{R6}) are checked using LM annotations as
discussed in the following sections.
For each of the annotation stages, we show real candidates drawn from the construction pipeline,
paired with the verdict the annotation panel returned for them. Three models
vote on each stage and a candidate advances only on unanimous acceptance, so a
single rejection is decisive.

\subsection{When is a bug discoverable?}
\label{appx:benchmark:discoverability}

Because agents receive no issue description, the expected behavior for each bug must be
recoverable from the base repository or from a failure that is unambiguously wrong. We use
the following two sources of evidence to make this distinction explicit.
Table~\ref{tab:lm-review-statistics-discoverable-bug} summarizes acceptance and agreement for this review.
Appendix~\ref{appx:prompts:is-discoverable-bug} gives the full prompt.

\textbf{Repository-defined contracts.}
The repository's own user-visible statements establish expected behavior. These
include documentation and repository content used to produce documentation,
such as docstrings and type annotations. A repository can also incorporate an
external source by reference: if a module states that it implements a named
standard, that standard becomes part of the contract. Likewise, code presented
as implementing a mathematical function or established scientific concept is
expected to follow the corresponding mathematical or scientific definition.

\textbf{Repository-independent failures.}
Some behavior is unambiguously undesirable without a repository-specific
contract, including segmentation faults, unintended deletion of user data, UI
freezes, and infinite loops. We apply this category conservatively: the expected
behavior must be sufficiently clear across reasonable interpretations that
documenting it explicitly would be superfluous.
In fact, only two bugs in the end solely rely on this category and they are listed below.

\textbf{Human validation.}
In a human study of \qHumanReviewBugCount{} sampled bugs from the final dataset (\S\ref{appx:benchmark:human-annotation}), a 3-of-4 majority of reviewers found that all \qHumanReviewDiscoverableMajorityAcceptCount{} bugs met the discoverability criterion and for all but four bugs, this vote was unanimous.
Each of the four bugs received a dissenting vote on discoverability from only one reviewer (\qHumanReviewDiscoverableRejectVerdictCount{} rejections and \qHumanReviewDiscoverableUnsureVerdictCount{} unsure vote).

\textbf{Agentic validation.}
In addition to the human validation, we also used agents to find repository contracts.
The corresponding validation study is described in \S\ref{appx:evidence_expected} and \S\ref{appx:benchmark:grounding-all-test-expectations} and corroborates the reliability of the filtering process.

\textbf{Annotation examples.} The examples below pair each source of evidence with a real accepted candidate;
each caption links to the evidence in the repository. We first show
repository-defined contracts, then repository-independent failures, and
close with rejected candidates.

\emph{Repository-defined contracts.}
The first set of examples consists of candidates that have clear repository-defined contracts.

The most direct evidence is documentation that users read.
In the first two examples, the API documentation and a docstring state the expected behavior, and the code does not follow it.

\annotationrow
  {\annotationexample[PromptGreen]{PythonMod}{Gold fix (truncated)}{Accepted: \reflink{https://github.com/aio-libs/aiohttp/pull/8745}{aiohttp~PR:\#8745}}
    {assets/annotation_example_is_discoverable_bug_accept.txt}
    {Documentation: the \reflink{https://github.com/aio-libs/aiohttp/blob/c7ff47b3916aaa4bb94074660e6969dbf1b77305/docs/client_reference.rst\#L2185-L2187}{API docs} define \texttt{status} as the HTTP response status.}}
  {\annotationexample[PromptGreen]{PythonMod}{Gold fix (excerpt)}{Accepted: \reflink{https://github.com/pandas-dev/pandas/pull/53709}{pandas~PR:\#53709}}
    {assets/annotation_example_is_discoverable_bug_accept_docstring.txt}
    {\reflink{https://github.com/pandas-dev/pandas/blob/30814ee4635c1055bb0ccb4d6c2648c783bc98b8/pandas/core/indexes/period.py\#L486-L487}{Docstring}: \texttt{period\_range} takes \texttt{freq} from a Period \texttt{start}/\texttt{end}; the code dropped the multiplier.}}

The contract does not have to be written in prose.
In sqlglot, a type annotation promises a return type.
In SciPy, an existing test already asserts the expected result.

\annotationrow
  {\annotationexample[PromptGreen]{PythonMod}{Gold fix (excerpt)}{Accepted: \reflink{https://github.com/tobymao/sqlglot/pull/4699}{sqlglot~PR:\#4699}}
    {assets/annotation_example_is_discoverable_bug_accept_type.txt}
    {Type annotation: \texttt{Table.name} is \reflink{https://github.com/tobymao/sqlglot/blob/b842d9383827d18482e36d6ea3041180a74d0abf/sqlglot/expressions.py\#L3252}{declared \texttt{-> str}} but can raise \texttt{AttributeError}.}}
  {\annotationexample[PromptGreen]{PythonMod}{Gold fix (excerpt)}{Accepted: \reflink{https://github.com/scipy/scipy/pull/12955}{SciPy~PR:\#12955}}
    {assets/annotation_example_is_discoverable_bug_accept_pretest.txt}
    {Pre-existing test: \reflink{https://github.com/scipy/scipy/blob/28c11c800cc3d1001736943aef0dfb8d57a8a43b/scipy/stats/tests/test_stats.py\#L790-L793}{\texttt{test\_weightedtau}} asserts the NaN-in-\texttt{y} result that the typo \texttt{isnan(sum(x))} breaks.}}

Expected behavior can also follow from how the code is used and named.
In Netty, the normal code path shows that the response must be passed on unreleased.
In click, the name \texttt{last\_option} tells the reader which option the variable must hold.

\annotationrow
  {\annotationexample[PromptGreen]{Java}{Gold fix (excerpt)}{Accepted: \reflink{https://github.com/netty/netty/pull/15602}{Netty~PR:\#15602}}
    {assets/annotation_example_is_discoverable_bug_accept_caller.txt}
    {Caller/example: this catch path releases a response that the \reflink{https://github.com/netty/netty/blob/2d24e1f27d00d13401d543a04beb5b76102dfd7d/codec-http/src/main/java/io/netty/handler/codec/http/HttpClientUpgradeHandler.java\#L209-L212}{normal path} forwards unreleased.}}
  {\annotationexample[PromptGreen]{PythonMod}{Gold fix (excerpt)}{Accepted: \reflink{https://github.com/pallets/click/pull/2816}{click~PR:\#2816}}
    {assets/annotation_example_is_discoverable_bug_accept_identifier.txt}
    {Identifier: \reflink{https://github.com/pallets/click/blob/8.0.1/src/click/shell_completion.py\#L466-L475}{\texttt{last\_option}} should hold the last option, but a missing \texttt{break} lets an earlier one overwrite it.}}

\begin{table}[tb]
  \centering
  \small
  \caption{\textbf{Acceptance and agreement for the discoverable-bug review.}
  Each percentage is a bug-level average over all 9k bug candidates that pass all non-LM checks.}
  \label{tab:lm-review-statistics-discoverable-bug}
  \begin{center}
      \begingroup
\setlength{\tabcolsep}{5pt}
\renewcommand{\arraystretch}{1.08}
\begin{tabular}{@{}lr@{}}
\multicolumn{2}{c}{\textbf{Agreement}} \\
\cmidrule(lr){1-2}
\textbf{Unanimous agreement} & \textbf{79\%} \\
Opus 5 (high) vs. Sol 5.6 (high) & 84\% \\
Opus 5 (high) vs. Gemini 3.1 Pro & 87\% \\
Sol 5.6 (high) vs. Gemini 3.1 Pro & 87\% \\
\end{tabular}
\hspace{4.5em}
\begin{tabular}{@{}lr@{}}
\multicolumn{2}{c}{\textbf{Acceptance rate}} \\
\cmidrule(lr){1-2}
\textbf{Unanimous accept} & \textbf{60\%} \\
Opus 5 (high) & 64\% \\
Sol 5.6 (high) & 73\% \\
Gemini 3.1 Pro & 74\% \\
\end{tabular}
\endgroup
   \end{center}
\end{table}

Other code in the same repository can also define the expected behavior.
In Jest, \texttt{fromJSON} must undo \texttt{toJSON}.
In dask, sibling functions show how \texttt{eye} should handle \texttt{dtype=None}.

\annotationrow
  {\annotationexample[PromptGreen]{JavaScript}{Gold fix (excerpt)}{Accepted: \reflink{https://github.com/jestjs/jest/pull/16353}{Jest~PR:\#16353}}
    {assets/annotation_example_is_discoverable_bug_accept_invariant.txt}
    {Internal invariant: a wrong array index breaks the \reflink{https://github.com/jestjs/jest/blob/9c9555f2a8adf4ffd374d0d78d5c3478b00f7107/packages/jest-haste-map/src/ModuleMap.ts\#L110-L126}{\texttt{fromJSON(toJSON(x))}} round trip of the \texttt{duplicates} map.}}
  {\annotationexample[PromptGreen]{PythonMod}{Gold fix (excerpt)}{Accepted: \reflink{https://github.com/dask/dask/commit/e25284dced9749f02bd5d8c80b6225153aa282d8}{dask~commit~e25284d}}
    {assets/annotation_example_is_discoverable_bug_accept_parallel.txt}
    {Parallel implementation: \reflink{https://github.com/dask/dask/blob/2.28.0/dask/array/creation.py\#L271}{\texttt{linspace}} and \reflink{https://github.com/dask/dask/blob/2.28.0/dask/array/creation.py\#L360}{\texttt{arange}} replace \texttt{dtype=None} with a default; \texttt{eye} crashed on it.}}

Last, a repository can point to a definition outside of it.
Luxon parses dates in the format of a named RFC, and the NumPy docstring gives the mathematical formula for the variance.

\annotationrow
  {\annotationexample[PromptGreen]{JavaScript}{Gold fix (excerpt)}{Accepted: \reflink{https://github.com/moment/luxon/pull/1782}{luxon~PR:\#1782}}
    {assets/annotation_example_is_discoverable_bug_accept_standard.txt}
    {Referenced standard: the \reflink{https://datatracker.ietf.org/doc/html/rfc5322\#section-4.3}{RFC~2822/5322} grammar allows the \texttt{UT} zone, but \texttt{fromRFC2822} returns an invalid date for it.}}
  {\annotationexample[PromptGreen]{PythonMod}{Gold fix (excerpt)}{Accepted: \reflink{https://github.com/numpy/numpy/pull/13177}{NumPy~PR:\#13177}}
        {assets/annotation_example_is_discoverable_bug_accept_math.txt}
        {Math definition: the \reflink{https://github.com/numpy/numpy/blob/51140bd0787127184f148cc6cc02a1196980f64e/numpy/core/fromnumeric.py\#L3398-L3399}{docstring} defines \texttt{var = mean(abs(x - x.mean())**2)}; \texttt{x*x} drops the \texttt{abs} for complex input.}}
\medskip
\noindent\emph{Repository-independent failures.}
A crash, hang, or data loss is unambiguously undesirable even with no repository
contract, so such a failure grounds discoverability on its own. It is usually
one of several sources of evidence for a bug; across the whole benchmark only
two bug instances rest on such a failure with no repository contract at all:
the go-git race and OpenCV out-of-bounds read in the final examples below.

\annotationrow
  {\annotationexample[PromptGreen]{Go}{Gold fix (excerpt)}{Accepted: \reflink{https://github.com/golang/go/commit/70e453b4361b80a85e6ebb37e4d43ec02db9a50a}{Go~commit~70e453b}}
    {assets/annotation_example_is_discoverable_bug_accept_crash.txt}
    {Crash: a \texttt{nil} value reaches \texttt{reflectlite.TypeOf(nil).String()}, which panics.}}
  {\annotationexample[PromptGreen]{Java}{Gold fix (excerpt)}{Accepted: \reflink{https://github.com/opensearch-project/OpenSearch/pull/11153}{OpenSearch~PR:\#11153}}
    {assets/annotation_example_is_discoverable_bug_accept_hang.txt}
    {Hang: a retry loop never incremented \texttt{retryCounter}, so a failing operation was retried forever.}}

The next two examples show data loss in Airflow and the first of the two bugs that have no repository contract at all, but are still considered discoverable.

\annotationrow
  {\annotationexample[PromptGreen]{PythonMod}{Gold fix (excerpt)}{Accepted: \reflink{https://github.com/apache/airflow/pull/63080}{Airflow~PR:\#63080}}
    {assets/annotation_example_is_discoverable_bug_accept_dataloss.txt}
    {Data loss: \texttt{session.expunge\_all()} also detaches rows that the caller queued but did not commit, so they are never written.}}
  {\annotationexample[PromptGreen]{Go}{Gold fix (excerpt)}{Accepted: \reflink{https://github.com/go-git/go-git/pull/1729}{go-git~PR:\#1729}}
    {assets/annotation_example_obviousonly_gogit.txt}
    {Crash (no repository contract): concurrent \texttt{FindHash} calls race on \texttt{offsetHash} and hit Go's fatal concurrent-map error.}}

Like the go-git race, the OpenCV bug below has no repository contract.
However, reading outside of a lookup table is a memory-safety error in any C++ program.

\annotationrow
  {\annotationexample[PromptGreen]{C++}{Gold fix (excerpt)}{Accepted: \reflink{https://github.com/opencv/opencv/pull/21257}{OpenCV~PR:\#21257}}
    {assets/annotation_example_obviousonly_opencv.txt}
    {Crash (no repository contract): an out-of-range value indexes the color LUT out of bounds.}}
  {\mbox{}}

Most non-discoverable candidates are dominated by changes that
\emph{add} a feature or capability rather than fix behavior the repository
already implies; a minority are performance tweaks or changes mislabeled as
fixes. The examples below were rejected at this stage.
In the first one, the tests expect gson to turn a JSON string into a number, but nothing in the repository promises this.
The quarkus change adds a new value to an enum, which is a new feature.

\annotationrow
  {\annotationexample[PromptRed]{Java}{Gold fix (excerpt)}{Rejected: \reflink{https://github.com/google/gson/pull/964}{gson~PR:\#964}}
    {assets/annotation_example_is_discoverable_bug_reject.txt}
    {Not discoverable: no documentation promises that a JSON string becomes a Number.}}
  {\annotationexample[PromptRed]{Java}{Gold fix (excerpt)}{Rejected: \reflink{https://github.com/quarkusio/quarkus/pull/51729}{quarkus~PR:\#51729}}
    {assets/annotation_example_nondiscoverable_quarkus.txt}
    {New feature: adds \texttt{PATCH} to the \texttt{HttpMethod} enum.}}

A change can also look like a fix but only add something that was missing.
The SymPy printer stated that \texttt{Min} and \texttt{Max} were not supported, and the OpenSearch PR is labeled as a bug fix but only adds a new getter.

\annotationrow
  {\annotationexample[PromptRed]{PythonMod}{Gold fix (excerpt)}{Rejected: \reflink{https://github.com/sympy/sympy/pull/22914}{SymPy~PR:\#22914}}
    {assets/annotation_example_nondiscoverable_sympy.txt}
    {New feature: adds printer support for \texttt{Min}/\texttt{Max}.}}
  {\annotationexample[PromptRed]{Java}{Gold fix (excerpt)}{Rejected: \reflink{https://github.com/opensearch-project/OpenSearch/pull/14495}{OpenSearch~PR:\#14495}}
    {assets/annotation_example_nondiscoverable_opensearch.txt}
    {Claim vs.\ code: only adds a \texttt{getKeepAlive()} getter.}}

The last two candidates do not change wrong behavior at all.
The Rails change makes a correct query faster, and the React change adds support for a new HTML attribute.

\annotationrow
  {\annotationexample[PromptRed]{Ruby}{Gold fix (excerpt)}{Rejected: \reflink{https://github.com/rails/rails/pull/58058}{Rails~PR:\#58058}}
    {assets/annotation_example_nondiscoverable_rails.txt}
    {Not a correctness bug: skips the query for a contradictory \texttt{WHERE}; the empty result was already correct.}}
  {\annotationexample[PromptRed]{JavaScript}{Gold fix (excerpt)}{Rejected: \reflink{https://github.com/facebook/react/pull/36148}{React~PR:\#36148}}
    {assets/annotation_example_nondiscoverable_react.txt}
    {New feature: adds \texttt{credentialless} to the known boolean HTML attributes.}}

\subsection{Hidden test quality}
\label{appx:benchmark:appropriate-tests}

\textbf{Hidden tests must not be overly specific.}
The \texttt{appropriate\_tests} annotation checks that the hidden tests are not
overly specific: any reasonable fix for the bug should pass them. For example,
a test should not check unrelated behavior or require the exact wording of an
error message unless that wording is part of the repository's contract.
Appendix~\ref{appx:prompts:appropriate-tests} gives the full prompt.
Table~\ref{tab:lm-review-statistics-appropriate-tests} summarizes acceptance and agreement for this review.

\textbf{Human validation.}
In a human study of \qHumanReviewBugCount{} sampled bugs from the final dataset (\S\ref{appx:benchmark:human-annotation}), a 3-of-4 majority of reviewers found that all \qHumanReviewAppropriateTestsMajorityAcceptCount{} bugs meet the appropriate-tests criterion, and all four reviewers found that \qHumanReviewAppropriateTestsUnanimousAcceptCount{} do.
Each of the remaining \qHumanReviewAppropriateTestsRejectVerdictCount{} bugs drew an objection from only one of the four reviewers.

\begin{table}[htb]
    \centering
    \small
    \caption{\textbf{Acceptance and agreement for the appropriate-tests review.}
    Each percentage is a bug-level average over all 9k bug candidates that pass all non-LM checks.}
    \label{tab:lm-review-statistics-appropriate-tests}
    \begin{center}
        \begingroup
\setlength{\tabcolsep}{5pt}
\renewcommand{\arraystretch}{1.08}
\begin{tabular}{@{}lr@{}}
\multicolumn{2}{c}{\textbf{Agreement}} \\
\cmidrule(lr){1-2}
\textbf{Unanimous agreement} & \textbf{86\%} \\
Opus 5 (high) vs. Sol 5.6 (high) & 88\% \\
Opus 5 (high) vs. Gemini 3.1 Pro & 93\% \\
Sol 5.6 (high) vs. Gemini 3.1 Pro & 90\% \\
\end{tabular}
\hspace{4.5em}
\begin{tabular}{@{}lr@{}}
\multicolumn{2}{c}{\textbf{Acceptance rate}} \\
\cmidrule(lr){1-2}
\textbf{Unanimous accept} & \textbf{76\%} \\
Opus 5 (high) & 88\% \\
Sol 5.6 (high) & 80\% \\
Gemini 3.1 Pro & 84\% \\
\end{tabular}
\endgroup
     \end{center}
\end{table}

\annotationrow
  {\annotationexample[PromptGreen]{PythonMod}{Test patch (excerpt)}{Accepted: \reflink{https://github.com/numpy/numpy/pull/17884}{NumPy~PR:\#17884}}
    {assets/annotation_example_appropriate_tests_accept.txt}
    {The sibling \texttt{np.put} already raises this \texttt{ValueError}, so any writeable-flag check in \texttt{putmask} passes.}}
  {\annotationexample[PromptRed]{PythonMod}{Test patch (truncated)}{Rejected: \reflink{https://github.com/scipy/scipy/pull/23734}{SciPy~PR:\#23734}}
    {assets/annotation_example_appropriate_tests_reject.txt}
    {Pins the exact error text the pull request itself introduced; an equally correct fix worded differently would fail.}}

Over-specificity takes forms other than error-message wording. The next two
rejections pin an internal artifact of one particular fix and an incidental
output shape, respectively.

\annotationrow
  {\annotationexample[PromptRed]{PythonMod}{Test patch (excerpt)}{Rejected: \reflink{https://github.com/mlflow/mlflow/pull/23179}{MLflow~PR:\#23179}}
    {assets/annotation_example_appropriate_tests_reject2.txt}
    {Imports and asserts on \texttt{\_r\_quote}, a helper this fix happens to introduce; an equally correct inline escape would fail.}}
  {\annotationexample[PromptRed]{PythonMod}{Test patch (excerpt)}{Rejected: \reflink{https://github.com/pypa/pip/pull/12281}{pip~PR:\#12281}}
    {assets/annotation_example_appropriate_tests_reject3.txt}
    {Pins each version string to an exact two-tuple; a fix that returned \texttt{(2, 18, 1)} for \texttt{2.18.1} would be equally correct and fail.}}

\textbf{Changes to tests must preserve existing test expectations.}
This rubric checks that the
test patch does not replace an expectation already asserted by the repository's
test suite with a contradictory one. Both the evaluation rule and the inference
prompt state that agents must not break existing base commit tests. A candidate
whose gold fix breaks such a test would therefore not be solvable within the framework of the benchmark and is rejected.
Similarly, bug instances that change test expectations might be more likely to conflict with each other.
Appendix~\ref{appx:prompts:preserves-existing-test-expectations} gives the full
prompt.
Table~\ref{tab:lm-review-statistics-preserves-existing-test-expectations} summarizes acceptance and agreement for this review.

\textbf{Human validation.}
In a human study of \qHumanReviewBugCount{} sampled bugs from the final dataset (\S\ref{appx:benchmark:human-annotation}), all four reviewers found that all \qHumanReviewPreservesExpectationsUnanimousAcceptCount{} bugs meet the criterion of preserving existing test expectations.

\textbf{Additional validation.}
As described in \S\ref{appx:benchmark:quality-control}, we also enforce \hyperlink{appx:benchmark:criterion-r5}{R5} through a deterministic check: for each candidate, we apply its gold fix and run all base commit tests, rejecting the candidate if any test fails.
\begin{table}[tbh]
    \centering
    \small
    \caption{\textbf{Acceptance and agreement for the preservation-of-expectations review.}
    Each percentage is a bug-level average over all 9k bug candidates that pass all non-LM checks.}
    \label{tab:lm-review-statistics-preserves-existing-test-expectations}
    \begin{center}
        \begingroup
\setlength{\tabcolsep}{5pt}
\renewcommand{\arraystretch}{1.08}
\begin{tabular}{@{}lr@{}}
\multicolumn{2}{c}{\textbf{Agreement}} \\
\cmidrule(lr){1-2}
\textbf{Unanimous agreement} & \textbf{98\%} \\
Opus 5 (high) vs. Sol 5.6 (high) & 98\% \\
Opus 5 (high) vs. Gemini 3.1 Pro & 98\% \\
Sol 5.6 (high) vs. Gemini 3.1 Pro & 99\% \\
\end{tabular}
\hspace{4.5em}
\begin{tabular}{@{}lr@{}}
\multicolumn{2}{c}{\textbf{Acceptance rate}} \\
\cmidrule(lr){1-2}
\textbf{Unanimous accept} & \textbf{86\%} \\
Opus 5 (high) & 87\% \\
Sol 5.6 (high) & 87\% \\
Gemini 3.1 Pro & 87\% \\
\end{tabular}
\endgroup
     \end{center}
\end{table}

\annotationrow
  {\annotationexample[PromptGreen]{Go}{Test patch (complete)}{Accepted: \reflink{https://github.com/golang/go/commit/18104621ce742af7be8d5049bb9aee588b562950}{Go~commit~1810462}}
    {assets/annotation_example_preserves_existing_test_expectations_accept1.txt}
    {One row appended to an existing test table, between untouched entries; no prior expectation is edited.}}
  {\annotationexample[PromptRed]{PythonMod}{Test patch (truncated)}{Rejected: \reflink{https://github.com/keras-team/keras/pull/22137}{Keras~PR:\#22137}}
    {assets/annotation_example_preserves_existing_test_expectations_reject1.txt}
    {An existing assertion is rewritten in place: the same inputs are now expected to cast to \texttt{float32} rather than \texttt{float16}.}}

\annotationrow
  {\annotationexample[PromptGreen]{Ruby}{Test patch (complete)}{Accepted: \reflink{https://github.com/ruby/ruby/pull/15792}{Ruby~PR:\#15792}}
    {assets/annotation_example_preserves_existing_test_expectations_accept2.txt}
    {One assertion appended to an existing test, tightening it without revising any previous expectation.}}
  {\annotationexample[PromptRed]{JavaScript}{Test patch (truncated)}{Rejected: \reflink{https://github.com/mui/material-ui/pull/44408}{MUI~PR:\#44408}}
    {assets/annotation_example_preserves_existing_test_expectations_reject2.txt}
    {The same conformance entry is edited: the rendered ref is now expected to be a button rather than a div.}}

\subsection{Additional quality checks}
\label{appx:benchmark:other-quality}

\textbf{Single unit of work}.
The bug instances are the smallest unit of scoring: an agent receives credit only when all of its hidden tests pass. Consequently, every instance must correspond to a single unit of work.
The \texttt{is\_single\_bug} annotation checks that a bug instance's hidden tests
correspond to a single root cause (that might or might not require multiple related changes to be fixed).

Appendix~\ref{appx:prompts:is-single-bug} gives the full prompt.
Table~\ref{tab:lm-review-statistics-single-bug} summarizes acceptance and agreement for this review.

\begin{table}[tb]
    \centering
    \small
    \caption{\textbf{Acceptance and agreement for the single-bug review.}
    Each percentage is a bug-level average over all 9k bug candidates that pass all non-LM checks.}
    \label{tab:lm-review-statistics-single-bug}
    \begin{center}
        \begingroup
\setlength{\tabcolsep}{5pt}
\renewcommand{\arraystretch}{1.08}
\begin{tabular}{@{}lr@{}}
\multicolumn{2}{c}{\textbf{Agreement}} \\
\cmidrule(lr){1-2}
\textbf{Unanimous agreement} & \textbf{93\%} \\
Opus 5 (high) vs. Sol 5.6 (high) & 94\% \\
Opus 5 (high) vs. Gemini 3.1 Pro & 96\% \\
Sol 5.6 (high) vs. Gemini 3.1 Pro & 95\% \\
\end{tabular}
\hspace{4.5em}
\begin{tabular}{@{}lr@{}}
\multicolumn{2}{c}{\textbf{Acceptance rate}} \\
\cmidrule(lr){1-2}
\textbf{Unanimous accept} & \textbf{90\%} \\
Opus 5 (high) & 93\% \\
Sol 5.6 (high) & 94\% \\
Gemini 3.1 Pro & 94\% \\
\end{tabular}
\endgroup
     \end{center}
\end{table}

Shown below are four examples of accepted and rejected bug instances based on this criterion.

\annotationrow
  {\annotationexample[PromptGreen]{PythonMod}{Gold fix (excerpt)}{Accepted: \reflink{https://github.com/keras-team/keras/pull/23375}{Keras~PR:\#23375}}
    {assets/annotation_example_is_single_bug_accept1.txt}
    {One key missing from \texttt{get\_config()}, so \texttt{sparse=True} is silently lost on a serialization round trip.}}
  {\annotationexample[PromptRed]{JavaScript}{Gold fix (excerpt, two files)}{Rejected: \reflink{https://github.com/sequelize/sequelize/pull/11454}{Sequelize~PR:\#11454}}
    {assets/annotation_example_is_single_bug_reject1.txt}
    {Two unrelated fixes in one patch: \texttt{NULL} comparator handling in the query generator, and \texttt{Where} support in an array helper.}}

\annotationrow
  {\annotationexample[PromptGreen]{Go}{Gold fix (excerpt)}{Accepted: \reflink{https://github.com/labstack/echo/pull/2385}{echo~PR:\#2385}}
    {assets/annotation_example_is_single_bug_accept2.txt}
    {One stray assignment copied the raw path into the context before matching; deleting it is the whole fix.}}
  {\annotationexample[PromptRed]{PythonMod}{Gold fix (excerpt)}{Rejected: \reflink{https://github.com/sympy/sympy/pull/19135}{SymPy~PR:\#19135}}
    {assets/annotation_example_is_single_bug_reject2.txt}
    {Two independent parser rules added at once; either could ship without the other.}}

\textbf{Human validation.}
In a human study of \qHumanReviewBugCount{} sampled bugs from the final dataset (\S\ref{appx:benchmark:human-annotation}), a 3-of-4 majority of reviewers found that all \qHumanReviewSingleBugMajorityAcceptCount{} bugs meet the single-bug criterion, and all four reviewers found that \qHumanReviewSingleBugUnanimousAcceptCount{} do.
The remaining bug drew an objection from only one of the four reviewers.

\textbf{Compatibility of all bug instances within a repository.}
We need to establish that all tests in the same repository can be satisfied together. We therefore run one
agent per repository using the same agent setup as described in \S\ref{appx:evidence_expected}. The judge receives the base
repository and an index of all selected test patches. It is instructed to inspect every patch
and report pairs that require mutually exclusive behavior, with exact quotes from both patches.
We verify that every bug identifier is returned, then compute the largest compatible subset from
the reported conflicts.
This procedure did not report any conflicting pairs.
Appendix~\ref{appx:prompts:task-compatibility} gives the complete prompt.

 \clearpage
\newpage
\section{Additional experimental results}
\label{appx:results}

This appendix provides additional performance metrics (\S\ref{appx:results:performance-metrics}) that supplement the main results in \S\ref{sec:results}.
Furthermore, \S\ref{appx:agent-behavior} significantly extends the trajectory findings in \S\ref{sec:analysis}.
Finally, \S\ref{appx:results:per-repository} provides per-repository results for all model variants.
\subsection{Performance metrics}
\label{appx:results:performance-metrics}
This section provides additional breakdowns of the main results. We compare model performance with token usage, examine different ways of averaging scores, break down results by programming language, and compare performance across repositories of different sizes.

\begin{figure}[h]
    \centering
    \includegraphics[width=0.50\linewidth]{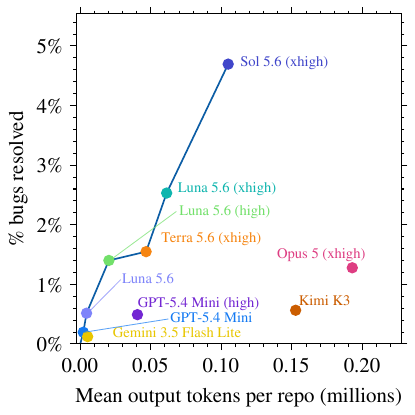}
    \caption{Percentage of bugs resolved versus mean total output tokens per repository; the line marks
    the Pareto frontier.}
    \label{fig:score-tokens-pareto}
\end{figure}

\textbf{Token usage.} Figure~\ref{fig:score-tokens-pareto} shows the percentage of bugs resolved versus total output tokens,
similar to the comparison in Figure~\ref{fig:score-cost-pareto}.
Bugs resolved rise with total output tokens, especially within a model family.

\textbf{Performance metrics and averaging.} Table~\ref{tab:score-averages} compares the percentage of bugs resolved across models.
No repository is fully repaired by any model, with the best repaired repository having around half of its bugs resolved.
The table also shows two different averaging methods to compute the same score:
The bug-micro averaging gives each evaluated bug equal weight, while the repo-macro averaging gives each repository equal weight.
Both metrics lead to similar rankings of the models.
We choose the micro-average, because it is conceptually easier, and equivalent to weighing the different repositories by the number of bugs they contain, which correlates heavily with difficulty (see Table~\ref{tab:score-by-bug-count}).

\textbf{Results by language.} Table~\ref{tab:language-compilation} compares results by language.
Performance is low across all language groups. Non-compiled repositories have
slightly higher average and union results than compiled repositories, while the differences among
individual language groups are larger.
Results for C/C++ repositories are the lowest overall, while JavaScript/TypeScript shows the highest results, most clearly for GPT-5.6 Sol.

\begin{figure}[p]
    \centering
    \includegraphics[width=0.7\linewidth]{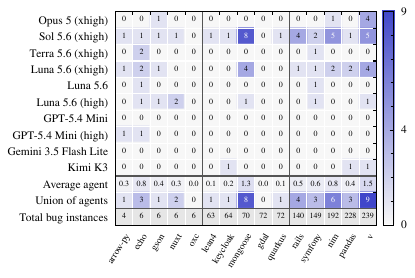}
    \caption{Number of bugs resolved for the five smallest repositories below 25 benchmark bugs,
    the five largest repositories with 25--74 bugs, and the five largest repositories
    with at least 75 bugs. Average agent is the mean across models; union of agents counts
    distinct bugs resolved by at least one model. The final row gives each repository's
    total benchmark bug instances.}
    \label{fig:task-score-heatmap}
\end{figure}

\textbf{Performance across repositories of different sizes.} Figure~\ref{fig:task-score-heatmap} compares resolved-bug counts across small,
medium, and large repositories.

\begin{table}[p]
    \centering
    \small
    \caption{Percentage of bugs resolved under bug-micro averaging, which gives
    every evaluated bug equal weight, and repo-macro averaging, which gives every
    repository equal weight; highest percentage resolved in one repository; and share of fully repaired repositories.}
    \label{tab:score-averages}
    \begin{center}
        \begin{tabular}{lrrrr}
\toprule
\textbf{Model} & \shortstack{\textbf{\% bugs resolved}\\\textbf{(bug-micro)}} & \shortstack{\textbf{\% bugs resolved}\\\textbf{(repo-macro)}} & \textbf{Best repo} & \textbf{Fully repaired} \\
\midrule
Sol 5.6 (xhigh) & 4.7\% & 10.3\% & 55.6\% & 0.0\% \\
Luna 5.6 (xhigh) & 2.5\% & 6.6\% & 44.4\% & 0.0\% \\
Terra 5.6 (xhigh) & 1.5\% & 4.3\% & 40.0\% & 0.0\% \\
Luna 5.6 (high) & 1.4\% & 4.8\% & 50.0\% & 0.0\% \\
Opus 5 (xhigh) & 1.3\% & 3.1\% & 33.3\% & 0.0\% \\
Kimi K3 & 0.6\% & 1.3\% & 20.0\% & 0.0\% \\
Luna 5.6 & 0.5\% & 1.1\% & 22.2\% & 0.0\% \\
GPT-5.4 Mini (high) & 0.5\% & 1.6\% & 25.0\% & 0.0\% \\
GPT-5.4 Mini & 0.2\% & 0.4\% & 11.1\% & 0.0\% \\
Gemini 3.5 Flash Lite & 0.1\% & 0.1\% & 5.3\% & 0.0\% \\
\bottomrule
\end{tabular}
 \end{center}
\end{table}

\begin{table}[p]
    \centering
    \small
    \caption{Results by language. Average is the mean across the three displayed agents;
    union counts each bug resolved by any included run of those agents.}
    \label{tab:language-compilation}
    \begin{center}
        \begin{tabular}{lrrrrrrr}
\toprule
\textbf{Language} & \textbf{Repos} & \textbf{Bugs} & \textbf{GPT-5.6 Sol} & \textbf{Opus 5} & \textbf{Luna 5.6} & \textbf{Average} & \textbf{Union} \\
\midrule
Compiled & 58 & 2,202 & 4.0\% & 1.2\% & 2.0\% & 2.4\% & 5.8\% \\
Non-compiled & 42 & 1,866 & 5.5\% & 1.4\% & 3.1\% & 3.3\% & 7.8\% \\
\midrule
Python & 25 & 1,201 & 4.5\% & 1.6\% & 2.7\% & 2.9\% & 7.2\% \\
JavaScript/TypeScript & 17 & 413 & 11.6\% & 1.9\% & 6.3\% & 6.6\% & 13.8\% \\
C/C++ & 14 & 488 & 2.0\% & 1.2\% & 0.8\% & 1.4\% & 3.1\% \\
Java & 10 & 363 & 6.9\% & 1.4\% & 3.3\% & 3.9\% & 8.5\% \\
Other & 34 & 1,603 & 3.4\% & 0.9\% & 1.8\% & 2.0\% & 5.2\% \\
\bottomrule
\end{tabular}
 \end{center}
\end{table}

\clearpage
\raggedbottom
\subsection{Agent behavior}
\label{appx:agent-behavior}

Expanding on \S\ref{sec:analysis}, our evaluation across \qDiscoverabilityOutcomeModelCount{} models and \qAgentBehaviorTaskCount{} repositories reveals a substantial gap between reading the repository and actual bug discovery and repair: agents navigate to bug-relevant files far more frequently than they successfully generate complete fixes. This discrepancy highlights why current models resolve only a small fraction of bugs. Crucially, trajectory analysis indicates that agents frequently regress during execution, undoing previously correct edits in later steps. Furthermore, model performance exhibits strong diversity: different models resolve distinct populations of bugs, while scaling repeated attempts across longer action trajectories and multiple agentic inferences yields diminishing, rapidly plateauing returns.

Our trajectory dataset encompasses \qAgentBehaviorRunCount{} runs across \qAgentBehaviorBugCount{} bug instances (\qAgentBehaviorUnknownFinalCount{} final outcomes were unavailable). 
Compared to the final benchmark dataset of 4,068 bugs used for all main results, the population used in this appendix included some bug instances that were later filtered out after all quality criteria were consistently applied. Since this corresponds to only around 2\% of the benchmark population, we believe this does not significantly impact the findings of this section.

We define a \textit{bug-relevant file} as any pre-existing source file modified by the gold fix. To track patch progression over time, evaluations are conducted at fixed checkpoints every 20 actions and upon run completion, totaling \qAgentBehaviorCheckpointCount{} checkpoints. When identical patches occur repeatedly within a trajectory, we record their last available evaluation.

\textbf{Models differ in how broadly they explore the repository.}
Reading makes up \qAgentBehaviorOpusReadActionShare{} of Opus actions and
\qAgentBehaviorTerraReadActionShare{} of Terra actions. Writes make up
\qAgentBehaviorOpusWriteActionShare{} of Opus actions and
\qAgentBehaviorTerraWriteActionShare{} of Terra actions, with Terra being the only model that writes more
often than it reads (Table~\ref{tab:agent-behavior-exploration},
Figure~\ref{fig:action-frequency}). \\
Models also differ in file breadth: Opus reads a mean of
\qAgentBehaviorOpusReadFiles{} distinct files per repository, more than three times Terra's
\qAgentBehaviorTerraReadFiles{}. Across models, agents often reach relevant code: Terra reads at
least one bug-relevant file in
\qAgentBehaviorTerraContactRuns{} of its \qAgentBehaviorTaskCount{} runs, compared with
\qAgentBehaviorSolContactRuns{} for Sol.

\par\smallskip
\noindent\begin{minipage}{\textwidth}
    \centering
    \small
    \setlength{\tabcolsep}{3pt}s
    \captionof{table}{How models explore the \qAgentBehaviorTaskCount{} repositories. Read, write,
    and execute are percentages of all actions; each action appears once. \textit{Files read} is the
    mean number of different files read per repository, and \textit{Reads per file} is the mean number
    of reads of each file. A run reaches a bug-relevant file when it reads any file changed by a gold
    fix.}
    \label{tab:agent-behavior-exploration}
    \begin{center}
        \begin{tabular}{@{}lrrrrrrr@{}}
\toprule
\textbf{Model} & \textbf{Read} & \textbf{Write} & \textbf{Execute} & \shortstack{\textbf{Files}\\\textbf{read}} & \shortstack{\textbf{Reads}\\\textbf{per file}} & \shortstack{\textbf{Runs reaching a}\\\textbf{bug-relevant file}} & \shortstack{\textbf{First contact}\\\textbf{progress}} \\
\midrule
Opus 5 & 52.5\% & 25.5\% & 18.8\% & 159.3 & 1.7 & 97/100 & 14.6\% \\
Sol 5.6 & 42.5\% & 31.7\% & 19.5\% & 113.4 & 2.0 & 98/100 & 12.6\% \\
Terra 5.6 & 35.9\% & 37.8\% & 19.4\% & 48.9 & 1.9 & 86/100 & 24.3\% \\
Luna 5.6 & 48.6\% & 28.0\% & 17.7\% & 89.5 & 2.3 & 96/100 & 10.9\% \\
\bottomrule
\end{tabular}
 \end{center}
\end{minipage}
\par\smallskip

\par\smallskip
\noindent\begin{minipage}{\textwidth}
    \centering
    \includegraphics[width=\linewidth]{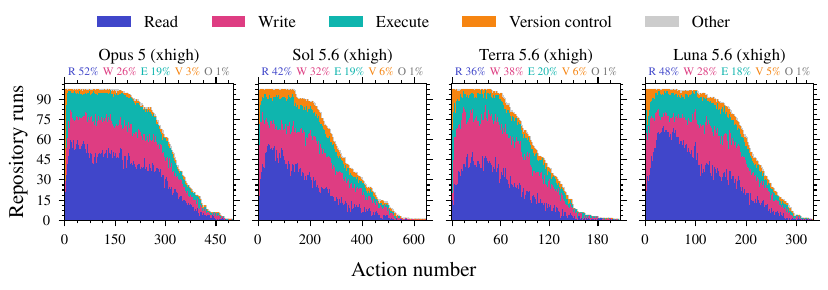}
    \captionof{figure}{Distribution of observable actions for each model across
    \qAgentBehaviorTaskCount{} repositories.
    Actions are ordered from the start to the end of each run. At every action number, the figure
    shows the number of runs with a read, write, execute, version-control, or other action. Not every
    run is present at later actions because runs use different numbers of actions; the number does
    not measure performance. Labels show the percentage across all actions.}
    \label{fig:action-frequency}
\end{minipage}
\par\smallskip

\textbf{Agents read a small share of the repository, and how small depends on its size.}
The median run reads \qCoverageTerraMedian{}--\qCoverageSolMedian{} of a repository's tracked files
(Table~\ref{tab:agent-behavior-coverage}). Repository size spans three orders of magnitude, so we
take the share within each repository rather than pooling. The distribution is skewed. Agents read
most of a small tree; Opus reads 45 of \texttt{testify}'s 55 files. On a large tree they read only
a few percent.

\par\smallskip
\noindent\begin{minipage}{\textwidth}
    \centering
    \small
    \setlength{\tabcolsep}{4pt}
    \captionof{table}{\textbf{Agents read a small share of each repository.} The share is the
    number of distinct files an agent explicitly read divided by the number of tracked files at the
    base commit, computed per repository. Two outlier Opus runs (\texttt{rails}, \texttt{nim}) are
    excluded: in each, a single glob command matched and counted every file in a directory as read,
    reaching 10.7 and 4.5 files per action against a cohort median of 0.28--0.40. This inflation also
    affects the file counts reported above, which are therefore upper bounds.}
    \label{tab:agent-behavior-coverage}
    \begin{center}
    \begin{tabular}{@{}lrrr@{}}
        \toprule
        \textbf{Model} & \textbf{Runs} & \shortstack{\textbf{Median share}\\\textbf{of repository read}} &
        \shortstack{\textbf{Largest share}\\\textbf{(repository)}} \\
        \midrule
        Opus 5    & \qCoverageOpusRuns{}  & \qCoverageOpusMedian{}  & \qCoverageOpusMax{} (\texttt{testify}) \\
        Sol 5.6   & \qCoverageSolRuns{}   & \qCoverageSolMedian{}   & \qCoverageSolMax{} (\texttt{formulajs}) \\
        Terra 5.6 & \qCoverageTerraRuns{} & \qCoverageTerraMedian{} & \qCoverageTerraMax{} (\texttt{testify}) \\
        Luna 5.6  & \qCoverageLunaRuns{}  & \qCoverageLunaMedian{}  & \qCoverageLunaMax{} (\texttt{formulajs}) \\
        \bottomrule
    \end{tabular}
    \end{center}
\end{minipage}
\par\smallskip

\textbf{Reading and editing bug-relevant files is far more common than resolving bugs.}
Across models, agents read at least one bug-relevant file for
\qAgentBehaviorTerraAnyRead{}--\qAgentBehaviorSolAnyRead{} of bugs, edit one for
\qAgentBehaviorOpusAnyEdit{}--\qAgentBehaviorSolAnyEdit{}, and resolve
\qAgentBehaviorOpusResolved{}--\qAgentBehaviorSolResolved{}
(Table~\ref{tab:agent-behavior-funnel}). Changes to the same lines as a gold fix are less
common still. Even when agents edit every bug-relevant file, most bugs are not resolved, showing that finding and changing relevant code does not always produce a correct repair.

\par\smallskip
\noindent\begin{minipage}{\textwidth}
    \centering
    \small
    \setlength{\tabcolsep}{4pt}
    \captionof{table}{Reading, editing, and bug resolution across all bug instances. ``Any bug-relevant
    file read'' means that an agent reads at least one file changed by a gold fix; ``Any bug-relevant
    file edited'' means that an agent changes at least one. ``Gold fix lines edited'' requires
    an agent change to include a line changed by that fix. The final column is for bugs with every
    bug-relevant file edited and hidden-test results available.}
    \label{tab:agent-behavior-funnel}
    \begin{center}
        \begin{tabular}{@{}lrrrrr@{}}
\toprule
\textbf{Model} & \shortstack{\textbf{Any bug-relevant}\\\textbf{file read}} & \shortstack{\textbf{Any bug-relevant}\\\textbf{file edited}} & \shortstack{\textbf{Gold fix}\\\textbf{lines edited}} & \textbf{Resolved} & \shortstack{\textbf{Resolved given all}\\\textbf{bug-relevant files edited}} \\
\midrule
Opus 5 & \shortstack{36.2\%\\1,506/4,165} & \shortstack{6.0\%\\249/4,165} & \shortstack{0.6\%\\27/4,165} & \shortstack{1.3\%\\55/4,120} & \shortstack{22.8\%\\46/202} \\
Sol 5.6 & \shortstack{38.5\%\\1,602/4,165} & \shortstack{21.5\%\\897/4,165} & \shortstack{3.7\%\\156/4,165} & \shortstack{5.0\%\\208/4,120} & \shortstack{24.4\%\\196/804} \\
Terra 5.6 & \shortstack{23.4\%\\973/4,165} & \shortstack{8.5\%\\353/4,165} & \shortstack{1.2\%\\50/4,165} & \shortstack{1.6\%\\68/4,135} & \shortstack{19.9\%\\62/312} \\
Luna 5.6 & \shortstack{37.9\%\\1,580/4,165} & \shortstack{17.1\%\\711/4,165} & \shortstack{2.0\%\\85/4,165} & \shortstack{2.7\%\\113/4,121} & \shortstack{16.4\%\\102/621} \\
\bottomrule
\end{tabular}
 \end{center}
\end{minipage}
\par\smallskip

\textbf{The edit-to-resolve gap holds regardless of what evidence a bug carries.}
Among \qDiscoverabilityOutcomeAnnotatedCount{} labeled bugs, at least one model edited a
bug-relevant file for \qDiscoverabilityOutcomeAnyEditCount{}
(\qDiscoverabilityOutcomeAnyEditShare{}). Yet at least one model resolved only
\qDiscoverabilityOutcomeAnyResolvedCount{} of the
\qDiscoverabilityOutcomeAnyResolvedKnownCount{} with available results
(\qDiscoverabilityOutcomeAnyResolvedShare{}). This gap holds both for bugs with repository
evidence only and for bugs with both sources (Table~\ref{tab:discoverability-outcomes}). The labels
show which evidence is available for each bug, not which evidence an agent used.

\par\smallskip
\noindent\begin{minipage}{\textwidth}
    \centering
    \small
    \setlength{\tabcolsep}{4pt}
    \captionof{table}{Evidence sources and results for labeled bugs. Each bug is present once. ``Any
    edit'' means that at least one of \qDiscoverabilityOutcomeModelCount{} models edited a file
    changed by a gold fix. The resolved percentage is for bugs resolved by at least one model.
    Bugs without available hidden-test results are not included in the resolved percentage, and
    bugs without evidence labels are not present.}
    \label{tab:discoverability-outcomes}
    \begin{center}
        \begin{tabular}{@{}lrrr@{}}
\toprule
\textbf{Evidence available} & \textbf{Bugs} & \textbf{Any edit} & \textbf{Resolved} \\
\midrule
Repository evidence only & 3,894 & 1,132/3,894 (29.1\%) & 243/3,849 (6.3\%) \\
Repository-independent failure only & 2 & 1/2 (50.0\%) & 0/2 (0.0\%) \\
Both & 225 & 42/225 (18.7\%) & 9/225 (4.0\%) \\
\midrule
All labeled bugs & 4,121 & 1,175/4,121 (28.5\%) & 252/4,076 (6.2\%) \\
\bottomrule
\end{tabular}
 \end{center}
\end{minipage}
\par\smallskip

\textbf{Models read before they edit or resolve bugs.}
Models read before they make any file edits. By the halfway mark (measured in total actions
performed), the number of bug-relevant files read is
\qAgentBehaviorMinHalfwayReadEndpoint{}--\qAgentBehaviorMaxHalfwayReadEndpoint{} of the final number,
the number edited is
\qAgentBehaviorMinHalfwayEditEndpoint{}--\qAgentBehaviorMaxHalfwayEditEndpoint{}, and the number of
bugs resolved is
\qAgentBehaviorMinHalfwayResolvedEndpoint{}--\qAgentBehaviorMaxHalfwayResolvedEndpoint{}
(Figures~\ref{fig:agent-behavior-progress-action}--\ref{fig:agent-behavior-progress-time}). Agents thus do more of their reading early in a run, while editing and bug recovery continue later.

\par\smallskip
\noindent\begin{minipage}{\textwidth}
    \centering
    \includegraphics[width=\linewidth]{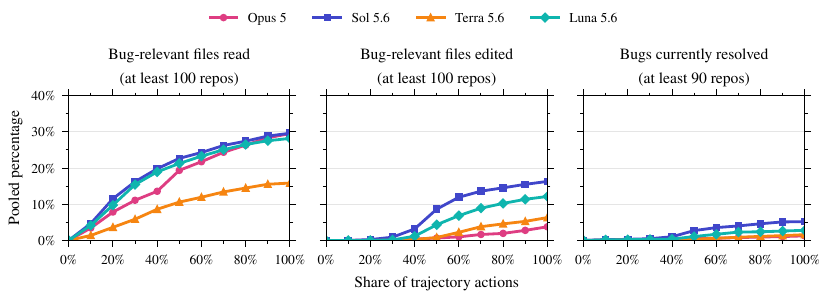}
    \captionof{figure}{Bug-relevant files read, bug-relevant files edited, and bugs resolved over
    the percentage of actions in each run. All three results use the same range from 0 to 40\%.
    Each curve is the percentage across all bugs in all repositories. Reading
    changes after each action; editing and resolution change only when a patch is evaluated. We do
    not count missing hidden-test results as failures. }
    \label{fig:agent-behavior-progress-action}
\end{minipage}
\par\smallskip

\par\smallskip
\noindent\begin{minipage}{\textwidth}
    \centering
    \includegraphics[width=\linewidth]{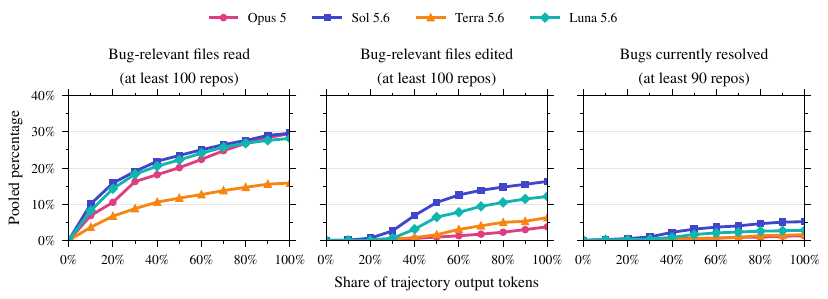}
    \captionof{figure}{The same reading, editing, and resolved-bug results with progress measured by
    output tokens. Progress is from 0\% to 100\% for every run. Reading changes after each
    action; editing and resolution change only when a patch is evaluated. We do not count missing
    hidden-test results as failures.}
    \label{fig:agent-behavior-progress-tokens}
\end{minipage}
\par\smallskip

\par\smallskip
\noindent\begin{minipage}{\textwidth}
    \centering
    \includegraphics[width=\linewidth]{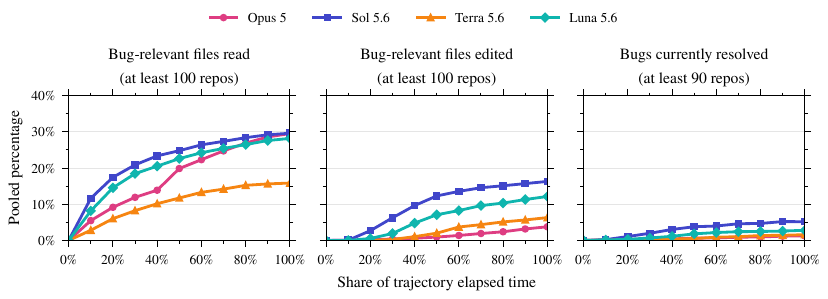}
    \captionof{figure}{The same reading, editing, and resolved-bug results with progress measured by
    elapsed time. Progress is from 0\% to 100\% within each run. Repository counts report how many runs have
    available time and hidden-test results. We do not count missing results as failures.}
    \label{fig:agent-behavior-progress-time}
\end{minipage}
\par\smallskip

\textbf{Agents change more lines across a run than the final patch contains.}
Across models, the median end-state patch ranges from \qAgentBehaviorMinMedianFinalLines{} to
\qAgentBehaviorMaxMedianFinalLines{} lines, while the median total number of lines changed across the trajectory
(i.e., total churn) is
\qAgentBehaviorMinMedianChurnLines{} to \qAgentBehaviorMaxMedianChurnLines{} lines
(Figure~\ref{fig:agent-behavior-patch-dynamics}, Table~\ref{tab:agent-behavior-dynamics}). Both
numbers are agent changes; neither is the gold fix.
We see that agents \emph{churn} frequently, changing many lines that are eventually not present in the final patch. Furthermore, most agent changes are not in the locations changed by a gold fix. In the submitted patches, only
\qTrajectorySolFilePrecision{}--\qTrajectoryOpusFilePrecision{} of
files changed by agents are bug-relevant. Of changed-line ranges, only
\qTrajectorySolRegionPrecision{}--\qTrajectoryOpusRegionPrecision{} include lines from a gold
fix. These percentages use one gold fix per bug and do not measure correctness,
because a valid repair may change other code: most changed files and line ranges are not part of that repair.

\par\smallskip
\noindent\begin{minipage}{\textwidth}
    \centering
    \includegraphics[width=\linewidth]{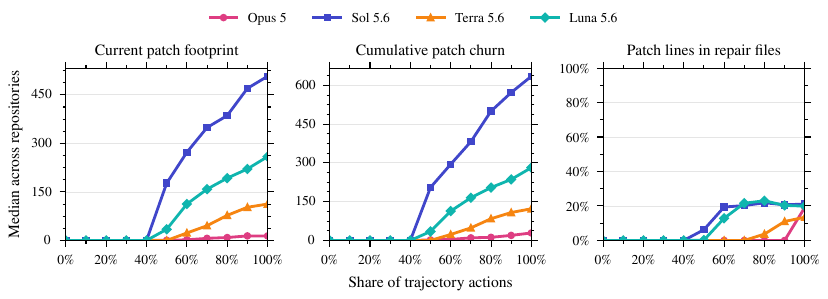}
    \captionof{figure}{Patch changes over the percentage of actions in each run. The first result
    shows the number of lines changed in the current patch. The second shows all line changes from
    one patch to the next across the run. The third shows the percentage of current changed lines
    in bug-relevant files. Each result is the median across \qAgentBehaviorTaskCount{} repositories.
    At 100\%, the patch is the submitted patch.}
    \label{fig:agent-behavior-patch-dynamics}
\end{minipage}
\par\smallskip

\textbf{Later work can undo correct changes, but most runs finish at their best checkpoint.}
Across the \qAgentBehaviorRunCount{} runs at increasing-time checkpoints, \qAgentBehaviorRegressionRuns{} lose at least one
previously resolved bug after the patch changes (Table~\ref{tab:agent-behavior-dynamics-main}): these patch changes undo \qAgentBehaviorRegressionEvents{} previously resolved bugs.
\qAgentBehaviorTransientBugs{} model-bug repairs pass at some point but not at the end, but only
\qAgentBehaviorRunsBelowPeak{}/\qAgentBehaviorRunCount{} runs finish below their best observed
checkpoint. A run's best checkpoint generally has the largest number of bugs resolved. Among runs with at
least one bug resolved, models spend a median
\qAgentBehaviorMinPostImprovement{}--\qAgentBehaviorMaxPostImprovement{} of their actions after the
last increase in bugs resolved. We count a regression only when the patch changes; repeated
evaluations of unchanged patches are reported below. Because we evaluate patches only at checkpoints, we
know the first checkpoint at which a bug is resolved, not the exact action that resolved it.

\par\smallskip
\noindent\begin{minipage}{\textwidth}
    \centering
    \small
    \setlength{\tabcolsep}{4pt}
    \captionof{table}{\textbf{Agents can undo correct repairs and continue working after their best
    checkpoint.} A regression is a bug resolved before a patch change but not resolved after it;
    this is about bugs resolved, not about the regression tests. ``Runs with regression'' counts
    runs with at least one regression.
    ``Resolved before but not at end'' counts bugs resolved at some checkpoint but not at the final
    checkpoint. ``Runs below best'' counts runs with more bugs resolved at a checkpoint than at the
    final checkpoint. ``Actions after last increase'' is the median remaining percentage
    of actions after the last increase in bugs resolved, among runs that resolve at least one bug.}
    \label{tab:agent-behavior-dynamics-main}
    \begin{center}
    \begin{tabular}{@{}lrrrr@{}}
        \toprule
        \textbf{Model} & \shortstack{\textbf{Runs with}\\\textbf{regression}} &
        \shortstack{\textbf{Resolved before}\\\textbf{but not at end}} &
        \shortstack{\textbf{Runs below}\\\textbf{best}} &
        \shortstack{\textbf{Actions after}\\\textbf{last increase}} \\
        \midrule
        Opus 5 & \qAgentBehaviorOpusRegressionRuns{}/\qAgentBehaviorTaskCount{} &
        \qAgentBehaviorOpusTransientBugs{} & \qAgentBehaviorOpusRunsBelowPeak{} &
        \qAgentBehaviorOpusPostImprovement{} \\
        Sol 5.6 & \qAgentBehaviorSolRegressionRuns{}/\qAgentBehaviorTaskCount{} &
        \qAgentBehaviorSolTransientBugs{} & \qAgentBehaviorSolRunsBelowPeak{} &
        \qAgentBehaviorSolPostImprovement{} \\
        Terra 5.6 & \qAgentBehaviorTerraRegressionRuns{}/\qAgentBehaviorTaskCount{} &
        \qAgentBehaviorTerraTransientBugs{} & \qAgentBehaviorTerraRunsBelowPeak{} &
        \qAgentBehaviorTerraPostImprovement{} \\
        Luna 5.6 & \qAgentBehaviorLunaRegressionRuns{}/\qAgentBehaviorTaskCount{} &
        \qAgentBehaviorLunaTransientBugs{} & \qAgentBehaviorLunaRunsBelowPeak{} &
        \qAgentBehaviorLunaPostImprovement{} \\
        \bottomrule
    \end{tabular}
    \end{center}
\end{minipage}
\par\smallskip

\par\smallskip
\noindent\begin{minipage}{\textwidth}
    \centering
    \scriptsize
    \setlength{\tabcolsep}{3pt}
    \captionof{table}{Additional results on patch changes and regressions by model. Patch lines and
    total changes show the median across repositories. For the same patch, we use its last
    available evaluation; we report a regression only when the patch changes. ``Known
    checkpoints'' is the number of checkpoints with an available evaluation out of all checkpoints
    for that model. ``Final patch lines'' is the size of the final patch, ``Maximum patch lines'' is
    the largest number of changed lines at any checkpoint, and ``Total churn'' is the total number
    of lines changed across all patch changes in a run. ``Runs with regression'' uses the definition
    above. ``Transient repairs'' is ``Resolved before but not at end,'' and ``Work after last
    improvement'' is ``Actions after last increase.''}
    \label{tab:agent-behavior-dynamics}
    \begin{center}
        \resizebox{\textwidth}{!}{\begin{tabular}{@{}lrrrrrrr@{}}
\toprule
\textbf{Model} & \shortstack{\textbf{Known}\\\textbf{checkpoints}} & \shortstack{\textbf{Final patch}\\\textbf{lines}} & \shortstack{\textbf{Maximum patch}\\\textbf{lines}} & \shortstack{\textbf{Total}\\\textbf{churn}} & \shortstack{\textbf{Runs with}\\\textbf{regression}} & \shortstack{\textbf{Transient}\\\textbf{repairs}} & \shortstack{\textbf{Work after last}\\\textbf{improvement}} \\
\midrule
Opus 5 & 1,526/1,670 & 14.0 & 22.0 & 29.0 & 6/100 & 5 & 25.9\% \\
Sol 5.6 & 1,590/1,756 & 507.5 & 557.5 & 635.0 & 17/100 & 12 & 24.8\% \\
Terra 5.6 & 563/615 & 112.5 & 114.0 & 122.5 & 5/100 & 5 & 25.9\% \\
Luna 5.6 & 1,030/1,132 & 259.0 & 267.0 & 281.5 & 14/100 & 6 & 37.6\% \\
\bottomrule
\end{tabular}
 }
\end{center}
\end{minipage}
\par\smallskip

\textbf{Agents stop because they judge their work done, not because they run out of budget.}
No run approaches either limit (Table~\ref{tab:agent-behavior-stopping}). Against a
\qStoppingStepLimit{}-action limit, the longest run uses \qStoppingMaxActions{}. Against a
\qStoppingWallHours{}-hour wall-clock limit, the longest takes \qStoppingMaxHours{}
(\qStoppingMaxWallShare{}). All \qStoppingSubmitted{}/\qAgentBehaviorRunCount{} runs end by
submitting. In their closing messages, \qStoppingBudgetCiting{} runs report an exhausted
search or budget. Stopping is deliberate: \qStoppingGitCheckRuns{} runs inspect their
diff or status in their last five commands. The task prompt asks agents to fix as many bugs as they
can, yet they stop well within their budget.

\par\smallskip
\noindent\begin{minipage}{\textwidth}
    \centering
    \small
    \setlength{\tabcolsep}{4pt}
    \captionof{table}{\textbf{No run exhausts its action or time budget.} Medians and maxima are
    taken over the \qAgentBehaviorTaskCount{} repositories for each model. ``At step limit'' counts
    runs reaching the \qStoppingStepLimit{}-step limit and ``at time limit'' counts runs reaching
    the \qStoppingWallHours{}-hour wall-clock limit. Actions are shell commands; a single model turn may
    issue more than one.}
    \label{tab:agent-behavior-stopping}
    \begin{center}
    \begin{tabular}{@{}lrrrrrr@{}}
        \toprule
        \textbf{Model} & \shortstack{\textbf{Median}\\\textbf{actions}} &
        \shortstack{\textbf{Most}\\\textbf{actions}} &
        \shortstack{\textbf{At step}\\\textbf{limit}} &
        \shortstack{\textbf{Median}\\\textbf{hours}} &
        \shortstack{\textbf{Most}\\\textbf{hours}} &
        \shortstack{\textbf{At time}\\\textbf{limit}} \\
        \midrule
        Opus 5    & \qStoppingOpusMedActions{}  & \qStoppingOpusMaxActions{}  & 0/\qAgentBehaviorTaskCount{} & \qStoppingOpusMedHours{}  & \qStoppingOpusMaxHours{}  & 0/\qAgentBehaviorTaskCount{} \\
        Sol 5.6   & \qStoppingSolMedActions{}   & \qStoppingSolMaxActions{}   & 0/\qAgentBehaviorTaskCount{} & \qStoppingSolMedHours{}   & \qStoppingSolMaxHours{}   & 0/\qAgentBehaviorTaskCount{} \\
        Terra 5.6 & \qStoppingTerraMedActions{} & \qStoppingTerraMaxActions{} & 0/\qAgentBehaviorTaskCount{} & \qStoppingTerraMedHours{} & \qStoppingTerraMaxHours{} & 0/\qAgentBehaviorTaskCount{} \\
        Luna 5.6  & \qStoppingLunaMedActions{}  & \qStoppingLunaMaxActions{}  & 0/\qAgentBehaviorTaskCount{} & \qStoppingLunaMedHours{}  & \qStoppingLunaMaxHours{}  & 0/\qAgentBehaviorTaskCount{} \\
        \bottomrule
    \end{tabular}
    \end{center}
\end{minipage}
\par\smallskip

\textbf{Agents search by writing throwaway programs, not only by reading files.}
For three of four models the second most common activity is piping a Python program into
\texttt{python3 -}, used both to search (ad-hoc tree-walkers scanning for suspect patterns) and to
edit (programs that rewrite source in place).

\textbf{Agents test with their own programs more than with the project's test suite.}
On the \qSearchImportSubcohort{} repositories where the graded code imports as a Python module,
\qSearchImportRateLow{}--\qSearchImportRateHigh{} of runs write a program that imports it directly.
This exercises the code rather than invoking the project as a user would. Running the project's own
test suite is a minority of execution actions
(\qSearchTestSuiteShareLow{}--\qSearchTestSuiteShareHigh{}).

\textbf{Different models resolve different bugs.}
Among bugs with results for all \qDiscoverabilityOutcomeModelCount{} models,
\qAgentBehaviorExactlyOneModelCount{} are resolved by
one model only, while only \qAgentBehaviorAllFourModelsCount{} are resolved by all
\qDiscoverabilityOutcomeModelCount{} models.
\qAgentBehaviorNoModelCount{} (\qAgentBehaviorNoModelShare{}) are not resolved by any model
(Table~\ref{tab:agent-behavior-overlap}). The final four rows show that every model resolves bugs
that no other model resolves. No model therefore resolves all bugs resolved by the other models.

\par\smallskip
\noindent\begin{minipage}{\textwidth}
    \centering
    \small
    \captionof{table}{Number of models resolving the same bug. Percentages use only bugs with
    available results for all \qDiscoverabilityOutcomeModelCount{} models. The last results show how
    many bugs are resolved by each model and no other model.}
    \label{tab:agent-behavior-overlap}
    \begin{center}
        \begin{tabular}{@{}lrr@{}}
\toprule
\textbf{Resolving models} & \textbf{Bugs} & \textbf{Share} \\
\midrule
No model & 3,849 & 93.4\% \\
Exactly one model & 159 & 3.9\% \\
Exactly two models & 60 & 1.5\% \\
Exactly three models & 44 & 1.1\% \\
All four models & 8 & 0.2\% \\
\midrule
Opus only & 20 & 0.5\% \\
Sol only & 104 & 2.5\% \\
Terra only & 7 & 0.2\% \\
Luna only & 28 & 0.7\% \\
\bottomrule
\end{tabular}
 \end{center}
\end{minipage}
\par\smallskip

\textbf{Repeated attempts recover additional bugs, but gains diminish.}
On a \qSequentialTaskCount{}-repository cohort, \qSequentialAgentCount{} sequential Luna
xhigh attempts raise bug recovery from \qSequentialLunaXHighOne{} to
\qSequentialLunaXHighFinal{}, while the union of \qSequentialAgentCount{} independent attempts
reaches \qIndependentLunaXHighFinal{} (Figure~\ref{fig:sequential-score-by-step}). With default
reasoning, sequential attempts raise recovery from \qSequentialLunaDefaultOne{} to
\qSequentialLunaDefaultFinal{}, while independent attempts reach
\qIndependentLunaDefaultFinal{}. A single Sol xhigh run resolves \qSequentialSolScore{} of bugs,
between the first and final sequential Luna xhigh results.
Independent runs resolve more with xhigh reasoning; sequential runs resolve more with default
reasoning. Both recover additional bugs, but gains are smaller for later attempts. 

\textbf{Reading a file does not mean reading all of its lines.}
By action \qBugProgressEndpointAction{}, Opus has read a
mean of \qBugProgressOpusEndpointFileCoverage{} of bug-relevant files per repository and a mean of
\qBugProgressOpusEndpointLocCoverage{} of all lines in those files. The corresponding Sol results
are \qBugProgressSolEndpointFileCoverage{} and \qBugProgressSolEndpointLocCoverage{}
(Figure~\ref{fig:agent-behavior-loc-sensitivity}). File breadth and line coverage therefore measure
different parts of exploration. Exact line ranges are available for
\qBugProgressOpusLineRecovery{} of Opus read requests and \qBugProgressSolLineRecovery{} of Sol
requests. The lower line result for Sol reflects both exploration and the smaller share of reads
with exact ranges. Because some reads do not have exact ranges, we use files for the primary
analysis and lines as an additional analysis.

\par\smallskip
\noindent\begin{minipage}{\textwidth}
    \centering
    \includegraphics[width=\linewidth]{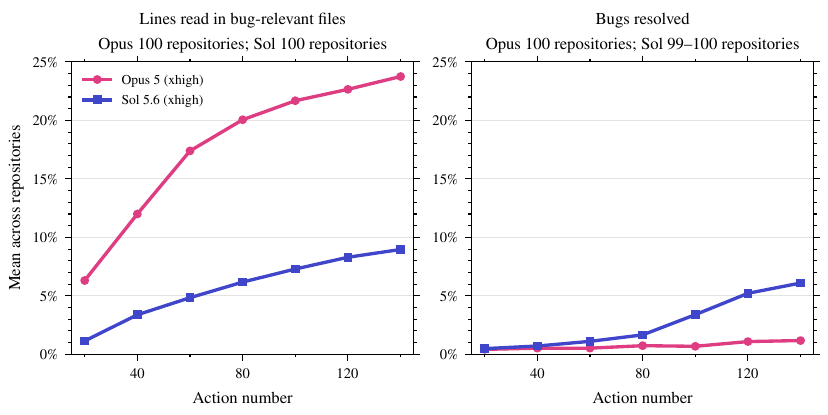}
    \captionof{figure}{Lines read in bug-relevant files and bugs resolved by action number for Opus and
    Sol. Each result is the mean across repositories. Lines are included only for reads with exact
    line ranges. Bug resolution is from the most recent evaluated patch.}
    \label{fig:agent-behavior-loc-sensitivity}
\end{minipage}
\par\smallskip

\clearpage
\subsection{Per-repository results}
\label{appx:results:per-repository}
Table~\ref{tab:task-results} reports the percentage of bugs resolved for all
\qMeasuredMainResultsVariantCount{} measured model variants across all
\qBenchmarkTaskCount{} benchmark repositories. 

\begingroup
\tiny
\setlength{\tabcolsep}{1.5pt}
\renewcommand{\arraystretch}{1.08}
\begin{longtable}{@{}lrrrrrrrrrr@{}}
\caption{Percentage of bugs resolved for all \qMeasuredMainResultsVariantCount{} measured model variants across all \qBenchmarkTaskCount{} benchmark repositories. Blank cells indicate unavailable evaluations.}\label{tab:task-results} \\
\toprule
\textbf{Repo} & \shortstack{\textbf{Opus 5}\\\textbf{(xhigh)}} & \shortstack{\textbf{Sol 5.6}\\\textbf{(xhigh)}} & \shortstack{\textbf{Terra 5.6}\\\textbf{(xhigh)}} & \shortstack{\textbf{Luna 5.6}\\\textbf{(xhigh)}} & \shortstack{\textbf{Luna 5.6}\\\textbf{(default)}} & \shortstack{\textbf{Luna 5.6}\\\textbf{(high)}} & \shortstack{\textbf{GPT-5.4 Mini}\\\textbf{(default)}} & \shortstack{\textbf{GPT-5.4 Mini}\\\textbf{(high)}} & \shortstack{\textbf{Gemini 3.5 Flash Lite}\\\textbf{(default)}} & \shortstack{\textbf{Kimi K3}\\\textbf{(max)}} \\
\midrule
\endfirsthead
\toprule
\textbf{Repo} & \shortstack{\textbf{Opus 5}\\\textbf{(xhigh)}} & \shortstack{\textbf{Sol 5.6}\\\textbf{(xhigh)}} & \shortstack{\textbf{Terra 5.6}\\\textbf{(xhigh)}} & \shortstack{\textbf{Luna 5.6}\\\textbf{(xhigh)}} & \shortstack{\textbf{Luna 5.6}\\\textbf{(default)}} & \shortstack{\textbf{Luna 5.6}\\\textbf{(high)}} & \shortstack{\textbf{GPT-5.4 Mini}\\\textbf{(default)}} & \shortstack{\textbf{GPT-5.4 Mini}\\\textbf{(high)}} & \shortstack{\textbf{Gemini 3.5 Flash Lite}\\\textbf{(default)}} & \shortstack{\textbf{Kimi K3}\\\textbf{(max)}} \\
\midrule
\endhead
\midrule
\multicolumn{11}{r}{\small Continued on next page} \\
\endfoot
\bottomrule
\endlastfoot
\texttt{agda} & 0.0\% & 0.0\% & 0.0\% & 0.0\% & 0.0\% & 0.0\% & 0.0\% & 0.0\% & 0.0\% & 0.0\% \\
\texttt{arrow} & 2.1\% & 2.1\% & 0.0\% & 0.0\% & 0.0\% & 0.0\% & 0.0\% & 0.0\% & 0.0\% & 0.0\% \\
\texttt{arrow-py} & 0.0\% & 25.0\% & 0.0\% & 25.0\% & 0.0\% & 0.0\% & 0.0\% & 25.0\% & 0.0\% & 0.0\% \\
\texttt{astropy} & 1.2\% & 2.5\% & 0.0\% & 0.0\% & 0.0\% & 0.0\% & 0.0\% & 0.0\% & 0.0\% & 0.0\% \\
\texttt{aws-cdk} & 2.7\% & 5.4\% & 2.7\% & 2.7\% & 0.0\% & 0.0\% & 0.0\% & 0.0\% & 0.0\% & 0.0\% \\
\texttt{babel} & 0.0\% & 3.2\% & 0.0\% & 0.0\% & 3.2\% & 0.0\% & 0.0\% & 0.0\% & 0.0\% & 0.0\% \\
\texttt{black} & 0.0\% & 9.5\% & 9.5\% & 9.5\% & 4.8\% & 9.5\% & 0.0\% & 4.8\% & 0.0\% & 4.8\% \\
\texttt{click} & 0.0\% & 46.2\% & 23.1\% & 15.4\% & 7.7\% & 7.7\% & 0.0\% & 7.7\% & 0.0\% & 0.0\% \\
\texttt{clippy} & 0.0\% & 0.0\% & 0.0\% & 0.0\% & 0.0\% & 0.0\% & 0.0\% & 0.0\% & 0.0\% & 0.0\% \\
\texttt{conan} & 4.2\% & 0.0\% & 8.3\% & 0.0\% & 0.0\% & 4.2\% & 0.0\% & 0.0\% & 0.0\% & 0.0\% \\
\texttt{cvc5} & 0.0\% & 2.6\% & 0.0\% & 0.0\% & 0.0\% & 0.0\% & 0.0\% & 0.0\% & 0.0\% & 0.0\% \\
\texttt{dart-sdk} & 2.0\% & 2.0\% & 2.0\% & 2.0\% & 2.0\% & 2.0\% & 2.0\% & 2.0\% & 2.0\% & 0.0\% \\
\texttt{dask} & 0.0\% & 4.8\% & 0.0\% & 0.0\% & 0.0\% & 0.0\% & 0.0\% & 0.0\% & 0.0\% & 0.0\% \\
\texttt{dataframes-jl} & 14.3\% & 28.6\% & 14.3\% & 14.3\% & 0.0\% & 28.6\% & 0.0\% & 0.0\% & 0.0\% & 14.3\% \\
\texttt{datafusion} & 0.0\% & 0.0\% & 2.6\% & 0.0\% & 0.0\% & 0.0\% & 0.0\% & 0.0\% & 0.0\% & 0.0\% \\
\texttt{dateutil} & 30.0\% & 40.0\% & 40.0\% & 10.0\% & 10.0\% & 10.0\% & 0.0\% & 20.0\% & 0.0\% & 20.0\% \\
\texttt{dayjs} & 18.2\% & 54.5\% & 27.3\% & 27.3\% & 0.0\% & 18.2\% & 0.0\% & 0.0\% & 0.0\% & 18.2\% \\
\texttt{distributions-jl} & 0.0\% & 0.0\% & 0.0\% & 11.1\% & 0.0\% & 0.0\% & 0.0\% & 11.1\% & 0.0\% & 0.0\% \\
\texttt{dmd} & 0.9\% & 0.0\% & 0.0\% & 0.0\% & 0.0\% & 0.0\% & 0.0\% & 0.0\% & 0.0\% & 0.0\% \\
\texttt{dolphinscheduler} & 10.0\% & 30.0\% & 0.0\% & 10.0\% & 0.0\% & 30.0\% & 0.0\% & 0.0\% & 0.0\% & 0.0\% \\
\texttt{duckdb} & 1.8\% & 1.8\% & 0.9\% & 0.9\% & 0.9\% & 0.9\% & 0.9\% & 0.0\% & 0.9\% & 0.0\% \\
\texttt{echo} & 0.0\% & 16.7\% & 33.3\% & 33.3\% & 16.7\% & 16.7\% & 0.0\% & 16.7\% & 0.0\% & 0.0\% \\
\texttt{elixir} & 0.0\% & 17.0\% & 4.3\% & 0.0\% & 2.1\% & 2.1\% & 0.0\% & 0.0\% & 0.0\% & 0.0\% \\
\texttt{fasthttp} & 0.0\% & 0.0\% & 11.1\% & 22.2\% & 0.0\% & 0.0\% & 0.0\% & 0.0\% & 0.0\% & 0.0\% \\
\texttt{formulajs} & 0.0\% & 55.6\% & 33.3\% & 33.3\% & 22.2\% & 33.3\% & 11.1\% & 11.1\% & 0.0\% & 0.0\% \\
\texttt{fsharp} & 0.0\% & 0.0\% & 0.0\% & 0.0\% & 0.0\% & 0.0\% & 0.0\% & 0.0\% & 0.0\% & 0.0\% \\
\texttt{gdal} & 0.0\% & 0.0\% & 0.0\% & 0.0\% & 0.0\% & 0.0\% & 0.0\% & 0.0\% & 0.0\% & 0.0\% \\
\texttt{gin} & 27.3\% & 27.3\% & 27.3\% & 9.1\% & 0.0\% & 9.1\% & 9.1\% & 9.1\% & 0.0\% & 0.0\% \\
\texttt{gitoxide} & 0.0\% & 0.0\% & 0.0\% & 0.0\% & 0.0\% & 0.0\% & 0.0\% & 0.0\% & 0.0\% & 0.0\% \\
\texttt{go} & 0.0\% & 3.1\% & 1.0\% & 0.0\% & 1.0\% & 0.0\% & 1.0\% & 0.0\% & 1.0\% & 0.0\% \\
\texttt{go-git} & 0.0\% & 21.7\% & 8.7\% & 13.0\% & 0.0\% & 0.0\% & 0.0\% & 0.0\% & 0.0\% & 0.0\% \\
\texttt{goawk} & 7.1\% & 35.7\% & 14.3\% & 35.7\% & 0.0\% & 28.6\% & 0.0\% & 0.0\% & 0.0\% & 14.3\% \\
\texttt{godot} & 0.0\% & 0.0\% & 0.0\% & 0.0\% & 0.0\% & 0.0\% & 0.0\% & 0.0\% & 0.0\% & 0.0\% \\
\texttt{gorm} & 0.0\% & 31.6\% & 0.0\% & 10.5\% & 5.3\% & 0.0\% & 0.0\% & 0.0\% & 0.0\% & 0.0\% \\
\texttt{gson} & 16.7\% & 16.7\% & 0.0\% & 16.7\% & 0.0\% & 16.7\% & 0.0\% & 0.0\% & 0.0\% & 0.0\% \\
\texttt{hdf5} & 0.0\% & 0.0\% & 0.0\% & 0.0\% & 0.0\% & 0.0\% & 0.0\% & 0.0\% & 0.0\% & 0.0\% \\
\texttt{idris2} & 0.0\% & 0.0\% & 0.0\% & 0.0\% & 0.0\% & 0.0\% & 0.0\% & 0.0\% & 0.0\% & 0.0\% \\
\texttt{isomorphic-git} & 0.0\% & 38.5\% & 15.4\% & 15.4\% & 0.0\% & 7.7\% & 0.0\% & 0.0\% & 0.0\% & 0.0\% \\
\texttt{isort} & 0.0\% & 27.3\% & 9.1\% & 9.1\% & 0.0\% & 18.2\% & 0.0\% & 0.0\% & 0.0\% & 0.0\% \\
\texttt{jerryscript} & 20.0\% & 0.0\% & 6.7\% & 6.7\% & 0.0\% & 0.0\% & 0.0\% & 0.0\% & 0.0\% & 13.3\% \\
\texttt{jest} & 0.0\% & 10.0\% & 0.0\% & 0.0\% & 0.0\% & 0.0\% & 0.0\% & 0.0\% & 0.0\% & 0.0\% \\
\texttt{keycloak} & 0.0\% & 1.6\% & 0.0\% & 0.0\% & 0.0\% & 0.0\% & 0.0\% & 0.0\% & 0.0\% & 1.6\% \\
\texttt{lean4} & 0.0\% & 1.6\% & 0.0\% & 0.0\% & 0.0\% & 0.0\% & 0.0\% & 0.0\% & 0.0\% & 0.0\% \\
\texttt{libexpat} & 0.0\% & 11.1\% & 0.0\% & 11.1\% & 0.0\% & 0.0\% & 0.0\% & 0.0\% & 0.0\% & 0.0\% \\
\texttt{luxon} & 28.6\% & 42.9\% & 21.4\% & 21.4\% & 0.0\% & 7.1\% & 0.0\% & 0.0\% & 0.0\% & 0.0\% \\
\texttt{material-ui} & 0.0\% & 5.1\% & 0.0\% & 2.6\% & 0.0\% & 2.6\% & 0.0\% & 0.0\% & 0.0\% & 0.0\% \\
\texttt{mathlib4} & 0.0\% & 0.0\% & 0.0\% & 0.0\% & 0.0\% & 0.0\% & 0.0\% & 0.0\% & 0.0\% & 0.0\% \\
\texttt{matplotlib} & 0.0\% & 1.3\% & 1.3\% & 0.0\% & 0.0\% & 0.0\% & 0.0\% & 0.0\% & 0.0\% & 0.0\% \\
\texttt{micropython} & 0.0\% & 0.0\% & 0.0\% & 0.0\% & 0.0\% & 0.0\% & 0.0\% & 0.0\% & 0.0\% & 0.0\% \\
\texttt{modelingtoolkit-jl} & 0.0\% & 0.0\% & 0.0\% & 0.0\% & 0.0\% & 0.0\% & 0.0\% & 0.0\% & 0.0\% & 0.0\% \\
\texttt{mongoose} & 0.0\% & 11.4\% & 0.0\% & 5.7\% & 0.0\% & 1.4\% & 0.0\% & 0.0\% & 0.0\% & 0.0\% \\
\texttt{mpmath} & 0.0\% & 18.8\% & 0.0\% & 12.5\% & 0.0\% & 0.0\% & 0.0\% & 0.0\% & 0.0\% & 0.0\% \\
\texttt{mypy} & 0.0\% & 0.0\% & 0.0\% & 1.1\% & 0.0\% & 0.0\% & 0.0\% & 0.0\% & 0.0\% & 1.1\% \\
\texttt{nacos} & 0.0\% & 16.7\% & 0.0\% & 8.3\% & 8.3\% & 8.3\% & 0.0\% & 0.0\% & 0.0\% & 0.0\% \\
\texttt{netty} & 0.0\% & 10.4\% & 2.1\% & 5.2\% & 1.0\% & 1.0\% & 0.0\% & 1.0\% & 1.0\% & 0.0\% \\
\texttt{networkx} & 0.0\% & 5.6\% & 0.0\% & 11.1\% & 5.6\% & 0.0\% & 0.0\% & 0.0\% & 0.0\% & 0.0\% \\
\texttt{nim} & 0.5\% & 2.6\% & 0.0\% & 1.0\% & 0.0\% & 0.0\% & 0.0\% & 0.0\% & 0.0\% & 0.0\% \\
\texttt{numpy} & 0.0\% & 0.8\% & 0.0\% & 0.0\% & 0.0\% & 0.0\% & 0.0\% & 0.0\% & 0.0\% & 0.0\% \\
\texttt{nuxt} & 0.0\% & 16.7\% & 0.0\% & 0.0\% & 0.0\% & 33.3\% & 0.0\% & 0.0\% & 0.0\% & 0.0\% \\
\texttt{openapi-generator} & 0.0\% & 0.0\% & 0.0\% & 0.0\% & 0.0\% & 0.0\% & 0.0\% & 0.0\% & 0.0\% & 0.0\% \\
\texttt{opencv} & 0.0\% & 0.0\% & 0.0\% & 2.1\% & 0.0\% & 0.0\% & 0.0\% & 0.0\% & 0.0\% & 0.0\% \\
\texttt{optuna} & 14.3\% & 0.0\% & 0.0\% & 14.3\% & 0.0\% & 14.3\% & 0.0\% & 0.0\% & 0.0\% & 0.0\% \\
\texttt{oxc} & 0.0\% & 0.0\% & 0.0\% & 0.0\% & 0.0\% & 0.0\% & 0.0\% & 0.0\% & 0.0\% & 0.0\% \\
\texttt{pandas} & 0.0\% & 0.4\% & 0.0\% & 0.9\% & 0.0\% & 0.0\% & 0.0\% & 0.0\% & 0.0\% & 0.4\% \\
\texttt{pendulum} & 33.3\% & 55.6\% & 33.3\% & 44.4\% & 0.0\% & 44.4\% & 0.0\% & 22.2\% & 0.0\% & 0.0\% \\
\texttt{php-src} & 0.0\% & 0.0\% & 0.0\% & 0.0\% & 0.0\% & 0.0\% & 0.0\% & 0.0\% & 0.0\% & 0.0\% \\
\texttt{phpspreadsheet} & 0.0\% & 0.0\% & 0.0\% & 11.1\% & 0.0\% & 0.0\% & 0.0\% & 0.0\% & 0.0\% & 0.0\% \\
\texttt{pint} & 7.7\% & 23.1\% & 0.0\% & 0.0\% & 0.0\% & 0.0\% & 0.0\% & 0.0\% & 0.0\% & 0.0\% \\
\texttt{prettier} & 0.0\% & 0.0\% & 0.0\% & 0.0\% & 0.0\% & 0.0\% & 0.0\% & 0.0\% & 0.0\% & 2.9\% \\
\texttt{proj} & 0.0\% & 9.5\% & 0.0\% & 0.0\% & 0.0\% & 0.0\% & 0.0\% & 0.0\% & 0.0\% & 0.0\% \\
\texttt{pulsar} & 0.0\% & 5.3\% & 0.0\% & 0.0\% & 0.0\% & 0.0\% & 0.0\% & 0.0\% & 0.0\% & 0.0\% \\
\texttt{quarkus} & 0.0\% & 1.4\% & 0.0\% & 0.0\% & 0.0\% & 0.0\% & 0.0\% & 0.0\% & 0.0\% & 0.0\% \\
\texttt{qutip} & 10.0\% & 10.0\% & 0.0\% & 0.0\% & 0.0\% & 0.0\% & 0.0\% & 0.0\% & 0.0\% & 0.0\% \\
\texttt{rails} & 0.0\% & 2.9\% & 0.0\% & 0.7\% & 0.0\% & 0.0\% & 0.0\% & 0.0\% & 0.0\% & 0.0\% \\
\texttt{react-router} & 0.0\% & 0.0\% & 0.0\% & 14.3\% & 0.0\% & 0.0\% & 0.0\% & 0.0\% & 0.0\% & 0.0\% \\
\texttt{roslyn} & 3.2\% & 0.0\% & 0.0\% & 0.0\% & 0.0\% & 0.0\% & 0.0\% & 0.0\% & 0.0\% & 0.0\% \\
\texttt{ruff} & 0.0\% & 0.0\% & 0.0\% & 0.0\% & 0.0\% & 0.0\% & 0.0\% & 0.0\% & 0.0\% & 0.0\% \\
\texttt{rxjava} & 7.7\% & 3.8\% & 15.4\% & 7.7\% & 7.7\% & 3.8\% & 7.7\% & 7.7\% & 0.0\% & 7.7\% \\
\texttt{scikit-learn} & 3.8\% & 3.8\% & 0.0\% & 2.5\% & 0.0\% & 1.2\% & 0.0\% & 1.2\% & 0.0\% & 0.0\% \\
\texttt{scipy} & 0.7\% & 2.2\% & 0.7\% & 0.7\% & 0.0\% & 0.0\% & 0.0\% & 0.7\% & 0.0\% & 0.7\% \\
\texttt{shapely} & 0.0\% & 55.6\% & 0.0\% & 44.4\% & 0.0\% & 33.3\% & 0.0\% & 0.0\% & 0.0\% & 11.1\% \\
\texttt{shardingsphere} & 0.0\% & 33.3\% & 0.0\% & 13.3\% & 0.0\% & 0.0\% & 0.0\% & 0.0\% & 0.0\% & 0.0\% \\
\texttt{sqlglot} & 0.0\% & 3.8\% & 3.1\% & 3.8\% & 0.8\% & 2.3\% & 0.0\% & 0.8\% & 0.0\% & 0.0\% \\
\texttt{stylelint} & 0.0\% & 3.7\% & 0.0\% & 1.9\% & 0.0\% & 1.9\% & 0.0\% & 0.0\% & 0.0\% & 0.0\% \\
\texttt{sunpy} & 6.7\% & 0.0\% & 0.0\% & 0.0\% & 0.0\% & 0.0\% & 0.0\% & 0.0\% & 0.0\% & 6.7\% \\
\texttt{svgo} & 9.1\% & 36.4\% & 27.3\% & 18.2\% & 0.0\% & 9.1\% & 0.0\% & 9.1\% & 0.0\% & 9.1\% \\
\texttt{swc} & 5.3\% & 5.3\% & 0.0\% & 5.3\% & 5.3\% & 5.3\% & 5.3\% & 5.3\% & 5.3\% & 5.3\% \\
\texttt{symfony} & 0.0\% & 1.3\% & 0.7\% & 0.7\% & 0.7\% & 0.7\% & 0.0\% & 0.0\% & 0.0\% & 0.0\% \\
\texttt{syn} & 0.0\% & 0.0\% & 0.0\% & 0.0\% & 0.0\% & 0.0\% & 0.0\% & 0.0\% & 0.0\% & 0.0\% \\
\texttt{tarantool} & 0.0\% & 7.5\% & 0.0\% & 0.0\% & 0.0\% & 0.0\% & 0.0\% & 0.0\% & 0.0\% & 0.0\% \\
\texttt{terser} & 0.0\% & 16.7\% & 33.3\% & 0.0\% & 0.0\% & 50.0\% & 0.0\% & 0.0\% & 0.0\% & 0.0\% \\
\texttt{testify} & 0.0\% & 0.0\% & 0.0\% & 0.0\% & 0.0\% & 0.0\% & 0.0\% & 0.0\% & 0.0\% & 0.0\% \\
\texttt{three-js} & 0.0\% & 11.1\% & 0.0\% & 44.4\% & 0.0\% & 11.1\% & 0.0\% & 0.0\% & 0.0\% & 0.0\% \\
\texttt{tinygrad} & 6.7\% & 0.0\% & 0.0\% & 0.0\% & 0.0\% & 0.0\% & 0.0\% & 0.0\% & 0.0\% & 0.0\% \\
\texttt{tokio} & 0.0\% & 16.7\% & 0.0\% & 8.3\% & 0.0\% & 0.0\% & 0.0\% & 0.0\% & 0.0\% & 0.0\% \\
\texttt{trino} & 2.3\% & 0.0\% & 0.0\% & 0.0\% & 0.0\% & 0.0\% & 0.0\% & 0.0\% & 0.0\% & 2.3\% \\
\texttt{v} & 1.7\% & 2.1\% & 0.0\% & 1.7\% & 0.0\% & 0.4\% & 0.0\% & 0.0\% & 0.0\% & 0.4\% \\
\texttt{webpack} & 0.0\% & 5.7\% & 0.0\% & 0.0\% & 0.0\% & 0.0\% & 0.0\% & 0.0\% & 0.0\% & 0.0\% \\
\texttt{xarray} & 6.5\% & 9.7\% & 0.0\% & 3.2\% & 3.2\% & 0.0\% & 0.0\% & 0.0\% & 0.0\% & 0.0\% \\
\texttt{z3} & 0.0\% & 0.0\% & 0.0\% & 0.0\% & 0.0\% & 0.0\% & 0.0\% & 0.0\% & 0.0\% & 0.0\% \\
\end{longtable}
\endgroup
  \clearpage
\flushbottom
\clearpage
\newpage
\section{Repository index and statistics}
\label{appx:task_index}
This appendix gives more details about the repositories in \bench{} and adds to the overview in \S\ref{sec:benchmark:statistics}.
\S\ref{appx:benchmark:additional-stats} gives more statistics about the repositories and their bugs.
\S\ref{appx:task_index:repositories} lists all repositories.
\S\ref{appx:task_index:szz} provides more plots related to the SZZ algorithm that we use to estimate bug lifetimes.

\subsection{Additional dataset statistics}
\label{appx:benchmark:additional-stats}
This section gives the full distributions behind the summary statistics in \S\ref{sec:benchmark:statistics}.
It shows how large the repositories and bugs are, which programming languages appear in the files changed by the gold fixes, and how often the gold fixes of two bugs change the same file.

\begin{figure}[htb]
    \centering
    \includegraphics[width=\linewidth]{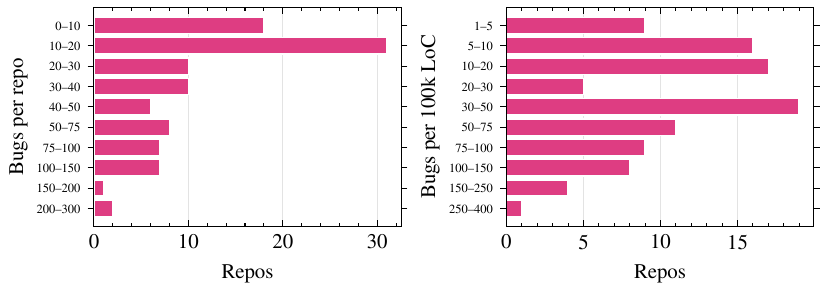}
    \caption{Binned repository-level distributions.}
    \label{fig:benchmark-task-distributions}
\end{figure}

\begin{figure}[htb]
    \centering
    \begin{minipage}[t]{0.32\textwidth}\includegraphics[width=\linewidth]{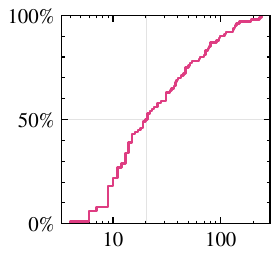}
        \centering\small (a) Bugs per repo
    \end{minipage}\hfill \begin{minipage}[t]{0.32\textwidth}\includegraphics[width=\linewidth]{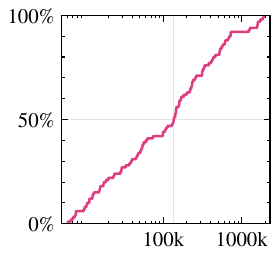}
        \centering\small (b) Graded-source LoC
    \end{minipage}\hfill \begin{minipage}[t]{0.32\textwidth}\includegraphics[width=\linewidth]{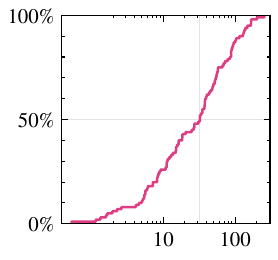}
        \centering\small (c) Bugs per 100k LoC
    \end{minipage}\caption{Empirical cumulative distributions of repository-level benchmark
    statistics. Faint-gray guides mark the median value and 50th percentile.}
    \label{fig:benchmark-repository-cumulative}
\end{figure}

\begin{figure}[htb]
    \centering
    \begin{minipage}[t]{0.24\textwidth}\includegraphics[width=\linewidth]{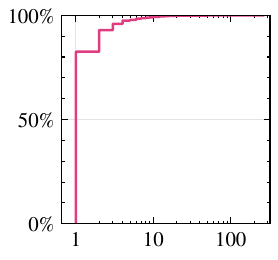}
        \centering\small (a) F2P tests
    \end{minipage}\hfill \begin{minipage}[t]{0.24\textwidth}\includegraphics[width=\linewidth]{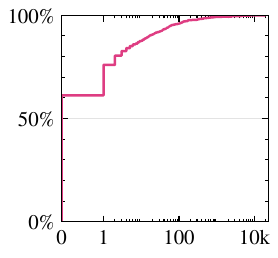}
        \centering\small (b) P2P tests
    \end{minipage}\hfill \begin{minipage}[t]{0.24\textwidth}\includegraphics[width=\linewidth]{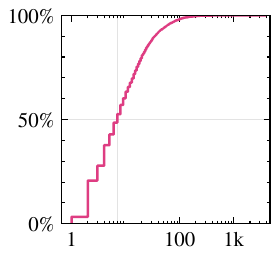}
        \centering\small (c) Gold fix size
    \end{minipage}\hfill \begin{minipage}[t]{0.24\textwidth}\includegraphics[width=\linewidth]{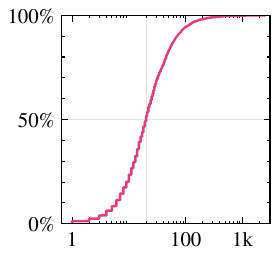}
        \centering\small (d) Test patch size
    \end{minipage}\caption{Empirical cumulative distributions of bug-level benchmark
    statistics. Faint-gray guides mark the median value and 50th
    percentile.}
    \label{fig:benchmark-bug-cumulative}
\end{figure}

Figure~\ref{fig:benchmark-task-distributions} gives binned distributions of repository
scale. Figures~\ref{fig:benchmark-repository-cumulative} and
\ref{fig:benchmark-bug-cumulative} give cumulative distributions of repository
scale and the bug-level statistics in Table~\ref{tab:benchmark-subtask-stats}.
Their means are \qBenchmarkMeanFailToPassTests{} F2P tests,
\qBenchmarkMeanPassToPassTests{} P2P tests,
\qBenchmarkMeanGoldPatchLines{} gold fix lines, and
\qBenchmarkMeanTestPatchLines{} test-patch lines per bug.

\begin{table}[htb]
    \centering
    \small
    \caption{\textbf{Gold fix overlap and separation.} The first row includes all
    benchmark bugs. The last two rows include pairs of bugs whose gold fixes each change
    exactly one existing source file, and both change the same file. Different content compares
    changed source lines after removing diff metadata and unchanged context. Distance is the
    minimum distance between the changed line ranges.}
    \label{tab:repair-file-overlap}
    \begin{center}
        \begin{tabular}{@{}lr@{}}
\toprule
\textbf{Measure} & \textbf{Share} \\
\midrule
Bugs sharing a bug-relevant file & \qBenchmarkSourceFileOverlapShare{} \\
Same-file repair pairs with different changed content &
\qBenchmarkDifferentSameFileRepairContentShare{} \\
Same-file repair pairs more than 100 lines apart &
\qBenchmarkDistantSameFileRepairShare{} \\
\bottomrule
\end{tabular}
     \end{center}
\end{table}

\begin{table}[htb]
    \centering
    \small
    \caption{Complete source-language breakdown for gold fixes underlying
    Figure~\ref{fig:subtask-language-treemap}.
    Bug counts can be fractional
    because a single gold fix can modify source files in multiple languages, resulting in fractional assignment of the bug to languages.}
    \label{tab:subtask-languages}
    \begin{center}
        \begin{tabularx}{0.49\linewidth}[t]{@{}l>{\raggedright\arraybackslash}Xrr@{}}
\toprule
\textbf{Language} & \textbf{File endings} & \textbf{Bugs} & \textbf{Share} \\
\midrule
Python & \texttt{.py}, \texttt{.pyi} & 1,058.7 & 26.02\% \\
C++ & \texttt{.cc}, \texttt{.cpp}, \texttt{.cxx}, \texttt{.h}, \texttt{.hh}, \texttt{.hpp}, \texttt{.hxx} & 385.8 & 9.48\% \\
Java & \texttt{.java}, \texttt{.template} & 381.5 & 9.38\% \\
JavaScript & \texttt{.cjs}, \texttt{.js}, \texttt{.jsx}, \texttt{.mjs} & 326.8 & 8.03\% \\
V & \texttt{.v} & 239.0 & 5.88\% \\
Rust & \texttt{.rs} & 226.5 & 5.57\% \\
Nim & \texttt{.nim} & 192.0 & 4.72\% \\
Go & \texttt{.go} & 189.0 & 4.65\% \\
C & \texttt{.c}, \texttt{.h} & 187.3 & 4.60\% \\
PHP & \texttt{.php} & 158.0 & 3.88\% \\
Ruby & \texttt{.rb} & 140.0 & 3.44\% \\
D & \texttt{.d} & 110.1 & 2.71\% \\
TypeScript & \texttt{.cts}, \texttt{.mts}, \texttt{.ts}, \texttt{.tsx} & 86.2 & 2.12\% \\
Lean & \texttt{.lean} & 71.0 & 1.75\% \\
Cython & \texttt{.in}, \texttt{.pxd}, \texttt{.pxi}, \texttt{.pyf}, \texttt{.pyx}, \texttt{.tp} & 58.7 & 1.44\% \\
Haskell & \texttt{.hs}, \texttt{.hs-boot} & 44.0 & 1.08\% \\
\bottomrule
\end{tabularx}
\hfill
\begin{tabularx}{0.49\linewidth}[t]{@{}l>{\raggedright\arraybackslash}Xrr@{}}
\toprule
\textbf{Language} & \textbf{File endings} & \textbf{Bugs} & \textbf{Share} \\
\midrule
Dart & \texttt{.dart} & 43.8 & 1.08\% \\
Elixir & \texttt{.ex} & 42.0 & 1.03\% \\
C\# & \texttt{.cs} & 29.5 & 0.73\% \\
F\# & \texttt{.fs}, \texttt{.fsi} & 28.0 & 0.69\% \\
Julia & \texttt{.jl} & 26.0 & 0.64\% \\
Idris & \texttt{.idr} & 13.0 & 0.32\% \\
Mustache & \texttt{.mustache} & 12.5 & 0.31\% \\
Erlang & \texttt{.erl}, \texttt{.yrl} & 5.0 & 0.12\% \\
Lua & \texttt{.lua} & 5.0 & 0.12\% \\
Fortran & \texttt{.f}, \texttt{.f77}, \texttt{.f90}, \texttt{.for} & 4.0 & 0.10\% \\
Visual Basic & \texttt{.vb} & 1.5 & 0.04\% \\
Alex & \texttt{.x} & 1.0 & 0.02\% \\
re & \texttt{.re} & 1.0 & 0.02\% \\
CMake & \texttt{CMakeLists.txt} & 0.7 & 0.02\% \\
Shell & \texttt{.sh} & 0.5 & 0.01\% \\
\midrule
Total & & 4,068.0 & 100.00\% \\
\bottomrule
\end{tabularx}
     \end{center}
\end{table}

\clearpage
\subsection{Repository index}
\label{appx:task_index:repositories}
Table~\ref{tab:repository-index} expands the benchmark overview in \S\ref{sec:benchmark:statistics}.
Repository names link to their upstream projects. License identifiers link to each
project's license declaration; multiple identifiers indicate alternative or file-scoped
terms.

\begingroup
\tiny
\setlength{\tabcolsep}{3pt}
\renewcommand{\arraystretch}{1.1}
\begin{longtable}{@{}llp{0.38\textwidth}p{0.17\textwidth}rr@{}}
\caption{Repository index.}\label{tab:repository-index} \\
\toprule
\textbf{Repository} & \textbf{Language} & \textbf{Description} & \textbf{License} & \textbf{Bugs} & \textbf{LoC} \\
\midrule
\endfirsthead
\toprule
\textbf{Repository} & \textbf{Language} & \textbf{Description} & \textbf{License} & \textbf{Bugs} & \textbf{LoC} \\
\midrule
\endhead
\midrule
\multicolumn{6}{r}{Continued on next page} \\
\endfoot
\bottomrule
\endlastfoot
\href{https://github.com/agda/agda}{\texttt{agda}} & \texttt{hs} & {\raggedright Dependently typed programming language and interactive theorem prover} & {\raggedright \href{https://github.com/agda/agda/blob/master/LICENSE}{\texttt{MIT}}} & 45 & 147k \\
\href{https://github.com/apache/arrow}{\texttt{arrow}} & \texttt{cpp} & {\raggedright Columnar data format and multi-language analytics libraries} & {\raggedright \href{https://github.com/apache/arrow/blob/main/LICENSE.txt}{\texttt{Apache-2.0}}} & 47 & 311k \\
\href{https://github.com/arrow-py/arrow}{\texttt{arrow-py}} & \texttt{py} & {\raggedright Date and time library for Python} & {\raggedright \href{https://github.com/arrow-py/arrow/blob/master/LICENSE}{\texttt{Apache-2.0}}} & 4 & 8k \\
\href{https://github.com/astropy/astropy}{\texttt{astropy}} & \texttt{py} & {\raggedright Core astronomy and astrophysics library for Python} & {\raggedright \href{https://github.com/astropy/astropy/blob/main/LICENSE.rst}{\texttt{BSD-3-Clause}}} & 80 & 229k \\
\href{https://github.com/aws/aws-cdk}{\texttt{aws-cdk}} & \texttt{ts} & {\raggedright Framework for defining cloud infrastructure in code} & {\raggedright \href{https://github.com/aws/aws-cdk/blob/main/LICENSE}{\texttt{Apache-2.0}}} & 37 & 200k \\
\href{https://github.com/babel/babel}{\texttt{babel}} & \texttt{js} & {\raggedright JavaScript compiler and transformation toolkit} & {\raggedright \href{https://github.com/babel/babel/blob/main/LICENSE}{\texttt{MIT}}} & 31 & 62k \\
\href{https://github.com/psf/black}{\texttt{black}} & \texttt{py} & {\raggedright Opinionated Python code formatter} & {\raggedright \href{https://github.com/psf/black/blob/main/LICENSE}{\texttt{MIT}}} & 21 & 15k \\
\href{https://github.com/pallets/click}{\texttt{click}} & \texttt{py} & {\raggedright Composable command-line interface toolkit for Python} & {\raggedright \href{https://github.com/pallets/click/blob/main/LICENSE.txt}{\texttt{BSD-3-Clause}}} & 13 & 10k \\
\href{https://github.com/rust-lang/rust-clippy}{\texttt{clippy}} & \texttt{rs} & {\raggedright Linter collection for Rust} & {\raggedright \href{https://github.com/rust-lang/rust-clippy/blob/master/LICENSE-APACHE}{\texttt{Apache-2.0}}; \href{https://github.com/rust-lang/rust-clippy/blob/master/LICENSE-MIT}{\texttt{MIT}}} & 76 & 135k \\
\href{https://github.com/conan-io/conan}{\texttt{conan}} & \texttt{py} & {\raggedright Package manager for C and C++} & {\raggedright \href{https://github.com/conan-io/conan/blob/develop2/LICENSE.md}{\texttt{MIT}}} & 24 & 45k \\
\href{https://github.com/cvc5/cvc5}{\texttt{cvc5}} & \texttt{cpp} & {\raggedright Automated SMT theorem prover} & {\raggedright \href{https://github.com/cvc5/cvc5/blob/main/COPYING}{\texttt{BSD-3-Clause}}} & 38 & 462k \\
\href{https://github.com/dart-lang/sdk}{\texttt{dart-sdk}} & \texttt{dart} & {\raggedright Dart language SDK, runtime, compilers, and libraries} & {\raggedright \href{https://github.com/dart-lang/sdk/blob/main/LICENSE}{\texttt{BSD-3-Clause}}} & 50 & 1,913k \\
\href{https://github.com/dask/dask}{\texttt{dask}} & \texttt{py} & {\raggedright Parallel and distributed computing library for Python} & {\raggedright \href{https://github.com/dask/dask/blob/main/LICENSE.txt}{\texttt{BSD-3-Clause}}} & 21 & 56k \\
\href{https://github.com/JuliaData/DataFrames.jl}{\texttt{dataframes-jl}} & \texttt{jl} & {\raggedright In-memory tabular data library for Julia} & {\raggedright \href{https://github.com/JuliaData/DataFrames.jl/blob/main/LICENSE.md}{\texttt{MIT}}} & 7 & 17k \\
\href{https://github.com/apache/datafusion}{\texttt{datafusion}} & \texttt{rs} & {\raggedright Extensible SQL query engine written in Rust} & {\raggedright \href{https://github.com/apache/datafusion/blob/main/LICENSE.txt}{\texttt{Apache-2.0}}} & 38 & 530k \\
\href{https://github.com/dateutil/dateutil}{\texttt{dateutil}} & \texttt{py} & {\raggedright Extensions to Python date and time handling} & {\raggedright \href{https://github.com/dateutil/dateutil/blob/master/LICENSE}{\texttt{Apache-2.0}}; \href{https://github.com/dateutil/dateutil/blob/master/LICENSE}{\texttt{BSD-3-Clause}}} & 10 & 6k \\
\href{https://github.com/iamkun/dayjs}{\texttt{dayjs}} & \texttt{js} & {\raggedright Lightweight immutable date and time library for JavaScript} & {\raggedright \href{https://github.com/iamkun/dayjs/blob/dev/LICENSE}{\texttt{MIT}}} & 11 & 7k \\
\href{https://github.com/JuliaStats/Distributions.jl}{\texttt{distributions-jl}} & \texttt{jl} & {\raggedright Probability distributions and statistical functions for Julia} & {\raggedright \href{https://github.com/JuliaStats/Distributions.jl/blob/master/LICENSE.md}{\texttt{MIT}}} & 9 & 16k \\
\href{https://github.com/dlang/dmd}{\texttt{dmd}} & \texttt{d} & {\raggedright Compiler for the D programming language} & {\raggedright \href{https://github.com/dlang/dmd/blob/master/LICENSE.txt}{\texttt{BSL-1.0}}} & 112 & 609k \\
\href{https://github.com/apache/dolphinscheduler}{\texttt{dolphinscheduler}} & \texttt{java} & {\raggedright Distributed workflow orchestration platform} & {\raggedright \href{https://github.com/apache/dolphinscheduler/blob/dev/LICENSE}{\texttt{Apache-2.0}}} & 10 & 183k \\
\href{https://github.com/duckdb/duckdb}{\texttt{duckdb}} & \texttt{cpp} & {\raggedright In-process analytical SQL database} & {\raggedright \href{https://github.com/duckdb/duckdb/blob/v2.0-cyanoptera/LICENSE}{\texttt{MIT}}} & 109 & 723k \\
\href{https://github.com/labstack/echo}{\texttt{echo}} & \texttt{go} & {\raggedright High-performance web framework for Go} & {\raggedright \href{https://github.com/labstack/echo/blob/master/LICENSE}{\texttt{MIT}}} & 6 & 8k \\
\href{https://github.com/elixir-lang/elixir}{\texttt{elixir}} & \texttt{ex} & {\raggedright Functional programming language for the Erlang VM} & {\raggedright \href{https://github.com/elixir-lang/elixir/blob/main/LICENSE}{\texttt{Apache-2.0}}} & 47 & 146k \\
\href{https://github.com/valyala/fasthttp}{\texttt{fasthttp}} & \texttt{go} & {\raggedright High-performance HTTP library for Go} & {\raggedright \href{https://github.com/valyala/fasthttp/blob/master/LICENSE}{\texttt{MIT}}} & 9 & 24k \\
\href{https://github.com/formulajs/formulajs}{\texttt{formulajs}} & \texttt{js} & {\raggedright JavaScript implementations of spreadsheet functions} & {\raggedright \href{https://github.com/formulajs/formulajs/blob/master/LICENSE}{\texttt{MIT}}} & 9 & 13k \\
\href{https://github.com/dotnet/fsharp}{\texttt{fsharp}} & \texttt{fs} & {\raggedright F\# compiler, core library, and development tools} & {\raggedright \href{https://github.com/dotnet/fsharp/blob/main/License.txt}{\texttt{MIT}}} & 28 & 336k \\
\href{https://github.com/OSGeo/gdal}{\texttt{gdal}} & \texttt{cpp} & {\raggedright Raster and vector geospatial data translation library} & {\raggedright \href{https://github.com/OSGeo/gdal/blob/master/LICENSE.TXT}{\texttt{MIT}}} & 72 & 1,731k \\
\href{https://github.com/gin-gonic/gin}{\texttt{gin}} & \texttt{go} & {\raggedright HTTP web framework for Go} & {\raggedright \href{https://github.com/gin-gonic/gin/blob/master/LICENSE}{\texttt{MIT}}} & 11 & 7k \\
\href{https://github.com/GitoxideLabs/gitoxide}{\texttt{gitoxide}} & \texttt{rs} & {\raggedright Git implementation written in Rust} & {\raggedright \href{https://github.com/GitoxideLabs/gitoxide/blob/main/LICENSE-APACHE}{\texttt{Apache-2.0}}; \href{https://github.com/GitoxideLabs/gitoxide/blob/main/LICENSE-MIT}{\texttt{MIT}}} & 9 & 160k \\
\href{https://github.com/golang/go}{\texttt{go}} & \texttt{go} & {\raggedright Go programming language and toolchain} & {\raggedright \href{https://github.com/golang/go/blob/master/LICENSE}{\texttt{BSD-3-Clause}}} & 98 & 1,612k \\
\href{https://github.com/go-git/go-git}{\texttt{go-git}} & \texttt{go} & {\raggedright Extensible Git implementation written in Go} & {\raggedright \href{https://github.com/go-git/go-git/blob/main/LICENSE}{\texttt{Apache-2.0}}} & 23 & 40k \\
\href{https://github.com/benhoyt/goawk}{\texttt{goawk}} & \texttt{go} & {\raggedright AWK interpreter with CSV support written in Go} & {\raggedright \href{https://github.com/benhoyt/goawk/blob/master/LICENSE.txt}{\texttt{MIT}}} & 14 & 10k \\
\href{https://github.com/godotengine/godot}{\texttt{godot}} & \texttt{cpp} & {\raggedright Cross-platform 2D and 3D game engine} & {\raggedright \href{https://github.com/godotengine/godot/blob/master/LICENSE.txt}{\texttt{MIT}}} & 17 & 1,275k \\
\href{https://github.com/go-gorm/gorm}{\texttt{gorm}} & \texttt{go} & {\raggedright Object-relational mapper for Go} & {\raggedright \href{https://github.com/go-gorm/gorm/blob/master/LICENSE}{\texttt{MIT}}} & 19 & 11k \\
\href{https://github.com/google/gson}{\texttt{gson}} & \texttt{java} & {\raggedright JSON serialization library for Java} & {\raggedright \href{https://github.com/google/gson/blob/main/LICENSE}{\texttt{Apache-2.0}}} & 6 & 17k \\
\href{https://github.com/HDFGroup/hdf5}{\texttt{hdf5}} & \texttt{c} & {\raggedright Library and file format for hierarchical scientific data} & {\raggedright \href{https://github.com/HDFGroup/hdf5/blob/develop/LICENSE}{\texttt{BSD-3-Clause}}} & 14 & 615k \\
\href{https://github.com/idris-lang/Idris2}{\texttt{idris2}} & \texttt{idr} & {\raggedright Dependently typed functional programming language} & {\raggedright \href{https://github.com/idris-lang/Idris2/blob/main/LICENSE}{\texttt{BSD-3-Clause}}} & 13 & 113k \\
\href{https://github.com/isomorphic-git/isomorphic-git}{\texttt{isomorphic-git}} & \texttt{js} & {\raggedright Git implementation for JavaScript runtimes} & {\raggedright \href{https://github.com/isomorphic-git/isomorphic-git/blob/main/LICENSE.md}{\texttt{MIT}}} & 13 & 15k \\
\href{https://github.com/pycqa/isort}{\texttt{isort}} & \texttt{py} & {\raggedright Import sorter for Python} & {\raggedright \href{https://github.com/PyCQA/isort/blob/main/LICENSE}{\texttt{MIT}}} & 11 & 11k \\
\href{https://github.com/jerryscript-project/jerryscript}{\texttt{jerryscript}} & \texttt{c} & {\raggedright Lightweight JavaScript engine for embedded systems} & {\raggedright \href{https://github.com/jerryscript-project/jerryscript/blob/master/LICENSE}{\texttt{Apache-2.0}}} & 15 & 134k \\
\href{https://github.com/jestjs/jest}{\texttt{jest}} & \texttt{ts} & {\raggedright JavaScript testing framework} & {\raggedright \href{https://github.com/jestjs/jest/blob/main/LICENSE}{\texttt{MIT}}} & 20 & 52k \\
\href{https://github.com/keycloak/keycloak}{\texttt{keycloak}} & \texttt{java} & {\raggedright Identity and access management server} & {\raggedright \href{https://github.com/keycloak/keycloak/blob/main/LICENSE.txt}{\texttt{Apache-2.0}}} & 64 & 573k \\
\href{https://github.com/leanprover/lean4}{\texttt{lean4}} & \texttt{lean} & {\raggedright Programming language and interactive theorem prover} & {\raggedright \href{https://github.com/leanprover/lean4/blob/master/LICENSE}{\texttt{Apache-2.0}}} & 63 & 233k \\
\href{https://github.com/libexpat/libexpat}{\texttt{libexpat}} & \texttt{c} & {\raggedright Streaming XML parser written in C} & {\raggedright \href{https://github.com/libexpat/libexpat/blob/master/COPYING}{\texttt{MIT}}} & 9 & 19k \\
\href{https://github.com/moment/luxon}{\texttt{luxon}} & \texttt{js} & {\raggedright Date and time library for JavaScript} & {\raggedright \href{https://github.com/moment/luxon/blob/master/LICENSE.md}{\texttt{MIT}}} & 14 & 7k \\
\href{https://github.com/mui/material-ui}{\texttt{material-ui}} & \texttt{js} & {\raggedright React component library implementing Material Design} & {\raggedright \href{https://github.com/mui/material-ui/blob/master/LICENSE}{\texttt{MIT}}} & 39 & 105k \\
\href{https://github.com/leanprover-community/mathlib4}{\texttt{mathlib4}} & \texttt{lean} & {\raggedright Mathematics library for the Lean theorem prover} & {\raggedright \href{https://github.com/leanprover-community/mathlib4/blob/master/LICENSE}{\texttt{Apache-2.0}}} & 14 & 1,865k \\
\href{https://github.com/matplotlib/matplotlib}{\texttt{matplotlib}} & \texttt{py} & {\raggedright Plotting library for Python} & {\raggedright \href{https://github.com/matplotlib/matplotlib/blob/main/LICENSE/LICENSE}{\texttt{PSF-2.0}}} & 78 & 142k \\
\href{https://github.com/micropython/micropython}{\texttt{micropython}} & \texttt{c} & {\raggedright Python implementation for microcontrollers} & {\raggedright \href{https://github.com/micropython/micropython/blob/master/LICENSE}{\texttt{MIT}}} & 22 & 423k \\
\href{https://github.com/SciML/ModelingToolkit.jl}{\texttt{modelingtoolkit-jl}} & \texttt{jl} & {\raggedright Symbolic modeling system for scientific computing in Julia} & {\raggedright \href{https://github.com/SciML/ModelingToolkit.jl/blob/master/LICENSE.md}{\texttt{MIT}}} & 10 & 23k \\
\href{https://github.com/Automattic/mongoose}{\texttt{mongoose}} & \texttt{js} & {\raggedright Object data modeling library for MongoDB and Node.js} & {\raggedright \href{https://github.com/Automattic/mongoose/blob/master/LICENSE.md}{\texttt{MIT}}} & 70 & 55k \\
\href{https://github.com/mpmath/mpmath}{\texttt{mpmath}} & \texttt{py} & {\raggedright Arbitrary-precision numerical computing library for Python} & {\raggedright \href{https://github.com/mpmath/mpmath/blob/master/LICENSE}{\texttt{BSD-3-Clause}}} & 16 & 48k \\
\href{https://github.com/python/mypy}{\texttt{mypy}} & \texttt{py} & {\raggedright Static type checker for Python} & {\raggedright \href{https://github.com/python/mypy/blob/master/LICENSE}{\texttt{MIT}}} & 91 & 137k \\
\href{https://github.com/alibaba/nacos}{\texttt{nacos}} & \texttt{java} & {\raggedright Service discovery and configuration management platform} & {\raggedright \href{https://github.com/alibaba/nacos/blob/develop/LICENSE}{\texttt{Apache-2.0}}} & 12 & 222k \\
\href{https://github.com/netty/netty}{\texttt{netty}} & \texttt{java} & {\raggedright Event-driven network application framework for Java} & {\raggedright \href{https://github.com/netty/netty/blob/4.2/LICENSE.txt}{\texttt{Apache-2.0}}} & 96 & 254k \\
\href{https://github.com/networkx/networkx}{\texttt{networkx}} & \texttt{py} & {\raggedright Graph analysis library for Python} & {\raggedright \href{https://github.com/networkx/networkx/blob/main/LICENSE.txt}{\texttt{BSD-3-Clause}}} & 18 & 98k \\
\href{https://github.com/nim-lang/Nim}{\texttt{nim}} & \texttt{nim} & {\raggedright Statically typed systems programming language} & {\raggedright \href{https://github.com/nim-lang/Nim/blob/devel/copying.txt}{\texttt{MIT}}} & 192 & 218k \\
\href{https://github.com/numpy/numpy}{\texttt{numpy}} & \texttt{py} & {\raggedright N-dimensional array library for numerical computing in Python} & {\raggedright \href{https://github.com/numpy/numpy/blob/main/LICENSE.txt}{\texttt{BSD-3-Clause}}} & 131 & 406k \\
\href{https://github.com/nuxt/nuxt}{\texttt{nuxt}} & \texttt{ts} & {\raggedright Full-stack web framework for Vue} & {\raggedright \href{https://github.com/nuxt/nuxt/blob/main/LICENSE}{\texttt{MIT}}} & 6 & 37k \\
\href{https://github.com/OpenAPITools/openapi-generator}{\texttt{openapi-generator}} & \texttt{mustache} & {\raggedright Generator for API clients, servers, and documentation from OpenAPI specs} & {\raggedright \href{https://github.com/OpenAPITools/openapi-generator/blob/master/LICENSE}{\texttt{Apache-2.0}}} & 31 & 434k \\
\href{https://github.com/opencv/opencv}{\texttt{opencv}} & \texttt{cpp} & {\raggedright Computer vision and image processing library} & {\raggedright \href{https://github.com/opencv/opencv/blob/5.x/LICENSE}{\texttt{Apache-2.0}}} & 47 & 553k \\
\href{https://github.com/optuna/optuna}{\texttt{optuna}} & \texttt{py} & {\raggedright Hyperparameter optimization framework} & {\raggedright \href{https://github.com/optuna/optuna/blob/master/LICENSE}{\texttt{MIT}}} & 7 & 28k \\
\href{https://github.com/oxc-project/oxc}{\texttt{oxc}} & \texttt{rs} & {\raggedright High-performance JavaScript and TypeScript toolchain} & {\raggedright \href{https://github.com/oxc-project/oxc/blob/main/LICENSE}{\texttt{MIT}}} & 6 & 521k \\
\href{https://github.com/pandas-dev/pandas}{\texttt{pandas}} & \texttt{py} & {\raggedright Data analysis library with labeled tabular structures} & {\raggedright \href{https://github.com/pandas-dev/pandas/blob/main/LICENSE}{\texttt{BSD-3-Clause}}} & 228 & 239k \\
\href{https://github.com/sdispater/pendulum}{\texttt{pendulum}} & \texttt{py} & {\raggedright Date and time library for Python} & {\raggedright \href{https://github.com/python-pendulum/pendulum/blob/master/LICENSE}{\texttt{MIT}}} & 9 & 12k \\
\href{https://github.com/php/php-src}{\texttt{php-src}} & \texttt{c} & {\raggedright PHP language interpreter} & {\raggedright \href{https://github.com/php/php-src/blob/master/LICENSE}{\texttt{BSD-3-Clause}}} & 26 & 1,570k \\
\href{https://github.com/PHPOffice/PhpSpreadsheet}{\texttt{phpspreadsheet}} & \texttt{php} & {\raggedright Spreadsheet reader and writer for PHP} & {\raggedright \href{https://github.com/PHPOffice/PhpSpreadsheet/blob/master/LICENSE}{\texttt{MIT}}} & 9 & 98k \\
\href{https://github.com/hgrecco/pint}{\texttt{pint}} & \texttt{py} & {\raggedright Physical quantities and unit conversion library for Python} & {\raggedright \href{https://github.com/hgrecco/pint/blob/master/LICENSE}{\texttt{BSD-3-Clause}}} & 13 & 11k \\
\href{https://github.com/prettier/prettier}{\texttt{prettier}} & \texttt{js} & {\raggedright Opinionated code formatter for web languages} & {\raggedright \href{https://github.com/prettier/prettier/blob/main/LICENSE}{\texttt{MIT}}} & 34 & 39k \\
\href{https://github.com/OSGeo/PROJ}{\texttt{proj}} & \texttt{cpp} & {\raggedright Cartographic projection and coordinate transformation library} & {\raggedright \href{https://github.com/OSGeo/PROJ/blob/master/COPYING}{\texttt{MIT}}} & 21 & 141k \\
\href{https://github.com/apache/pulsar}{\texttt{pulsar}} & \texttt{java} & {\raggedright Distributed publish-subscribe messaging platform} & {\raggedright \href{https://github.com/apache/pulsar/blob/master/LICENSE}{\texttt{Apache-2.0}}} & 19 & 379k \\
\href{https://github.com/quarkusio/quarkus}{\texttt{quarkus}} & \texttt{java} & {\raggedright Kubernetes-native Java application framework} & {\raggedright \href{https://github.com/quarkusio/quarkus/blob/main/LICENSE}{\texttt{Apache-2.0}}} & 72 & 670k \\
\href{https://github.com/qutip/qutip}{\texttt{qutip}} & \texttt{py} & {\raggedright Quantum dynamics simulation library for Python} & {\raggedright \href{https://github.com/qutip/qutip/blob/master/LICENSE.txt}{\texttt{BSD-3-Clause}}} & 10 & 74k \\
\href{https://github.com/rails/rails}{\texttt{rails}} & \texttt{rb} & {\raggedright Web application framework for Ruby} & {\raggedright \href{https://github.com/rails/rails/blob/main/MIT-LICENSE}{\texttt{MIT}}} & 140 & 157k \\
\href{https://github.com/remix-run/react-router}{\texttt{react-router}} & \texttt{ts} & {\raggedright Routing library for React} & {\raggedright \href{https://github.com/remix-run/react-router/blob/main/LICENSE.md}{\texttt{MIT}}} & 14 & 29k \\
\href{https://github.com/dotnet/roslyn}{\texttt{roslyn}} & \texttt{cs} & {\raggedright C\# and Visual Basic compilers and code analysis APIs} & {\raggedright \href{https://github.com/dotnet/roslyn/blob/main/License.txt}{\texttt{MIT}}} & 31 & 1,591k \\
\href{https://github.com/astral-sh/ruff}{\texttt{ruff}} & \texttt{rs} & {\raggedright Python linter and code formatter written in Rust} & {\raggedright \href{https://github.com/astral-sh/ruff/blob/main/LICENSE}{\texttt{MIT}}} & 52 & 318k \\
\href{https://github.com/reactivex/rxjava}{\texttt{rxjava}} & \texttt{java} & {\raggedright Reactive programming library for the JVM} & {\raggedright \href{https://github.com/ReactiveX/RxJava/blob/4.x/LICENSE}{\texttt{Apache-2.0}}} & 26 & 166k \\
\href{https://github.com/scikit-learn/scikit-learn}{\texttt{scikit-learn}} & \texttt{py} & {\raggedright Machine learning library for Python} & {\raggedright \href{https://github.com/scikit-learn/scikit-learn/blob/main/COPYING}{\texttt{BSD-3-Clause}}} & 80 & 145k \\
\href{https://github.com/scipy/scipy}{\texttt{scipy}} & \texttt{py} & {\raggedright Scientific computing library for Python} & {\raggedright \href{https://github.com/scipy/scipy/blob/main/LICENSE.txt}{\texttt{BSD-3-Clause}}} & 136 & 517k \\
\href{https://github.com/shapely/shapely}{\texttt{shapely}} & \texttt{py} & {\raggedright Geometric object manipulation and analysis library for Python} & {\raggedright \href{https://github.com/shapely/shapely/blob/main/LICENSE.txt}{\texttt{BSD-3-Clause}}} & 9 & 20k \\
\href{https://github.com/apache/shardingsphere}{\texttt{shardingsphere}} & \texttt{java} & {\raggedright Distributed SQL middleware for data sharding and governance} & {\raggedright \href{https://github.com/apache/shardingsphere/blob/master/LICENSE}{\texttt{Apache-2.0}}} & 15 & 328k \\
\href{https://github.com/tobymao/sqlglot}{\texttt{sqlglot}} & \texttt{py} & {\raggedright SQL parser, optimizer, and transpiler for Python} & {\raggedright \href{https://github.com/tobymao/sqlglot/blob/main/LICENSE}{\texttt{MIT}}} & 130 & 51k \\
\href{https://github.com/stylelint/stylelint}{\texttt{stylelint}} & \texttt{js} & {\raggedright Configurable linter for CSS and related languages} & {\raggedright \href{https://github.com/stylelint/stylelint/blob/main/LICENSE}{\texttt{MIT}}} & 54 & 54k \\
\href{https://github.com/sunpy/sunpy}{\texttt{sunpy}} & \texttt{py} & {\raggedright Solar physics data analysis library for Python} & {\raggedright \href{https://github.com/sunpy/sunpy/blob/main/LICENSE.rst}{\texttt{BSD-2-Clause}}} & 15 & 46k \\
\href{https://github.com/svg/svgo}{\texttt{svgo}} & \texttt{js} & {\raggedright SVG optimizer} & {\raggedright \href{https://github.com/svg/svgo/blob/main/LICENSE}{\texttt{MIT}}} & 11 & 13k \\
\href{https://github.com/swc-project/swc}{\texttt{swc}} & \texttt{rs} & {\raggedright JavaScript and TypeScript compiler written in Rust} & {\raggedright \href{https://github.com/swc-project/swc/blob/main/LICENSE}{\texttt{Apache-2.0}}} & 19 & 233k \\
\href{https://github.com/symfony/symfony}{\texttt{symfony}} & \texttt{php} & {\raggedright Web application framework for PHP} & {\raggedright \href{https://github.com/symfony/symfony/blob/8.2/LICENSE}{\texttt{MIT}}} & 149 & 1,228k \\
\href{https://github.com/dtolnay/syn}{\texttt{syn}} & \texttt{rs} & {\raggedright Parser for Rust source code} & {\raggedright \href{https://github.com/dtolnay/syn/blob/master/LICENSE-APACHE}{\texttt{Apache-2.0}}} & 13 & 33k \\
\href{https://github.com/tarantool/tarantool}{\texttt{tarantool}} & \texttt{c} & {\raggedright In-memory database and application server} & {\raggedright \href{https://github.com/tarantool/tarantool/blob/master/LICENSE}{\texttt{BSD-2-Clause}}} & 40 & 312k \\
\href{https://github.com/terser/terser}{\texttt{terser}} & \texttt{js} & {\raggedright JavaScript parser, mangler, and compressor} & {\raggedright \href{https://github.com/terser/terser/blob/master/LICENSE}{\texttt{BSD-2-Clause}}} & 6 & 29k \\
\href{https://github.com/stretchr/testify}{\texttt{testify}} & \texttt{go} & {\raggedright Assertions and mocking toolkit for Go tests} & {\raggedright \href{https://github.com/stretchr/testify/blob/master/LICENSE}{\texttt{MIT}}} & 9 & 10k \\
\href{https://github.com/mrdoob/three.js}{\texttt{three-js}} & \texttt{js} & {\raggedright 3D graphics library for JavaScript} & {\raggedright \href{https://github.com/mrdoob/three.js/blob/dev/LICENSE}{\texttt{MIT}}} & 9 & 159k \\
\href{https://github.com/tinygrad/tinygrad}{\texttt{tinygrad}} & \texttt{py} & {\raggedright Lightweight deep learning framework} & {\raggedright \href{https://github.com/tinygrad/tinygrad/blob/master/LICENSE}{\texttt{MIT}}} & 15 & 170k \\
\href{https://github.com/tokio-rs/tokio}{\texttt{tokio}} & \texttt{rs} & {\raggedright Asynchronous runtime for Rust} & {\raggedright \href{https://github.com/tokio-rs/tokio/blob/master/LICENSE}{\texttt{MIT}}} & 12 & 108k \\
\href{https://github.com/trinodb/trino}{\texttt{trino}} & \texttt{java} & {\raggedright Distributed SQL query engine} & {\raggedright \href{https://github.com/trinodb/trino/blob/master/LICENSE}{\texttt{Apache-2.0}}} & 43 & 724k \\
\href{https://github.com/vlang/v}{\texttt{v}} & \texttt{v} & {\raggedright Compiled programming language focused on simplicity and speed} & {\raggedright \href{https://github.com/vlang/v/blob/master/LICENSE}{\texttt{MIT}}} & 239 & 262k \\
\href{https://github.com/webpack/webpack}{\texttt{webpack}} & \texttt{js} & {\raggedright Module bundler for JavaScript and web assets} & {\raggedright \href{https://github.com/webpack/webpack/blob/main/LICENSE}{\texttt{MIT}}} & 35 & 130k \\
\href{https://github.com/pydata/xarray}{\texttt{xarray}} & \texttt{py} & {\raggedright Labeled multidimensional array library for Python} & {\raggedright \href{https://github.com/pydata/xarray/blob/main/LICENSE}{\texttt{Apache-2.0}}} & 31 & 60k \\
\href{https://github.com/Z3Prover/z3}{\texttt{z3}} & \texttt{cpp} & {\raggedright SMT theorem prover} & {\raggedright \href{https://github.com/Z3Prover/z3/blob/master/LICENSE.txt}{\texttt{MIT}}} & 11 & 691k \\
\end{longtable}
\endgroup
 \clearpage
\newpage
\subsection{SZZ coverage curves for all repositories}
\label{appx:task_index:szz}
\begin{figure}[h]
    \centering
    \includegraphics[width=\textwidth]{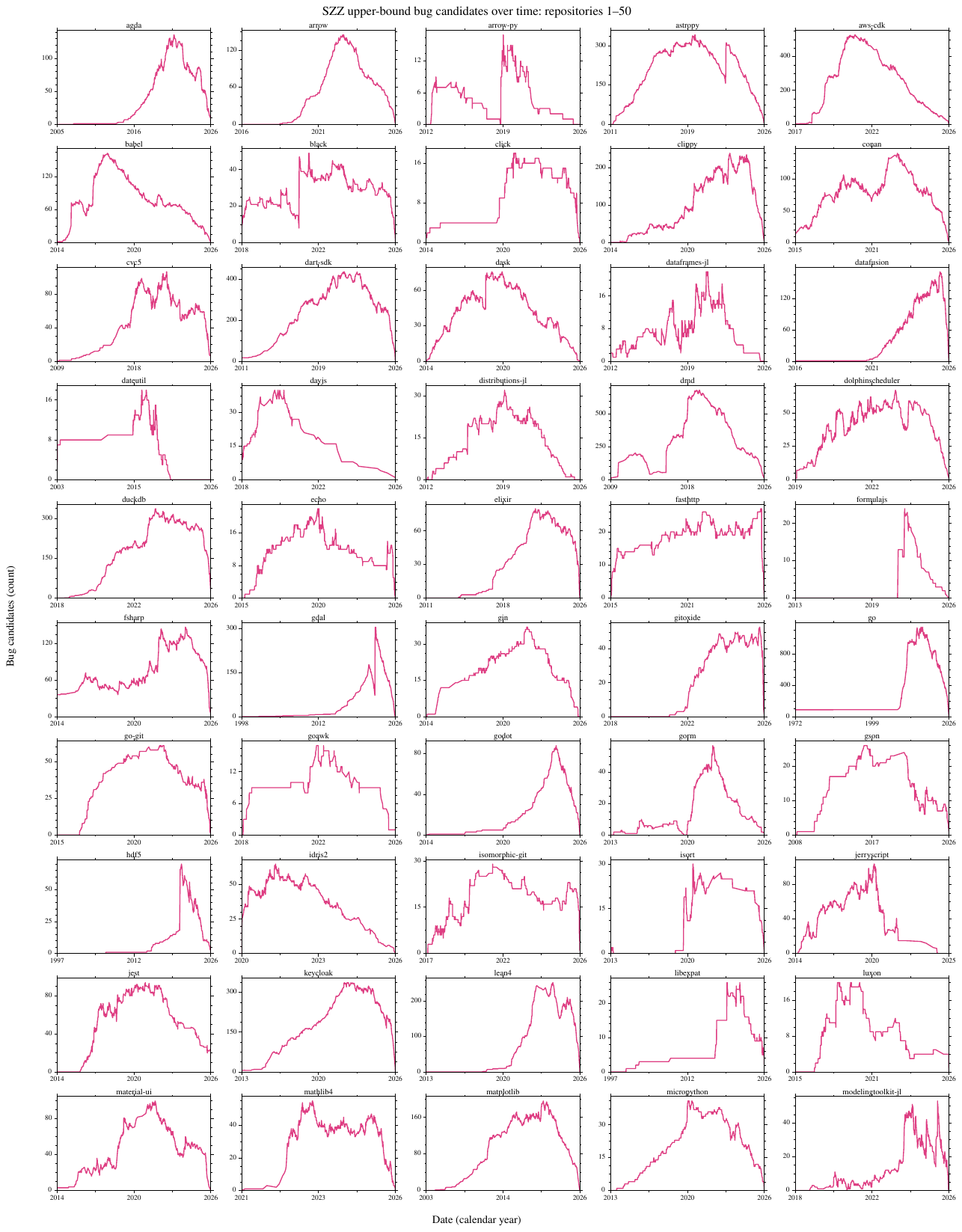}
    \caption{Estimates of the number of concurrent bug instance candidates across the histories of
    benchmark repositories 1--50, in the order listed in Table~\ref{tab:repository-index}.}
    \label{fig:repository-szz-01-50}
\end{figure}

\begin{figure}[h]
    \centering
    \includegraphics[width=\textwidth]{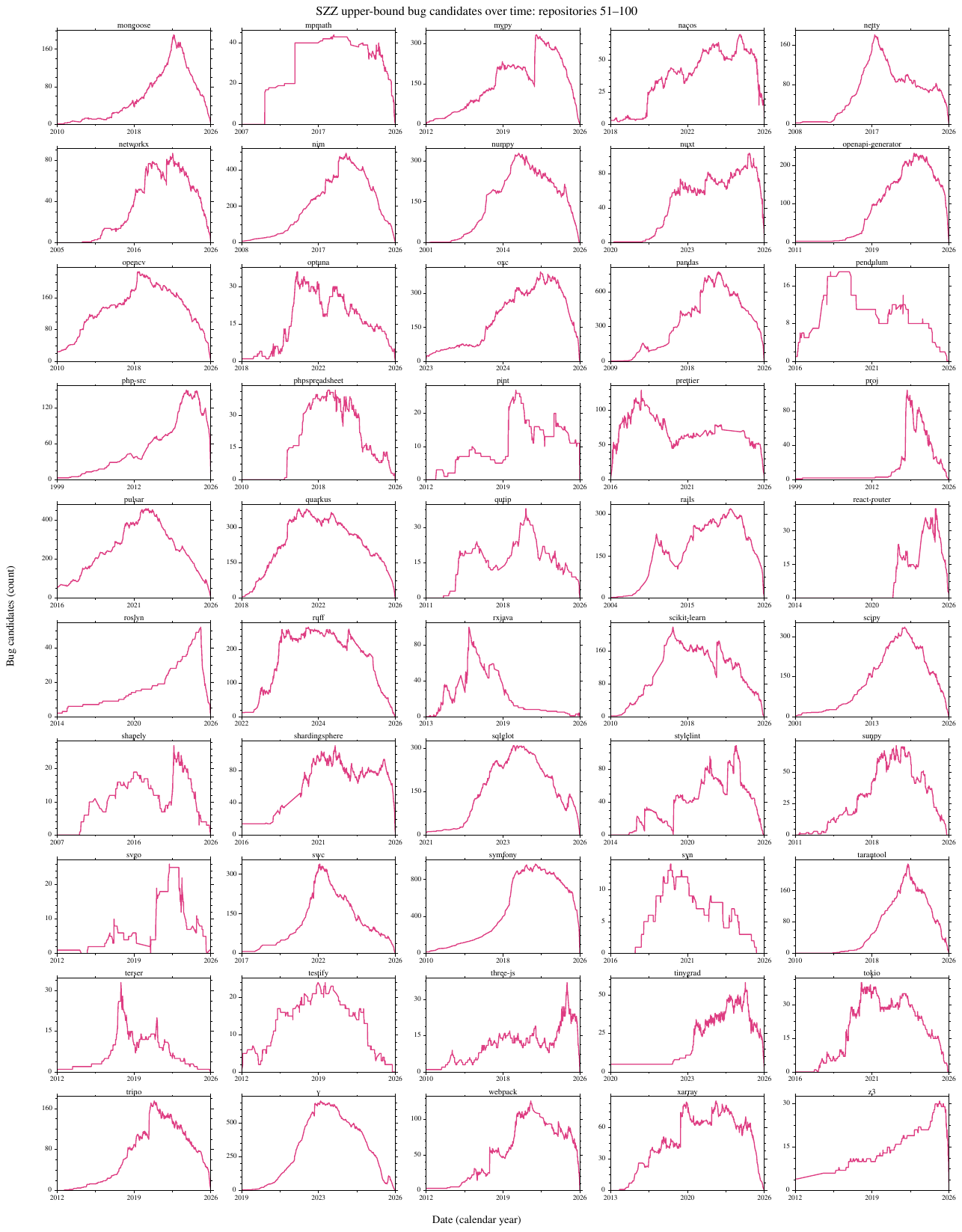}
    \caption{Estimates of the number of concurrent bug instance candidates across the histories of
    benchmark repositories 51--100, in the order listed in Table~\ref{tab:repository-index}.}
    \label{fig:repository-szz-51-100}
\end{figure}

Figures~\ref{fig:repository-szz-01-50} and~\ref{fig:repository-szz-51-100} show the coverage curves from the SZZ algorithm (\S\ref{sec:benchmark:construction}) for all benchmark repositories.

 \clearpage
\newpage
\section{Prompts}
\label{appx:prompts}

This appendix reproduces the prompts used to construct the benchmark, instruct the
benchmark agent, and support the validation studies reported in the paper. Text marked
\texttt{[SYSTEM]} and \texttt{[USER TEMPLATE]} identifies message roles and was not
sent as part of the message. Double-braced fields are template variables replaced with
candidate-specific content at runtime.

\subsection{Inference prompt}
\label{appx:prompts:inference}

The no-information condition underlies the main results in \S\ref{sec:results}; Appendix~\ref{appx:exp_setup}
describes the surrounding inference environment. The
square-bracketed role labels are editorial and were not sent to the model. The values
shown as \verb|<repository path>| and in the \verb|<system_information>| block vary
with the execution environment; all other text, capitalization, and emphasis are
preserved from the prompt.

\verbpromptboxfile{assets/inference_prompt_no_info.txt}

\subsubsection{Ablation-specific changes}

These are the ablations reported in \S\ref{sec:results:ablations}. All use the same system message. The snippets below show only how the user
message differs from the no-information condition.

\verbpromptboxfile{assets/inference_prompt_count_snippet.txt}

The \texttt{counts} condition stops there. The remaining conditions append one of the
following snippets.

\textbf{File names (\texttt{filenames}).}

\verbpromptboxfile{assets/inference_prompt_filenames_snippet.txt}

\textbf{Line ranges (\texttt{line\_ranges}).}

\verbpromptboxfile{assets/inference_prompt_line_ranges_snippet.txt}

\textbf{Issue reports (\texttt{issue\_texts}).}

\verbpromptboxfile{assets/inference_prompt_issue_texts_snippet.txt}

\textbf{Line ranges and issue reports (\texttt{all}).}

\verbpromptboxfile{assets/inference_prompt_all_snippet.txt}

The \texttt{line\_ranges\_separate} and \texttt{all\_separate} variants use the
corresponding snippets above but launch one run per bug, so each prompt contains only
that bug's hints. The \texttt{no\_info\_sequential} variant uses exactly the
no-information user message shown above; it instead changes execution by launching ten
fresh agents in sequence on the same working tree.

\subsection{Benchmark-construction annotation prompts}
\label{appx:prompts:annotation}

The first prompt is a high-recall pre-filter. A candidate rejected here does not
proceed through the remaining construction pipeline. The other four prompts assess
reproduced candidates; each must receive unanimous acceptance from the annotation
panel. Together, these prompts enforce the LM-assessed parts of the criteria in
Appendix~\ref{appx:benchmark:subtask-criteria}; Appendix~\ref{appx:benchmark:quality-control}
describes how they fit into the construction pipeline.

\subsubsection{Pre-filtering for PRs that fix bugs (\texttt{is\_bug\_lm})}

\verbpromptboxfile{assets/annotation_prompt_is_bug_lm.txt}

\subsubsection{Ensuring that a bug candidate corresponds to a single unit of work (\texttt{is\_single\_bug})}
\label{appx:prompts:is-single-bug}

\verbpromptboxfile{assets/annotation_prompt_is_single_bug.txt}

\subsubsection{Ensuring that a bug candidate is discoverable (\texttt{is\_discoverable\_bug})}
\label{appx:prompts:is-discoverable-bug}

\verbpromptboxfile{assets/annotation_prompt_is_discoverable_bug.txt}

\subsubsection{Ensuring that a bug instance candidate has appropriate tests (\texttt{appropriate\_tests})}
\label{appx:prompts:appropriate-tests}

\verbpromptboxfile{assets/annotation_prompt_appropriate_tests.txt}

\subsubsection{Ensuring that a PR leaves existing test expectations unchanged
(\texttt{preserves\_existing\_test\_expectations})}
\label{appx:prompts:preserves-existing-test-expectations}

\verbpromptboxfile{assets/annotation_prompt_preserves_existing_test_expectations.txt}

\subsection{Validation-study annotation prompts}
\label{appx:prompts:validation}

These reviews were run after benchmark selection. They support the descriptive and
feasibility analyses in Appendix~\ref{appx:benchmark:validation}. Each reviewer
saw the base repository and a blind input packet, without the issue, gold fix,
history, or earlier annotations.

\subsubsection{Bug characteristics and evidence}
\label{appx:prompts:bug-characteristics}

\verbpromptboxfile{assets/annotation_prompt_bug_characteristics.txt}

\subsubsection{Support for test expectations}
\label{appx:prompts:test-expectation-support}

\verbpromptboxfile{assets/annotation_prompt_test_expectation_support.txt}

\subsubsection{Compatibility within a repository}
\label{appx:prompts:task-compatibility}

\verbpromptboxfile{assets/annotation_prompt_task_compatibility.txt}

\subsection{Additional annotation examples}
\label{appx:prompts:annotation-examples}

These examples show real candidates at the high-recall bug-fix pre-filter.
The examples for the four annotations that determine final benchmark membership
appear with the corresponding criteria in
Appendices~\ref{appx:benchmark:discoverability}--\ref{appx:benchmark:other-quality}.

\subsubsection{Bug fix pre-filter (\texttt{is\_bug\_lm})}
\annotationrow
  {\annotationexample[PromptGreen]{PlainText}{Pull request text (truncated)}{Accepted: Mastodon~\#31144}
    {assets/annotation_example_is_bug_lm_accept1.txt}
    {The title names the exact runtime exception and the body links the issue that reported it.}}
  {\annotationexample[PromptRed]{PlainText}{Commit message (truncated)}{Rejected: LLVM~\#102723}
    {assets/annotation_example_is_bug_lm_reject1.txt}
    {The commit message describes new feature work rather than a repair of broken behavior.}}

\annotationrow
  {\annotationexample[PromptGreen]{PlainText}{Pull request text (truncated)}{Accepted: NumPy~\#12566}
    {assets/annotation_example_is_bug_lm_accept2.txt}
    {An explicit \texttt{BUG:} prefix, a named crash, and a linked tracker issue leave nothing to interpret.}}
  {\annotationexample[PromptRed]{PlainText}{Commit message (truncated)}{Rejected: LLVM~\#115851}
    {assets/annotation_example_is_bug_lm_reject2.txt}
    {Framed as an annotation that consumers ``would benefit'' from, not as a fix for a wrong result.}}
 
\end{document}